\documentclass[3p,twocolumn]{elsarticle}

\usepackage{lineno,hyperref}
\usepackage{subfigure}
\usepackage{amsmath}

\usepackage{amssymb}
\usepackage{amsthm}
\usepackage{epsfig}
\usepackage[usenames]{color}
\usepackage{algorithm}
\usepackage{algorithmic}

\usepackage{bm}
\usepackage{array}

\modulolinenumbers[5]

\journal{International Journal of Solids and Structures}

\graphicspath{{fig/}{../fig/}}

\newcommand{\mb}[1]{\ensuremath{\mathbf{#1}}}
\newcommand{\mbd}[1]{\textsf{\textbf{#1}}}
\newcommand{\mbds}[1] {\pmb{\bm{#1}}}

\begin{document}

\begin{frontmatter}

\title{Geometrically exact beam elements and smooth contact schemes for the modeling of fiber-based materials and structures}

\author[mechanosynth,lnm]{Christoph Meier\corref{cor1}}
\ead{meier@lnm.mw.tum.de}
\author[lnm]{Maximilian J. Grill}
\author[lnm]{Wolfgang A. Wall}
\author[lnm]{Alexander Popp}

\address[mechanosynth]{Mechanosynthesis Group, Massachusetts Institute of Technology, 77 Massachusetts Avenue, Cambridge, 02139, MA, USA}
\address[lnm]{Institute for Computational Mechanics, Technical University of Munich, Boltzmannstrasse 15, D--85748 Garching b. M\"unchen, Germany}

\cortext[cor1]{Corresponding author}

\begin{abstract}
This work focuses on finite element formulations for the accurate modeling and efficient simulation of the implicit dynamics of slender fiber- or rod-like components and their contact interaction when being embedded in complex systems of fiber-based materials and structures. Recently, the authors have proposed a novel all-angle beam contact (ABC) formulation that combines the advantages of existing point and line contact models in a variationally consistent manner. However, the ABC formulation has so far only been applied in combination with a special torsion-free beam model, which yields a very simple and efficient finite element formulation, but which is restricted to initially straight beams with isotropic cross-sections. In order to abstain from these restrictions, the current work combines the ABC formulation with a geometrically exact Kirchhoff-Love beam element formulation that is capable of treating even the most general cases of slender beam problems in terms of initial geometry and external loads. While the neglect of shear deformation that is inherent to this formulation has been shown to provide considerable numerical advantages in the range of high beam slenderness ratios, alternative shear-deformable beam models are required for examples with thick beams. For that reason, the current contribution additionally proposes a novel geometrically exact beam element based on the Simo-Reissner theory. Similar to the torsion-free and the Kirchhoff-Love beam elements, also this Simo-Reissner element is based on a $C^1$-continuous Hermite interpolation of the beam centerline, which will allow for smooth contact kinematics. For this Hermitian Simo-Reissner element, a consistent spatial convergence behavior as well as the successful avoidance of membrane and shear locking will be demonstrated numerically. All in all, the combination of the ABC formulation with these different beam element variants (i.e.~the torsion-free element, the Kirchhoff-Love element and the Simo-Reissner element) results in a very flexible and modular simulation framework that allows to choose the optimal element formulation for any given application in terms of accuracy, efficiency and robustness. Based on several practically relevant examples, the different variants are compared numerically, and, eventually, a general recommendation concerning the optimal choice of beam elements is made.
\end{abstract}


\end{frontmatter}


\section{Introduction}
\label{sec:intro}

Highly slender fiber- or rod-like components represent essential constituents of mechanical systems in countless fields of application and scientific disciplines such as mechanical engineering, biomedical engineering, materials science and bio- or molecular physics. Examples are high-tensile industrial webbings, fiber-reinforced composite materials,  fibrous materials with tailored porosity, synthetic polymer materials or also cellulose fibers determining the characteristics of paper~\cite{durville2010,borodulina_and_kulachenko2012,kulachenko2012b,gadot_and_orgeas2015,rodney_and_orgeas2016}. On entirely different time and length scales, such slender components are relevant when analyzing the supercoiling process of DNA strands, the characteristics of carbon nanotubes or the Brownian dynamics within the cytoskeleton of biological cells~\cite{mueller2015,wang2007,yang1993}. Often, these slender components can be modeled as 1D Cosserat continua based on a geometrically nonlinear beam theory. In all mentioned cases, mechanical contact interaction crucially influences the overall system behavior. The current work focuses on the development of finite element formulations that are capable of accurately modeling the dynamics of slender components and their contact interaction and that allow for an efficient and robust numerical simulation of complex systems of slender fibers.

Geometrically nonlinear beam finite elements are an efficient and accurate tool for modeling and solving mechanical problems of the category mentioned above. In the recent contributions~\cite{romero2008,bachau2014}, different types of nonlinear beam element formulations have been evaluated and compared, and the so-called geometrically exact beam formulations (see e.g \cite{crisfield1999, eugster2013, jelenic1999, romero2004, romero2002, simo1985, simo1986, sonneville2014, zupan2003}) have been recommended in terms of model accuracy and computational efficiency as compared with alternative formulations such as absolute nodal coordinate (ANC) or solid beam elements (see e.g. \cite{shabana2001,shabana1998,bathe1979,frischkorn2013}). The vast majority of existing geometrically exact beam element formulations are based on the Simo-Reissner beam theory of thick rods incorporating the modes of axial tension, shear, torsion and bending. On the contrary, in the authors' recent contributions~\cite{meier2014,meier2015,meier2016} the first geometrically exact beam element formulations based on the Kirchhoff-Love theory of thin rods have been proposed that are capable of modeling general beam geometries with arbitrary initial curvatures and anisotropic cross-section shapes and that preserve important mechanical properties such as objectivity and path-independence. According to the Kirchhoff-Love theory, these formulations abstain from the representation of shear deformation. In~\cite{meier2016}, it has been shown that the avoidance of the very high stiffness contributions resulting from shear modes leads to considerable numerical advantages in the range of high beam slenderness ratios as compared with existing formulations of Simo-Reissner type. Concretely, a lower discretization error level per degree of freedom as well as a considerably decreased number of accumulated Newton iterations, by a factor of almost two orders of magnitude for high beam slenderness ratios in the range of $\zeta\!=\!10^4$, could be achieved by the proposed Kirchhoff-Love elements as compared to existing Simo-Reissner elements from the literature. Since such high beam slenderness ratios are typically prevalent in most of the applications mentioned above, the novel Kirchhoff-Love elements seem to be the ideal numerical tool for such scenarios. In the mentioned reference~\cite{meier2016}, four different Kirchhoff-Love element variants have been proposed. They basically differ in the applied rotation interpolation, either based on a strong or a weak enforcement of the Kirchhoff constraint, as well as in the parametrization of nodal rotations, either based on nodal rotation vectors or on nodal tangent vectors. Besides these general Kirchhoff-Love beam element formulations, reduced models leading to special torsion-free beam element formulations have been proposed in~\cite{meier2015}. There, it has been shown that under certain restrictions concerning the initial beam geometry (straight beams with isotropic/circular cross-sections) and external loads (no torsional components of external moments), the (static) Kirchhoff-Love theory yields solutions with vanishing torsion even for arbitrarily large displacements and rotations. This finding justified the development of torsion-free beam element formulations that inherit the high accuracy well-known for geometrically exact beam element formulations, while simultaneously avoiding any rotational degrees of freedom typical for geometrically exact formulations. In turn, this leads to considerably simplified and consequently more efficient finite element formulations, characterized e.g.~by symmetric stiffness matrices as well as symmetric and constant mass matrices. For all of the (general and reduced) Kirchhoff-Love element formulations mentioned so far, essential properties such as objectivity, path-independence, consistent convergence behavior, the avoidance of locking effects or the conservation of energy and momentum by the employed spatial discretizations have been shown analytically and numerically. Moreover, all of these formulations have been based on a $C^1$-continuous beam centerline representation, which enables smooth kinematics in the context of beam-to-beam contact schemes.

In the mechanical modeling and numerical simulation of beam-to-beam contact interaction basically two different types of approaches can be distinguished: Point-to-point contact models (see e.g.~\cite{wriggers1997,zavarise2000,konjukhov2010,litewka2002,litewka2002b,litewka2007,litewka2005,kulachenko2012,neto2015,litewka2013,litewka2015}) consider a discrete contact force at the closest point of the beams, while line-to-line contact models (see e.g.~\cite{durville2004,durville2010,durville2012,vu_and_durville2015,chamekh2009,chamekh2014,meier2015b}) assume distributed contact forces along the beams. In the authors' recent work~\cite{meier2015b,meier2015c}, it has been shown that line contact formulations applied to slender beams provide very accurate and robust mechanical models in the range of small contact angles, whereas the computational efficiency considerably decreases with increasing contact angles. On the other hand, point contact formulations serve as sufficiently accurate and very elegant and efficient models in the regime of large contact angles, while they are not applicable for small contact angles as a consequence of non-unique closest point projections. In order to combine the advantages of these two basic types of formulations, while abstaining from their disadvantages, a novel all-angle beam contact (ABC) formulation has been proposed in~\cite{meier2015c}. This formulation applies a point contact formulation in the range of large contact angles and a line contact formulation in the range of small contact angles, the two being smoothly connected by means of a variationally consistent model transition approach. However, since only the formulation of beam-to-beam contact schemes themselves was in the focus of that work, all examples presented there have employed the reduced torsion-free beam element formulation~\cite{meier2015} mentioned above. Undoubtedly, the assumptions underlying this reduced model are fulfilled by many of the practically relevant applications mentioned above. Nevertheless, there is also a range of applications involving highly slender fibers where mechanical torsion plays an important role, e.g. all kinds of initially curved fiber geometries or fibers loaded by torsional moments.

In order to face this requirement, several original scientific contributions are proposed in the present work. One aim of this contribution is to combine the all-angle beam contact formulation proposed in~\cite{meier2015c} with the different types of general Kirchhoff-Love beam element formulations proposed in~\cite{meier2016}. Especially the combination with the rotation vector-based parametrization of nodal rotations as proposed in~\cite{meier2016} will allow for a simple realization of complex beam structures with beam-to-beam joints while still allowing for a $C^1$-continuous geometry representation and smooth contact kinematics. On the other hand, this beam element variant will require a proper transformation of the beam-to-beam contact residual and stiffness contributions derived in~\cite{meier2015c}. While in~\cite{meier2016}, Kirchhoff-Love finite elements have been identified as the formulations of choice in the regime of high beam slenderness ratios, it is beyond all question that finite element formulations of Simo-Reissner type should be preferred for thick beam geometries where shear deformation may play an important role. Unfortunately, there are only a few very recent contributions considering Simo-Reissner type formulations with smooth geometry representation (see e.g. ~\cite{weeger2016}), a property that can be regarded as highly beneficial for beam-to-beam contact formulations. To the best of the authors' knowledge, none of these few existing formulations are based on a Hermite interpolation of the beam centerline similar to~\cite{meier2014,meier2015,meier2016}. Compared to alternative $C^1$-interpolations, such a Hermite representation comprises the advantage of a simple representation of nodal orientations and beam-to-beam joints via nodal rotation / tangent vectors. Moreover, it allows for a modular beam-to-beam contact framework since, on the one hand, the contact interaction is entirely determined by the beam centerline representation, and, on the other hand, the centerline discretization of the considered (general and torsion-free) Kirchhoff-Love elements is based on the same Hermite polynomials. For these reasons, a novel geometrically exact Simo-Reissner beam element formulation based on a third-order Hermite centerline interpolation will also be proposed in this work. For this formulation, optimal spatial convergence rates and the avoidance of membrane and shear locking will be shown numerically. Eventually, a series of practically relevant applications that pose important challenges to the applied beam element formulations will be investigated. Therein, the performance of the different beam element formulations under consideration, i.e. of the torsion-free Kirchhoff-Love formulations, the general Kirchhoff-Love formulations as well as the Simo-Reissner formulations, will be compared and a recommendation will be made concerning the optimal choice of beam elements.

The remainder of this work is organized as follows: In Section~\ref{sec:beams}, the most important aspects of the torsion-free beam element formulation proposed in~\cite{meier2015}, the Kirchhoff-Love beam element formulations proposed in~\cite{meier2016} as well as the underlying Hermite interpolation of the beam centerline are briefly recapitulated. Moreover, the novel Hermite interpolation-based Simo-Reissner beam element is presented in this section, and important properties such as optimal spatial convergence behavior and the avoidance of membrane locking effects are verified numerically. In Section~\ref{sec:contact}, the basics of the ABC formulation first presented in~\cite{meier2015c} are repeated. Thanks to the modularity of the proposed framework, most of the investigated finite element formulations can directly be combined with the ABC formulation. Only for the Kirchhoff-Love beam elements with rotation vector-based parametrization of nodal rotations, a transformation of the resulting residual and stiffness contributions is required, which will also be considered in Section~\ref{sec:contact}. Subsequently, Section~\ref{sec:examples} contains a series of practically relevant numerical examples that are intended to investigate and compare the performance of the different finite element variants. Eventually, in Section~\ref{sec:conclusion}, a summary of the most important results and a recommendation concerning the optimal choice of beam elements will be given.

\section{Geometrically exact beam elements}
\label{sec:beams}

\begin{figure*}[t!!!]
 \centering
  \includegraphics[width=0.90\textwidth]{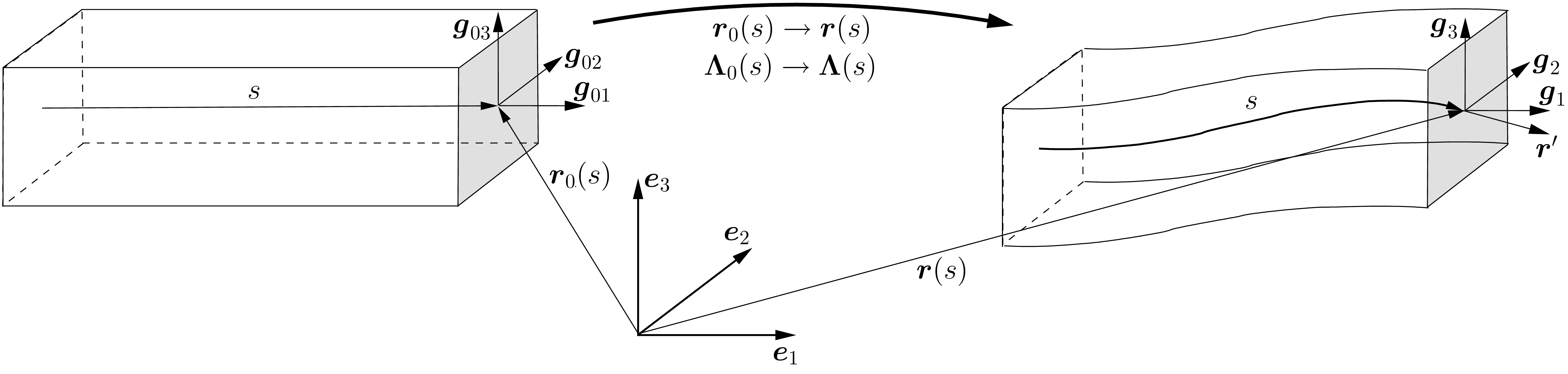}
  \caption{Kinematic quantities defining the initial and deformed configuration of the geometrically exact beam.}
  \label{fig:cpt1_beamkinematics1}
\end{figure*}

The geometrically exact beam theory considered in this work is based on the Bernoulli hypothesis of undeformable cross-sections. Consequently, the configuration of the beam is uniquely defined by the beam centerline curve $s,t \rightarrow \mb{r}(s,t) \in \Re^3$ representing the line that connects the cross-section centroids and by a field of right-handed orthonormal triads $s,t \rightarrow \mb{\Lambda}(s,t):=(\mb{g}_{1}(s,t), \mb{g}_{2}(s,t), \mb{g}_{3}(s,t))_{\mb{e}_i} \in SO^3$ rigidly attached to these cross-sections and determining their orientation. Here, the rotation tensor $\mb{\Lambda}(s,t)$ describes the rotation from the global Cartesian frame $\mb{e}_{1},\mb{e}_{2},\mb{e}_{3}$ onto the local cross-section frame $\mb{g}_{1}(s,t),\mb{g}_{2}(s,t),\mb{g}_{3}(s,t)$ according to $\mb{g}_{j}(s,t)=\mb{\Lambda}(s,t) \mb{e}_{j}(s,t)$ for $j=1,2,3$ and $SO^3$ represents the so-called special orthogonal group of 3D rotations. Furthermore, $t$ is the time and $s \in [0,l]=:\Omega_l \subset \Re$ is an arc-length parametrization of the initial centerline curve $\mb{r}_0(s)$ and $l \in \Re$ the beam length in the initial configuration. Here and in the following, the index $0$ of a quantity refers to the unstressed initial configuration. Throughout this work, the index near a matrix $(.)_{\mb{e}_{i}}$ denotes the basis in which the associated tensor is represented. The kinematic quantities describing the initial and deformed configuration of the beam are illustrated in Figure~\ref{fig:cpt1_beamkinematics1}. Throughout this work, the beam slenderness ratio $\zeta\!:=\! l/R$, defined as the ratio of the beam length $l$ and the cross-section radius $R$, will often be employed as tool of characterization.

The following spatial discretization will exclusively be performed in the context of the finite element method (FEM). The discretization of the rotation field $\mb{\Lambda}(s,t)$ usually represents the major complexity in the development of geometrically exact beam element formulations, since the underlying configuration space $SO^3$ represents a nonlinear manifold lacking standard vector space properties such as additivity or commutativity for its elements. However, the \textit{beam-to-beam contact schemes} considered in this contribution are based on the additional assumption of circular cross-sections. Even though the rotation interpolation is still a crucial constituent of the beam element formulation, the beam contact interaction is entirely described by the beam centerline configuration in this case. For that reason, the challenging topic of interpolating the rotation field $\mb{\Lambda}(s,t)$ will not be further detailed here. Instead, the focus lies on the beam centerline interpolation and the associated nodal degrees of freedom. For simplicity, the arguments $s$ and $t$ will often be omitted in the following.

For conservative systems, the beam problems under consideration can be described by energy contributions $\Pi_{kin},\Pi_{int},\Pi_{ext}$ and $\Pi_{con}$ resulting from kinetic, internal, external and contact forces. Based on the Hamilton principle, the strong form of the dynamic balance equations within a considered time interval $t \in [0,T]$ can be derived via the following variational problem statement:
\begin{align}
\label{hamilton_principle}
  \delta \int \limits_{t=0}^T \! \! \mathcal{L} \, dt =0 \quad \text{with} \quad \left[\delta \mb{r}\!=\!\mb{0}, \, \delta \mb{\Lambda} \!=\!\mb{0}\right]_{t=0}^{t=T}.
\end{align}
The Lagrangian $\mathcal{L}$ occurring in~\eqref{hamilton_principle} is defined as
\begin{align}
\label{hamilton_lagrangian}
\begin{split}
  \mathcal{L}\!&=\! {\Pi}_{kin}\!-\! {\Pi}_{int} \!-\! {\Pi}_{ext}\!-\! {\Pi}_{con} \\
 					   &=\! {\Pi}_{kin}\!-\! {\Pi}_{int} \!+\! {W}_{ext}\!-\! {\Pi}_{con},
\end{split}
\end{align}
where the alternative variant employed in the second line of~\eqref{hamilton_lagrangian} is based on the work contribution ${W}_{ext}\!=\!-{\Pi}_{ext}$ of the external forces. On the other hand, the weak form of balance equations required for the FEM discretization can be derived by applying the method of weighted residuals to the corresponding strong form, a more general procedure that is also valid for non-conservative problems. Inserting the test and trial functions underlying the finite element discretization into the weak form of the dynamic balance equations typically yields a global system of equations of the form
\begin{align}
\label{global_system}
\begin{split}
\!\!\!\!\!\!\mbd{R}_{tot}\!&=\!\mbd{R}_{kin}(\mbd{X},\dot{\mbd{X}},\ddot{\mbd{X}})\!+\!\mbd{R}_{int}(\mbd{X})\!-\!\mbd{R}_{ext}(\mbd{X})\\
&+ \mbd{R}_{con}(\mbd{X})\!=\!\mbd{0}.\!\!\!\!\!\!
\end{split}
\end{align}
The global residual vector~\eqref{global_system} represents the spatially discretized weak form of mechanical equilibrium, where $\mbd{X}$ is the assembled global vector of time-continuous primary variables containing the nodal degrees of freedom of the finite element discretization and $\dot{(.)}$ as well as $\ddot{(.)}$ represent the first and second time derivative. How to express these derivatives via displacements of the current and preceding time steps based on a proper temporal discretization scheme will be discussed in Section~\ref{sec:beams_temporal}. The global residual contributions of internal, kinetic, external and contact forces result from a proper assembly of the corresponding element-wise contributions $\mbd{r}_{int}(\hat{\mbd{x}}), \mbd{r}_{kin}(\hat{\mbd{x}}), \mbd{r}_{ext}(\hat{\mbd{x}})$ and $\mbd{r}_{con}(\hat{\mbd{x}})$, which will be considered in this section and the subsequent Section~\ref{sec:contact}. Here, the vector $\hat{\mbd{x}}$ summarizes all nodal degrees of freedom associated with one finite element. The system of equations resulting from~\eqref{global_system} after temporal discretization will depend in a nonlinear manner on the vector of nodal unknowns $\mbd{X}_{n}$ at a considered (discrete) time step $n$. Within this work, a Newton-Raphson scheme will be employed in order to solve this nonlinear system of equations. Element-local as well as assembled global contributions to the linearization of~\eqref{global_system} will be denoted as $\mbd{k}_{int}, \mbd{k}_{kin},\mbd{k}_{ext},\mbd{k}_{con}$ as well as $\mbd{K}_{int}, \mbd{K}_{kin},\mbd{K}_{ext},\mbd{K}_{con}$ in the following, with $\mbd{k}_{...}:=d \mbd{r}_{...} / d \hat{\mbd{x}}$ and $\mbd{K}_{...}:=d \mbd{R}_{...} / d \mbd{X}$.

\subsection{Beam centerline interpolation}
\label{sec:beams_centerline}

All beam element formulations considered in this work share the same $C^1$-continuous centerline interpolation based on third-order Hermite polynomials as proposed in~\cite{meier2014}. Concretely, a Bubnov-Galerkin approach is followed in this contribution leading to a discretized beam centerline given by:
\begin{align}
\label{interpolation}
\begin{split}
\!\!\!\!\!\!\mb{r}(\xi) \! \approx \! \mb{r}_h(\xi)  \! &= \!\! \sum_{i=1}^{2} N^i_{d}(\xi) \mb{d}^i \!+\! \frac{l}{2} \!\sum_{i=1}^{2} N^i_{t}(\xi) \mb{t}^i, \\
\delta \mb{r}(\xi) \! \approx \! \delta \mb{r}_{h}(\xi) \! &= \!\! \sum_{i=1}^{2} N^i_{d}(\xi) \delta \mb{d}^i \!+\! \frac{l}{2} \!\sum_{i=1}^{2} N^i_{t}(\xi) \delta \mb{t}^i\!.\!\!\!\!\!\!
\end{split}
\end{align}
Here, $\mb{d}^i, \mb{t}^i \! \in \! \Re^3$ are position and tangent vectors at the two element nodes $i=1,2$, $\delta \mb{d}^i, \delta \mb{t}^i \! \in \! \Re^3$ represent their variations and $\xi \in [-1;1]$ is an element parameter coordinate that can explicitly be related to the arc-length coordinate $s$ according to $(.)_{,s}=(.)_{,\xi}\cdot J(\xi)$, with the element Jacobian $J(\xi)=||\mb{r}_{0,\xi}(\xi)||$. Similar to the abbreviation $(.)^{\prime}\!=\!(.)_{,s}$ for the arc-length derivative, we will use the notation $(.)^{\shortmid}\!=\!(.)_{,\xi}$ for the derivative with respect to the parameter coordinate. Here and in the following, the index $h$ denotes the spatially discretized version of a quantity, but this index will often be omitted in the following when there is no danger of confusion. The third-order Hermite shape functions $N^i_{d}(\xi)$ and $N^i_{t}(\xi)$ (see \cite{meier2014} for the properties of these polynomials) are defined as
\begin{align}
\label{shapefunctions}
\begin{split}
\!\!\!\!\!\! N^1_{d} \!&=\! \frac{1}{4}(2\!+\!\xi)(1\!-\!\xi)^2\!, \,\, N^2_{d} \!=\! \frac{1}{4}(2\!-\!\xi)(1\!+\!\xi)^2\!, \\
N^1_{t} \!&=\! \frac{1}{4}(1\!+\!\xi)(1\!-\!\xi)^2\!, \,\, N^2_{t} \!=\! -\frac{1}{4}(1\!-\!\xi)(1\!+\!\xi)^2\!,\!\!\!\!\!\!
\end{split}
\end{align}
and provide a $C^1$-continuous beam centerline representation, thus enabling smooth contact kinematics for all investigated beam element variants.

\subsection{Simo-Reissner beam element}
\label{sec:beams_SR}

\subsubsection{Space-continuous problem}
\label{sec:beams_SR_continous}

The Simo-Reissner theory allows for shear-deformation, i.e. the cross-section orientation is completely independent from the beam centerline curve and can in general be described by three degrees of freedom, e.g. in terms of a rotation vector field $\boldsymbol{\psi}(s) \in \Re^3$ according to $\mb{\Lambda}(s)=\mb{\Lambda}(\boldsymbol{\psi}(s))$. For further details on large rotations and how such a rotation vector parametrization can be realized by means of the so-called Rodrigues formula, the interested reader is exemplarily referred to~\cite{argyris1982,simo1985,simo1986,cardona1988,ibrahimbegovic1995b}. In a next step, the simplest possible case is assumed for the constitutive behavior of the beam, i.e.~that it can be described by means of a length-specific hyper-elastic stored energy function as proposed in~\cite{simo1985}:
\begin{align}
\label{storedenergysimoreissner}
\!\!\!\!\!\!\tilde{\Pi}_{int}(\mb{\Omega},\mb{\Gamma}) \!=\! \frac{1}{2} \mb{\Omega}^T \! \mb{C}_M \mb{\Omega} \!+\! \frac{1}{2} \mb{\Gamma}^T \mb{C}_F \mb{\Gamma}. \!\!\!\!\!\!
\end{align}
In the following, length-specific $\tilde{\Pi}_{...}$ and integrated ${\Pi}_{...}$ energy contributions are related according to:
\begin{align}
\label{storedenergysimoreissner}
\!\!\!\!\!\! {\Pi}_{...} \!=\! \int \limits_0^l \tilde{\Pi}_{...} ds. \!\!\!\!\!\!
\end{align}
The material deformation measure $\mb{S}(\mb{\Omega})\!:=\!\mb{\Lambda}^T\!\mb{\Lambda}^{\prime}$ represents torsion and bending, while $\boldsymbol{\Gamma} \!:=\! \mb{\Lambda}^T\!\mb{r}^{\prime}  \!-\! \mb{e}_1$ represents axial tension and shear. Moreover, $\mb{S}(\mb{a})$ denotes the unique skew-symmetric tensor representing the vector product $\mb{S}(\mb{a})\mb{b} \!=\! \mb{a}\!\times\!\mb{b} \,\, \text{for} \,\, \mb{a},\mb{b}  \!\in\! \Re^3$. The constitutive matrices $\mb{C}_M$ and $\mb{C}_F$ read
\begin{align}
\begin{split}
\label{storedenergyfunctionkirchhoff2} 
\mb{C}_M\!&=\! \textbf{diag}\big[GI_T,EI_2,EI_3\big]_{\mb{e}_{i}}, \\ 
\mb{C}_F\!&=\! \textbf{diag}\big[EA,GA_2,GA_3\big]_{\mb{e}_{i}},
\end{split}
\end{align}
where $E$ and $G$ are the Young's modulus and the shear modulus, $A,A_2,A_3$ are the cross-section area and the reduced cross-section values, $I_2$ and $I_3$ are the two principal moments of inertia and $I_T$ is the torsional moment of inertia. Similarly, the length-specific kinetic energy can be formulated as:
\begin{align}
\label{kineticenergy}
\!\!\!\!\tilde{\Pi}_{kin}(\mb{w},\dot{\mb{r}}) \!=\! \frac{1}{2} \mb{w}^T \! \mb{c}_{\rho} \mb{w} \!+\! \frac{1}{2} \rho A \dot{\mb{r}}^T \! \dot{\mb{r}}.\!\!\!\!
\end{align}
Equivalently to~\eqref{storedenergyfunctionkirchhoff2}, the inertia tensor $\mb{c}_{\rho}$ is given:
\begin{align}
\label{kineticenergy2}
\mb{c}_{\rho}\!=\!\textbf{diag}\big[\rho (\underbrace{I_2\!+\!I_3}_{=:I_P}),\rho I_2,\rho I_3 \big]_{\mb{g}_{i}}.
\end{align}
Here, $\rho$ is the mass density, $\dot{\mb{r}}$ the centerline velocity vector and $\mb{S}(\mb{w})\!:=\! \dot{\mb{\Lambda}}\mb{\Lambda}^T$ defines the spatial angular velocity vector $\mb{w}$. Based on these energies and the consideration of external forces acting on the beam domain and boundary, the weak form of the balance equations can be derived (see e.g.~\cite{simo1985}):
\begin{align}
\label{weakformspatial}
\begin{split}
\!\!\!\!\!\!G\!&=\! \int \limits_0^l \!\bigg( \underbrace{\delta \boldsymbol{\theta}^{\prime T}  \mb{m}}_{\delta_{o} \boldsymbol{\omega}^T  \mb{m}}  +  \underbrace{( \delta \mb{r}^{\prime} - \delta \boldsymbol{\theta} \times
\mb{r}^{\prime} )^T\mb{f}}_{\delta_{o} \boldsymbol{\gamma}^T\mb{f}} \bigg) ds\\
\!&-\! \int \limits_0^l \!\bigg(\delta \boldsymbol{\theta}^T  (\tilde{\mb{m}}+\mb{m}_{\rho}) + \delta \mb{r}^T (\tilde{\mb{f}} + \mb{f}_{\rho}) \bigg) ds\\
&-\Big[\delta \mb{r}^T  \mb{f}_{\sigma} \Big]_{\varGamma_{\sigma}} -  \Big[\delta \boldsymbol{\theta}^T  \mb{m}_{\sigma} \Big]_{\varGamma_{\sigma}}\dot{=}\,0.\!\!\!\!\!\!
\end{split}
\end{align}
In~\eqref{weakformspatial}, the distributed external forces $\tilde{\mb{f}}$ and moments $\tilde{\mb{m}}$, the discrete forces  $\mb{f}_{\sigma}$ and moments $\mb{m}_{\sigma}$ at the Neumann boundary $\varGamma_{\sigma}$, the vector of (additive) virtual displacements $\delta \boldsymbol{r}(s) \!\in\! \Re^3$ as well as the vector of (multiplicative) virtual rotations $\delta \boldsymbol{\theta}(s) \!\in\! \Re^3$, also denoted as spin vector, can be identified. The identity of the objective variations $\delta_o(.)\!:=\!\delta(.)\!-\!\delta \boldsymbol{\theta}\!\times\!(.)$ of the spatial deformation measures $\boldsymbol{\omega}\!:=\!\mb{\Lambda}\boldsymbol{\Omega}$ and $\boldsymbol{\gamma}\!:=\!\mb{\Lambda}\boldsymbol{\Gamma}$ with the curley bracket terms has been shown in~\cite{simo1985} via work-pairing. Furthermore, the force and moment stress resultants $\mb{f}$ and $\mb{m}$ as well as the distributed inertia forces and moments $\mb{f}_{\rho}$ and $\mb{m}_{\rho}$ are defined according to
\begin{align}
\label{stress_resultants_and_inertia}
\begin{split}
\!\!\!\!\!\! 
\mb{f}\!&=\! \mb{\Lambda}\mb{C}_F \mb{\Gamma}, \,\, 
\mb{m}\!=\! \mb{\Lambda}\mb{C}_M \mb{\Omega}, \\
\mb{f}_{\rho} \!&=\! -\rho A \ddot{\mb{r}}, \,\, 
\mb{m}_{\rho}\!=\!-(\mb{S}(\mb{w}) \mb{c}_{\rho} \mb{w} \!+\! \mb{c}_{\rho} \mb{a}),
\!\!\!\!\!\!
\end{split}
\end{align}
where $\ddot{\mb{r}}$ is the centerline acceleration vector and $\mb{a}\!:=\! \dot{\mb{w}}$ the spatial angular acceleration vector. Finally, the problem formulation has to be completed by proper boundary conditions on the Neumann and Dirichlet boundaries $\varGamma_{\sigma}$ and $\varGamma_{u}$ and proper initial conditions at $t=0$ in order to end up with a well-defined initial boundary value problem:
\begin{align}
\label{reissner_boundaryconditions}
\begin{split}
\!\!\!\!\!\!
&\mb{r} \!=\! \mb{r}_{u}, \, \boldsymbol{\Lambda} \!=\! \boldsymbol{\Lambda}_{u}  \,\, \text{on} \,\, \varGamma_{u},  \\
&\mb{f} \!=\! \mb{f}_{\sigma}, \, \mb{m} \!=\! \mb{m}_{\sigma} \, \, \text{on} \,\, \varGamma_{\sigma},\\
&\mb{r}\!=\!\mb{r}_0, \, \dot{\mb{r}}\!=\!\mb{v}_0, \, 
\boldsymbol{\Lambda}\!=\!\boldsymbol{\Lambda}_0, \, \mb{w}\!=\!\mb{w}_0 \,\,
\text{at} \,\, t\!=\!0.\!\!\!\!\!\!
\end{split}
\end{align}
Based on a proper trial space $(\mb{r}(s,t),\boldsymbol{\Lambda}(s,t)) \!\in\! \mathcal{U}$ of functions satisfying~\eqref{reissner_boundaryconditions} and a proper weighting space $(\delta \mb{r}(s), \delta \boldsymbol{\theta}(s)) \!\in\! \mathcal{V}$ of functions satisfying $\delta \mb{r}\!=\!\mb{0}, \, \delta \boldsymbol{\theta} \!=\! \mb{0} \, \text{on} \, \varGamma_{u}$, the space-continuous Simo-Reissner beam problem is completely defined.

\subsubsection{Spatial discretization}
\label{sec:beams_SR_discrete}

\begin{figure}[ht]
 \centering
  \includegraphics[width=0.45\textwidth]{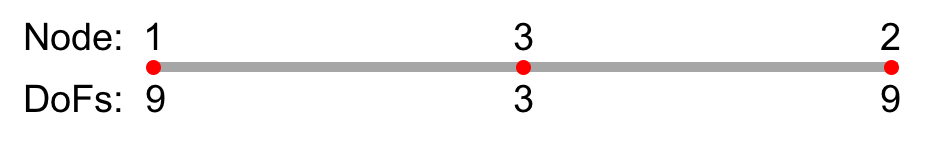}
  \caption{Degrees of freedom of Simo-Reissner element}
  \label{fig:fig_nodedof_SR}
\end{figure}

The essential step in deriving the element residual vector from this weak form lies in the spatial discretization of the centerline curve $\mb{r}$ and the rotation field $\mb{\Lambda}$, i.e.~in replacing the trial and test spaces $\mathcal{U}$ and $\mathcal{V}$ by suitable finite-dimensional subspaces $\mathcal{U}_h \!\subset\! \mathcal{U}$ and $\mathcal{V}_h \!\subset\! \mathcal{V}$. Here, the centerline discretization will be given by the $C^1$-continuous Hermite interpolation~\eqref{interpolation} specifiing $\mb{r}_h(\xi)$ as well as $\delta \mb{r}_h(\xi)$. On the other hand, the interpolation of the rotation field is given by the three-noded variant of the objective interpolation scheme proposed in~\cite{crisfield1999}. Assuming that each of the three nodal triads $\mb{\Lambda}^i$ for $i\!=\!1,2,3$ is uniquely defined by proper nodal primary degrees of freedom, this interpolation can be formulated in the following abstract manner:
\begin{align}
\label{abstract_triad_interpolation}
\!\!\!\!\!\!\mb{\Lambda}(\xi) \!\approx\! \mb{\Lambda}_h(\xi) \!=\! \text{nl}_{SR}(\mb{\Lambda}^1,\mb{\Lambda}^2,\mb{\Lambda}^3,\xi).\!\!\!\!\!\!
\end{align}
The expression $\text{nl}_{SR}(.)$ in~\eqref{abstract_triad_interpolation} should represent an arbitrary function that depends on its arguments in a nonlinear manner. Following the approach in~\cite{crisfield1999}, the field of virtual rotations $\delta \boldsymbol{\theta}(s)$ is discretized in a Petrov-Galerkin manner based on third-order Lagrange polynomials $L^i(\xi)$ according to:
\begin{align}
\label{petrovspininterpolation}
\delta \boldsymbol{\theta}_h(\xi) = \sum_{i=1}^{3} L^i(\xi) \delta \hat{\boldsymbol{\theta}}^i.
\end{align}
The specific analytical expression for the interpolation~\eqref{abstract_triad_interpolation} as well as the resulting element residual and stiffness matrices are given in~\ref{anhang:sr_element}. It is assumed that the nodal primary degrees of freedom describing the nodal triads $\mb{\Lambda}^i(\boldsymbol{\psi}^i)$ are given by nodal rotation vectors $\boldsymbol{\psi}^i \in \Re^3$ for $i\!=\!1,2,3$. Additionally, the nodal primary degrees of freedom $\hat{\mb{d}}^1,\hat{\mb{t}}^1,\hat{\mb{d}}^2,\hat{\mb{d}}^2$ at the element boundary nodes $1$ and $2$ are employed in order to define the Hermite centerline interpolation~\eqref{interpolation}. Here and in the following, nodal primary variables of the finite element discretization will be marked by a hat $\hat{(.)}$. All in all, it can be concluded that the configuration of one finite element is completely defined by a set $\hat{\mbd{x}}_{SR}$ of nodal degrees of freedom uniquely determining the discrete beam centerline curve~\eqref{interpolation} and the rotation field~\eqref{abstract_triad_interpolation}.  Since the beam-to-beam contact schemes considered in this work solely depend on the beam centerline configuration, an additional element-wise subset of nodal degrees of freedom $\hat{\mbd{d}}_{SR}$ sufficient to describe the centerline curve is introduced at this point.  These two sets $\hat{\mbd{x}}_{SR}$ and $\hat{\mbd{d}}_{SR}$ shall briefly be summarized for the proposed smooth Simo-Reissner element, in the following denoted as SR element:
\begin{align}
\label{lambda_reissner}
\begin{split}
& \boldsymbol{\Lambda}^1(\hat{\boldsymbol{\psi}}^{1}), \,\,
\boldsymbol{\Lambda}^2(\hat{\boldsymbol{\psi}}^{2}), \,\,
\boldsymbol{\Lambda}^3(\hat{\boldsymbol{\psi}}^{3}), \\
&\mb{d}^1=\hat{\mb{d}}^1, \,\, \mb{d}^2=\hat{\mb{d}}^2, \\
&\mb{t}^1=\hat{\mb{t}}^1, \,\,\,\,\, \mb{t}^2=\hat{\mb{t}}^2, \\
& \hat{\mbd{x}}_{SR}\!:=\!(\hat{\mb{d}}^{1T}\!,\hat{\mb{t}}^{1T}\!,\hat{\boldsymbol{\psi}}^{1T}\!,\hat{\mb{d}}^{2T}\!,\hat{\mb{t}}^{2T}\!,\hat{\boldsymbol{\psi}}^{2T},\hat{\boldsymbol{\psi}}^{3T})^T, \\
& \hat{\mbd{d}}_{SR}\!:=\!(\hat{\mb{d}}^{1T}\!,\hat{\mb{t}}^{1T}\!,\hat{\mb{d}}^{2T}\!,\hat{\mb{t}}^{2T}\!)^T.
\end{split}
\end{align}
In Figure~\ref{fig:fig_nodedof_SR}, the number of degrees of freedom associated with the element boundary nodes $1$ and $2$ and the element mid node $3$ are illustrated. 

As argued in~\ref{anhang:sr_element}, the employed interpolation schemes will result in a finite element formulation that preserves objectivity and path-independence, a crucial property for geometrically exact beam elements. Furthermore, the spatial discretization allows for an exact conservation of linear and angular momentum. Moreover, a reduced Gauss-Lobatto integration scheme will be applied for integration of the weak form contributions associated with the deformation measure $\boldsymbol{\Gamma}$ in order to avoid membrane and shear locking.  In the next section, this latter statement will be verified numerically by investigating the spatial convergence based on a benchmark test from the literature.

\subsubsection{Verification of spatial convergence}
\label{sec:beams_SR_example}

The initial geometry is represented by a $45^{\circ}$-degree circular arc-segment with curvature radius $r_0\!=\!100$ that lies completely in the global $x$-$y$-plane and that is clamped at one end. The section constitutive parameters of the beam result from a quadratic cross-section shape with side length $R\!=\!1$ and a Young's modulus of $E\!=\!10^7$ as well as a shear modulus of $G\!=\!0.5 \cdot 10^7$. This initial geometry is loaded by an out-of-plane force $\mb{f}\!=\!(0,0,f_z)^T$ in global $z$-direction with magnitude $f_z\!=\!600$. This example has initially been proposed by Bathe and Bolourchi~\cite{bathe1979} and can meanwhile be considered as standard benchmark test for geometrically exact beam element formulations that has been investigated by many authors (see e.g.~\cite{simo1986,cardona1988,ibrahimbegovic1995a,dvorkin1988,crisfield1990,jelenic1999,crivelli1993,schulz2001,eugster2013,romero2004,romero2008,avello1991}). While the original definition of the slenderness ratio yields a value of $\zeta\!=\!l/R\!=\!100\pi/4$ for this example, a slightly modified definition of the slenderness ratio according to $\tilde{\zeta}\!=\!r_0/R\!=\!100$ is employed here.

\begin{figure}[htp]
\centering
    \includegraphics[width=0.42\textwidth]{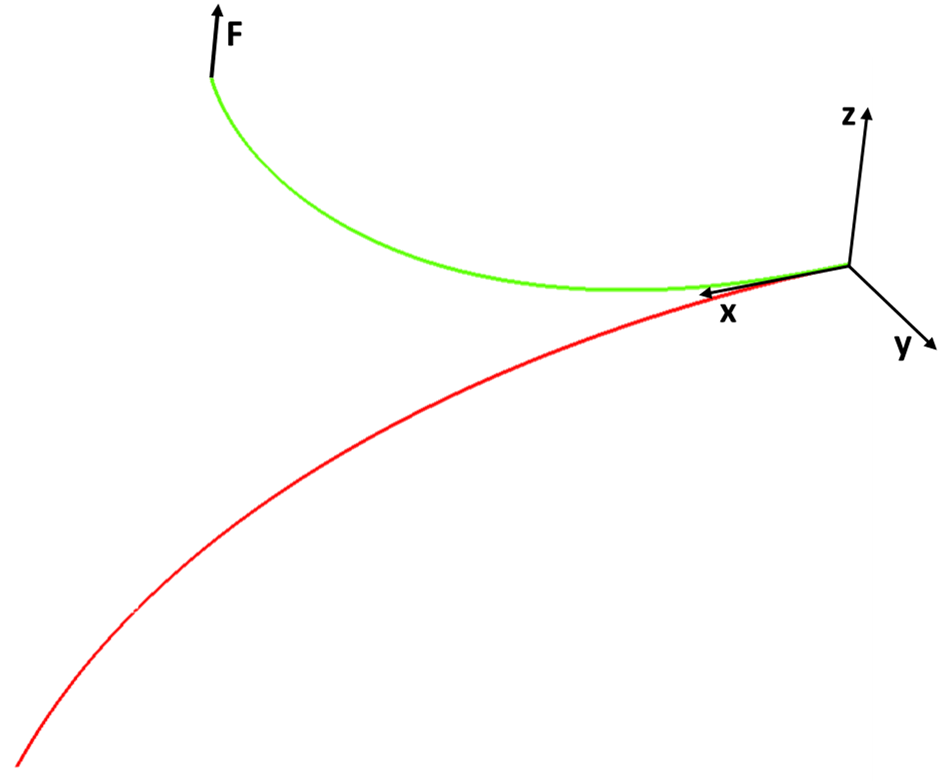}
\caption{Initial (red) and final (green) geometry.}
\label{fig:arcsegment_geometry}
\end{figure}

\begin{figure}[htp]
 \centering
    \includegraphics[width=0.45\textwidth]{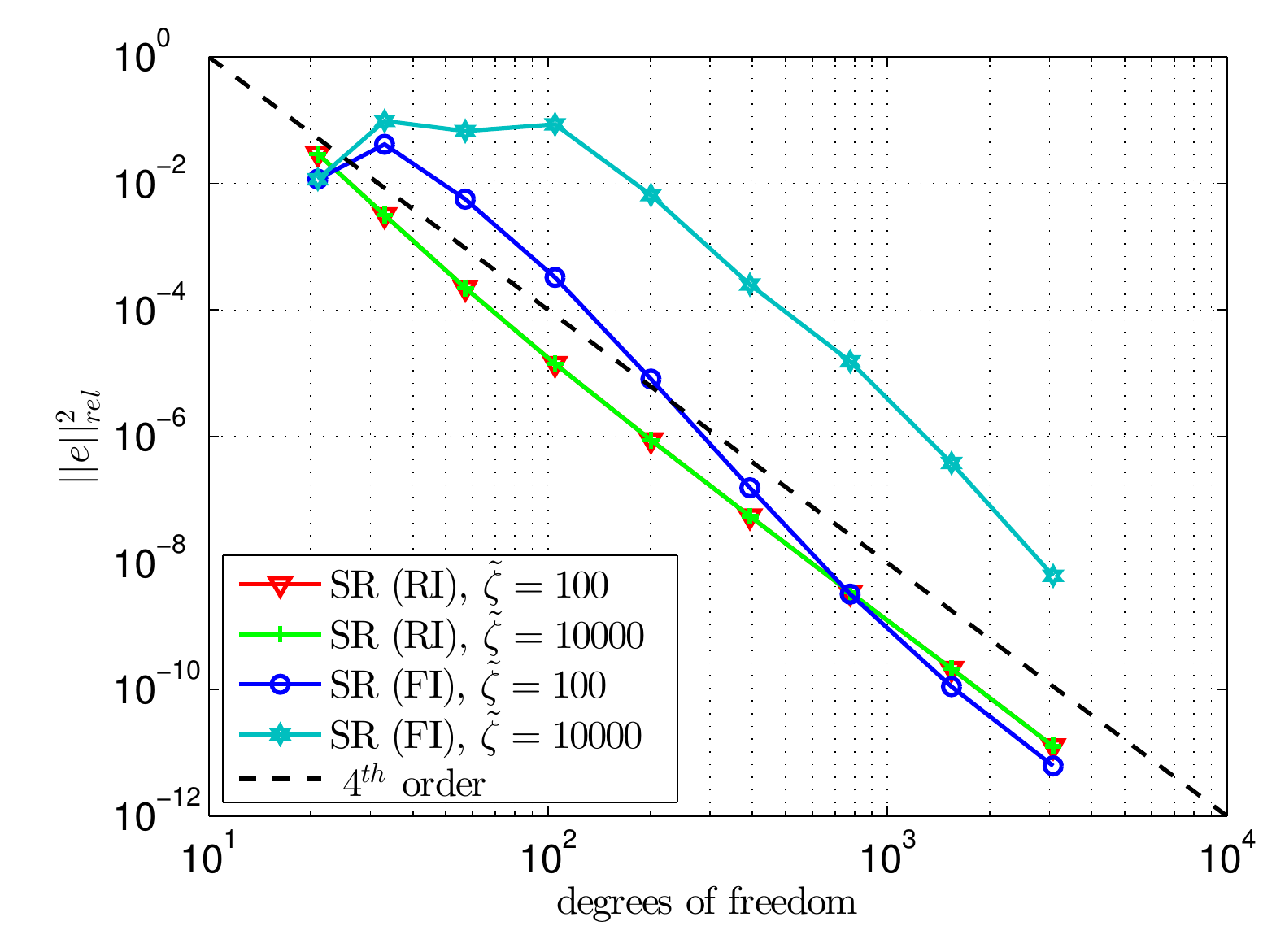}
\caption{$L^2$-error for two slenderness ratios~$\tilde\zeta$ and full/reduced integration (FI/RI). Ref.: $512$ SR elements.}
\label{fig:examples_circlesegment}
\end{figure}

For comparison reasons, also a second variant of this example with increased slenderness ratio $\tilde{\zeta}\!=\!r_0/R\!=\!10000$, i.e. $R\!=\!0.01$, and adapted force $f_z\!=\!6\cdot 10^{-6}$ will be investigated. The initial and deformed geometry are illustrated in Figure~\ref{fig:arcsegment_geometry}. In Figure~\ref{fig:examples_circlesegment}, the relative $L^2$-error (see e.g.~\cite{meier2014} for the exact definition of the applied error norm) has been plotted over the total number of degrees of freedom for the two different slenderness ratios  $\tilde{\zeta}\!=\!100$ and $\tilde{\zeta}\!=\!10000$ as well as for discretizations with $1,2,4,8,16,32,64$ and $128$ Simo-Reissner (SR) elements. In order to investigate the effectiveness of the employed reduced integration procedure (RI) based on a 3-point Gauss-Lobatto integration scheme (integration points at the element boundary nodes and the element mid-point), also the $L^2$-error resulting from a full integration (FI) by means of a 4-point Gauss-Legendre integration scheme is presented in Figure~\ref{fig:examples_circlesegment}. Similar to the observations made in~\cite{meier2015}, the full integration leads to a clearly visible decline in the spatial convergence rate that can be attributed to membrane and shear locking. The deterioration of the convergence order increases with increasing beam slenderness ratio~$\tilde{\zeta}$ and decreases with mesh refinement as consequence of an decreasing element slenderness ratio (see also~\cite{meier2015}). On the contrary, the proposed reduced Gauss-Lobatto integration scheme successfully avoids membrane and shear locking, thus leading to the optimal convergence order of four for both investigated beam slenderness ratios. Furthermore, it has been verified numerically (not plotted in Figure~\ref{fig:examples_circlesegment}) that a reduced integration approach based on a 3-point Gauss-Legendre integration scheme (3 integration points in the elements interior) hardly improves the spatial convergence behavior as compared to the 4-point scheme plotted in Figure~\ref{fig:examples_circlesegment}. This observation underlines the importance of the special choice of Gauss-Lobatto integration points, which are identical to the collocation points applied by the MCS method as proposed in~\cite{meier2015} for the avoidance of membrane locking.

\subsection{Kirchhoff-Love beam element}
\label{sec:beams_KL}

\subsubsection{Space-continuous problem}
\label{sec:beams_KL_continous}

In contrast to the developments of the last section, the Kirchhoff-Love theory is based on the additional constraint of vanishing shear strains, requiring that the cross-section triads remain perpendicular to the beam centerline tangent $\mb{t}(s)\!:=\!\mb{r}^{\prime}(s)$:
\begin{align}
\label{kirchhoffconstraints}
\mb{g}_2(s) \cdot \mb{t}(s) \equiv 0 \quad \text{and} \quad \mb{g}_3(s) \cdot \mb{t}(s) \equiv 0.
\end{align}
Due to this constraint, the cross-section orientation is not described by three independent parameters $\boldsymbol{\psi}(s) \in \Re^3$ anymore. Instead, it is defined by the tangent vector $\mb{t}(s)$ and one additional scalar degrees of freedom $\varphi(s)$ representing twist rotations with respect to the tangent vector. Analogously, the spin vector is subject to the Kirchhoff constraint and can be uniquely defined by an additional scalar variational degree of freedom $\delta \Theta_1$:
\begin{align}
\label{lambda_kirchhoff}
\begin{split}
\!\!\!\!\!\!
\boldsymbol{\Lambda}(s)&\!=\!\boldsymbol{\Lambda}(\mb{t}(s),\varphi(s)) \quad \text{with} \quad \mb{r}^{\prime}\!=\! \mb{t}, \,\, \mb{g}_1\!=\!\frac{\mb{t}}{|| \mb{t} ||} \!\!\!\!\!\! \\
\delta \boldsymbol{\theta} \!&=\! \delta \boldsymbol{\theta}_{\parallel} \!+\! \delta \boldsymbol{\theta}_{\perp} \!=\! \delta \Theta_1 \mb{g}_1 \!+\! \frac{\mb{S}(\mb{t}) \delta \mb{t}}{|| \mb{t} ||^2}.\!\!\!\!\!\!
\end{split}
\end{align}
Basically, the length-specific kinetic and hyper-elastic energies as well as the weak form of the Kirchhoff-Love theory can be derived from the corresponding quantities of the Simo-Reissner theory by inserting the constraints~\eqref{lambda_kirchhoff}. For the length-specific hyper-elastic stored energy function, this procedure shall briefly be demonstrated
\begin{align}
\label{storedenergyfunctionkirchhoff}
\!\!\!\!\!\!\tilde{\Pi}_{int}(\mb{\Omega},\epsilon) \!=\! \frac{1}{2} \mb{\Omega}^T \! \mb{C}_M \mb{\Omega} \!+\! \frac{1}{2} EA \epsilon^2, \!\!\!\!\!\!
\end{align}
while the kinetic energy remains unchanged as compared to~\eqref{kineticenergy}. In~\eqref{storedenergyfunctionkirchhoff}, the deformation measure $\mb{\Omega}$ associated with torsion and bending is defined as in the Simo-Reissner case, but with $\boldsymbol{\Lambda}(\mb{t}(s),\varphi(s))$ instead of $\boldsymbol{\Lambda}(\boldsymbol{\psi}(s))$, while $\epsilon\!=\! ||\mb{r}^{\prime}||\!-\!1$ represents the axial tension of the beam centerline. The energy contribution of the shear modes vanishes as consequence of the Kirchhoff constraint~\eqref{kirchhoffconstraints}. Consequently, the weak form~\eqref{weakformspatial} can be simplified by replacing the term $( \delta \mb{r}^{\prime} - \delta \boldsymbol{\theta} \times \mb{r}^{\prime} )^T\mb{f}$ with $\delta \epsilon EA \epsilon \!=:\! \delta \epsilon F_1$. Or in other words, the restriction of the arbitrary variations $\delta \boldsymbol{\theta}$ by the Kirchhoff-constraint~\eqref{kirchhoffconstraints} eliminates the shear force contributions $\mb{f}_{\perp}\!=\!\mb{f}\!-\! F_1 \mb{g}_1$ from the weak form. However, the choice of a suitable, singularity-free parametrization and interpolation of the constrained rotation field $\boldsymbol{\Lambda}(\mb{t}(s),\varphi(s))$ that preserves essential properties such as objectivity and path-independence is a non-trivial task, which makes the finite element realization of Kirchhoff-Love formulations often more challenging than for Simo-Reissner formulations. This issue will not be further detailed in this work, and the interested reader is examplarily referred to~\cite{meier2014,meier2016} instead. In the following section, only the most important information on spatial discretization that is required for the beam-to-beam contact formulations in Section~\ref{sec:contact} will be presented.

\subsubsection{Spatial discretization}
\label{sec:beams_KL_discrete}

\begin{figure}[ht]
 \centering
  \includegraphics[width=0.45\textwidth]{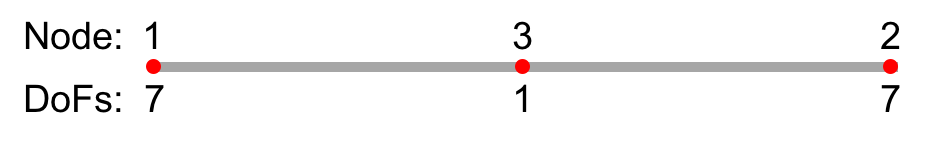}
  \caption{Degrees of freedom of Kirchhoff-Love element}
  \label{fig:fig_nodedof_KL}
\end{figure}

Also the Kirchhoff-Love element formulations presented in this section will rely on a smooth Hermite interpolation of the beam centerline according to~\eqref{interpolation}. Furthermore, similar to the Simo-Reissner case above, the rotation interpolation is based on three nodal triads $\mb{\Lambda}^1,\mb{\Lambda}^2$ and $\mb{\Lambda}^3$. However, due to the Kirchhoff constraint of vanishing shear deformation, only one additional scalar degree of freedom (DoF) $\hat{\varphi}^i$ is required at each node in order to determine the orientation of the nodal triads:
\begin{align}
\label{lambda_kirchhoff_tan}
\begin{split}
& \boldsymbol{\Lambda}^1(\hat{\mb{t}}^1,\hat{\varphi}^1), \,\,
\boldsymbol{\Lambda}^2(\hat{\mb{t}}^2,\hat{\varphi}^2), \,\,
\boldsymbol{\Lambda}^3(\mb{t}(\xi^3),\hat{\varphi}^3), \\
&\mb{d}^1=\hat{\mb{d}}^1, \,\, \mb{d}^2=\hat{\mb{d}}^2, \\
&\mb{t}^1=\hat{\mb{t}}^1, \,\,\,\,\, \mb{t}^2=\hat{\mb{t}}^2, \\
& \hat{\mbd{x}}_{KL-TAN}\!:=\!(\hat{\mb{d}}^{1T}\!,\hat{\mb{t}}^{1T}\!,\hat{\varphi}^1,\hat{\mb{d}}^{2T}\!,\hat{\mb{t}}^{2T}\!,\hat{\varphi}^2,\hat{\varphi}^3)^T, \\
& \hat{\mbd{d}}_{KL-TAN}\!:=\!(\hat{\mb{d}}^{1T}\!,\hat{\mb{t}}^{1T}\!,\hat{\mb{d}}^{2T}\!,\hat{\mb{t}}^{2T}\!)^T.
\end{split}
\end{align}
Is is emphasized, that the tangent vectors at the element boundary nodes $1$ and $2$ are given by the nodal tangents $\hat{\mb{t}}^1$ and $\hat{\mb{t}}^2$ representing primary variables of the Hermite interpolation~\eqref{interpolation}, while the tangent vector at the element mid-node $3$ is determined by the arc-length derivative $\mb{t}(\xi^3)=\mb{r}^{\prime}(\xi^3)$ of the Hermite interpolation~\eqref{interpolation} evaluated at the mid-node coordinate $\xi^3=0$. Consequently, this mid-triad depends on all four nodal vectors $\hat{\mb{d}}^1, \hat{\mb{d}}^2, \hat{\mb{t}}^1$ and $\hat{\mb{t}}^2$ defining the beam centerline. Details, how the scalar degrees of freedom $\hat{\varphi}^i$ are actually defined and the triad orientation is "measured" can be found in~\cite{meier2016} and will not be further specified here. Also the vectors $\hat{\mbd{x}}_{KL-TAN}$ and $\hat{\mbd{d}}_{KL-TAN}$, containing all element DoFs and the element centerline DoFs, respectively, have been summarized in~\eqref{lambda_kirchhoff}. Accordingly, the proposed elements consist of two boundary nodes with seven DoFs as well as one mid-node with one DoF (see Figure~\ref{fig:fig_nodedof_KL}). 

So far, the four degrees of freedom $(\hat{\mb{t}}^i,\hat{\varphi}^i)$ for $i=1,2$ have been applied in order to determine the orientation of the nodal triads $\boldsymbol{\Lambda}^i=\boldsymbol{\Lambda}(\xi^i)$ as well as the orientation and norm of the non-unit tangent vectors $\mb{t}^i=\mb{t}(\xi^i)$ at the element boundary nodes $1$ and $2$. Due to the Kirchhoff constraint $\mb{t}^i=||\mb{t}^i|| \mb{g}_1^i$ and the definition of the rotation tensor $\mb{g}_1^i=\boldsymbol{\Lambda}^i \mb{e}_1$, these nodal quantities can alternatively also be parametrized by the set $(\hat{\boldsymbol{\psi}}^i,\hat{t}^i)$ for $i=1,2$, where the $\hat{\boldsymbol{\psi}}^i$ represent nodal rotation vectors as introduced in Section~\ref{sec:beams_SR} and $\hat{t}^i:=||\mb{t}^i||$ is defined as the norm of the nodal tangent vectors. This second, rotation vector-based variant leads to the following set of nodal primary degrees of freedom:
\begin{align}
\label{lambda_kirchhoff_rot}
\begin{split}
& \boldsymbol{\Lambda}^1(\hat{\boldsymbol{\psi}}^1), \,\,
\boldsymbol{\Lambda}^2(\hat{\boldsymbol{\psi}}^2), \,\,
\boldsymbol{\Lambda}^3(\mb{t}(\xi^3),\hat{\varphi}^3), \\
&  \mb{d}^1= \hat{\mb{d}}^1, \,\,
    \,\,\,\,\,\,\,\,\,\,\,\,\,\,\,\,\,\,\,\,\,\, \mb{d}^2 =  \hat{\mb{d}}^2, \\
& \mb{t}^1=\hat{t}^1 \boldsymbol{\Lambda}^1(\hat{\boldsymbol{\psi}}^1) \mb{e}_1, \,\,
  \mb{t}^2=\hat{t}^2 \boldsymbol{\Lambda}^2(\hat{\boldsymbol{\psi}}^2) \mb{e}_1, \\
& \hat{\mbd{x}}_{KL-ROT}\!:=\!(\hat{\mb{d}}^{1T}\!,\hat{\boldsymbol{\psi}}^{1T}\!,\hat{t}^1,\hat{\mb{d}}^{2T}\!,\hat{\boldsymbol{\psi}}^{2T}\!,\hat{t}^2,\hat{\varphi}^3)^T, \\
& \hat{\mbd{d}}_{KL-ROT}\!:=\!(\hat{\mb{d}}^{1T}\!,\hat{\boldsymbol{\psi}}^{1T}\!,\hat{t}^1,\hat{\mb{d}}^{2T}\!,\hat{\boldsymbol{\psi}}^{2T}\!,\hat{t}^2)^T.
\end{split}
\end{align}
It is emphasized that the finite element formulation resulting from this choice of nodal primary degrees of freedom is the only one considered in this work for which the nodal tangent vectors appearing in the Hermite interpolation~\eqref{interpolation} are no primary variables of the discretization, but depend on the alternative set $(\hat{\boldsymbol{\psi}}^i,\hat{t}^i)$. Consequently, this will be the only element formulation where the beam-to-beam contact schemes considered in Section~\ref{sec:contact} have to be adapted as compared to the original works~\cite{meier2015b,meier2015c}. It has been derived in~\cite{meier2016} that the finite element formulations resulting from the tangent vector-based rotation parametrization~\eqref{lambda_kirchhoff_tan} and from the rotation vector-based rotation parametrization~\eqref{lambda_kirchhoff_rot} will yield identical finite element solutions. However, it has also been shown that the variant~\eqref{lambda_kirchhoff_tan} typically leads to a better performance of the nonlinear solver, while the variant~\eqref{lambda_kirchhoff_rot} simplifies the modeling of complex Dirichlet boundary conditions and joints. This latter aspect will be demonstrated in the numerical example of Section~\ref{sec:examples_micro}. Throughout this work, the general Kirchhoff-Love elements presented in this section will be denoted as KL elements. More specifically, the tangent vector-based variant~\eqref{lambda_kirchhoff_tan} and the rotation vector-based variant~\eqref{lambda_kirchhoff_rot} will be distinguished by the notations KL-TAN and KL-ROT when necessary.

Besides the two different parametrizations of nodal triads considered so far, also two different interpolation schemes for the rotation field $\boldsymbol{\Lambda}(s)$ have been considered in~\cite{meier2016}. The first one is based on a strong enforcement of the Kirchhoff constraint and shall be noted in an abstract manner as follows:
\begin{align}
\label{abstract_triad_interpolation2}
\!\!\!\!\!\!\mb{\Lambda}(\xi) \!\approx\! \mb{\Lambda}_h(\xi) \!=\! \text{nl}_{SKL}(\mb{\Lambda}^1,\mb{\Lambda}^2,\mb{\Lambda}^3,\mb{t}(\xi),\xi).\!\!\!\!\!\!
\end{align}
Here, the notation $\text{nl}_{SKL}(.)$ again represents a nonlinear function in its arguments. Since the Kirchhoff constraint has to be fulfilled in a strong manner, the interpolation scheme~\eqref{abstract_triad_interpolation2} does not only depend on the nodal triads $\mb{\Lambda}^i$, but also on the arc-length derivative $\mb{t}(\xi)$ of the discrete centerline curve~\eqref{interpolation}. The second interpolation scheme proposed in~\cite{meier2016} fulfills the Kirchhoff constraint only at the three element nodes. Consequently, an arbitrary rotation interpolation scheme can be chosen between the nodal triads $\mb{\Lambda}^i$. In~\cite{meier2016}, the same rotation interpolation scheme~\eqref{abstract_triad_interpolation} as for the Simo-Reissner case has been employed:
\begin{align}
\label{abstract_triad_interpolation3}
\begin{split}
\!\!\!\!\!\!\mb{\Lambda}(\xi) \!\approx\! \mb{\Lambda}_h(\xi) \!&=\! \text{nl}_{WKL}(\mb{\Lambda}^1,\mb{\Lambda}^2,\mb{\Lambda}^3,\xi) \\
\!&=\! \text{nl}_{SR}(\mb{\Lambda}^1,\mb{\Lambda}^2,\mb{\Lambda}^3,\xi).\!\!\!\!\!\!
\end{split}
\end{align}
The indices SKL and WKL stand for a strong and weak enforcement of the Kirchhoff constraint, respectively. Both variants apply the Hermite interpolation~\eqref{interpolation} for the beam centerline. All in all, four different finite element variants SKL-TAN, SKL-ROT, WKL-TAN, WKL-ROT result from a combination of the two interpolation schemes~\eqref{abstract_triad_interpolation2} and~\eqref{abstract_triad_interpolation3} as well as the two sets of nodal rotation parametrizations~\eqref{lambda_kirchhoff_tan} and~\eqref{lambda_kirchhoff_rot}. However, only the latter aspect will be important for the considered beam-to-beam contact schemes, whereas a detailed description and comparison of the two different rotation interpolations can be found in~\cite{meier2016}. Consequently, no distinction will be made between the SKL and WKL variants in the following, and the additional indices S and W will be omitted. The element residual and stiffness contributions of all these four variants are again summarized in~\cite{meier2016}.  Moreover, in that reference, it has been shown that the Kirchhoff-Love element formulations considered in this section typically yield a lower discretization error level per degree of freedom as comparable Simo-Reissner elements from the literature. Also the observed nonlinear solver performance clearly advocates the application of Kirchhoff-Love formulations in the range of high beam slenderness ratios. Concretely, savings in the total number of Newton iterations up to two orders of magnitude could be achieved for slenderness ratios in the range of $\zeta=10^4$ as compared to the investigated Simo-Reissner formulations. Subsequent numerical examples will confirm this trend.

\subsection{Torsion-free beam element}
\label{sec:beams_TF}

\begin{figure}[ht]
 \centering
  \includegraphics[width=0.45\textwidth]{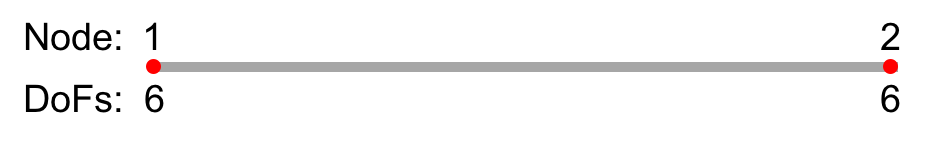}
  \caption{Degrees of Freedom of Torsion-Free element}
  \label{fig:fig_nodedof_TF}
\end{figure}

The torsion-free beam element formulation considered in the following has originally been proposed in~\cite{meier2015}. There, it has been shown that under certain restrictions concerning initial beam geometry (straight beams with isotropic/circular cross-sections) and external loads (no torsional components of external moments), expressed by
\begin{align}
\label{torsionless}
\begin{split}
\!\!\!\!\!\!
&I_2=I_3=:I, \quad 
\mb{\kappa}_0 \equiv 0, \\
&\mb{g}_1(s,t) \!\cdot\! \mb{\tilde{m}}(s,t) \!\equiv\! 0 ,\quad  
\Big[ \mb{g}_1(t) \!\cdot\! \mb{m}_{\sigma}(t) \Big]_{\varGamma_{\sigma}} \!\!\!\!\!=\!0,\!\!\!\!\!\!
\end{split}
\end{align}
 the Kirchhoff-Love theory yields solutions with vanishing torsion even for arbitrarily large displacements and rotations. Here, the Frenet-Serret vector
 \begin{align}
\label{kappa}
 \mb{\kappa}:= \frac{\mb{S}(\mb{r}^{\prime}) \mb{r}^{\prime \prime}}{|| \mb{r}^{\prime} ||^2}
\end{align}
describes the curvature of the beam centerline. This finding of vanishing torsion justified the development of a static torsion-free beam element formulation in~\cite{meier2015}. In~\cite{meier2015b}, the formulation has been extended to dynamics, yielding the following length-specific kinetic and hyper-elastic stored energies:
\begin{align}
\label{storedenergyfunction}
\!\tilde{\Pi}_{kin}\!:=\! \frac{1}{2}\rho A v^2, \,\, \tilde{\Pi}_{int}\!:=\! \frac{1}{2}\! \left[EA \epsilon^2 \!+\! EI \kappa^2\right].\!\!\!\!\!\!
\end{align}
Consequently, the deformation is entirely described by the modes of axial tension $\epsilon$ and (isotropic) bending $\kappa:=||\mb{\kappa}||$, while the considered inertia effects solely depend on the centerline velocity field $v\!:=\!||\dot{\mb{r}}||$. Deriving the weak form of balance equations from these energy expressions yields the following very compact result: 
\begin{align}
\label{torsionfree_weakform}
\begin{split}
\!\!\!\!\!\!
 0\!=\!\!\!
  & \int \limits_0^l
  \Bigg[
  \delta \epsilon EA \epsilon 
 +\delta \boldsymbol{\kappa} EI \boldsymbol{\kappa} 
 +\delta \mb{r}^T \!\! \rho A \ddot{\boldsymbol{r}}
  \Bigg] ds \\
  - \!\! & \int \limits_0^l \!\!
  \Bigg[ \!
 \delta \mb{r}^T \mb{\tilde{f}} \!+\! \delta \boldsymbol{\theta^T_{\perp}} \! \mb{\tilde{m}_{\perp}}
  \! \Bigg] ds
 \!-\!\Bigg[ \! 
 \delta \mb{r}^T \bar{\mb{f}} \!+\! \delta \boldsymbol{\theta^T_{\perp}} \! \bar{\mb{m}}_{\perp} 
 \! \Bigg]_{\Gamma_{\sigma}} \hspace{-0.35cm}. \!\!\!\!\!\!
\end{split}
\end{align}
As indicated by the subscript $(.)_{\perp}$, the torsion-free beam theory is only applicable if the external moment vectors contain no components parallel to the centerline tangent. In~\cite{meier2015}, the result that restrictions~\eqref{torsionless} lead to a state of (exactly) vanishing torsion for large-deformation Kirchhoff-Love beam problems has only been derived for static problems. However, the observations made in numerical investigations suggest that these restrictions typically lead to very small, although not exactly vanishing, torsion values also for dynamic problems, which justifies the application of the torsion-free beam element, in the following denoted as TF element, also in such scenarios. A thorough mechanical derivation of this result for the general case of dynamic Kirchhoff-Love beam problems will be addressed in a forthcoming contribution. From~\eqref{torsionfree_weakform}, it can readily be seen that the torsion-free beam problem is completely defined by the configuration of the beam centerline, which will be discretized by means of~\eqref{interpolation} also for the TF beam element formulation. Thus, the challenging and computationally involved discretization of the rotation field can completely be avoided, which results in twelve degrees of freedom entirely describing the element state:
\begin{align}
\label{lambda_torsionfree}
\begin{split}
\!\!\!\!\!\!&\mb{d}^1=\hat{\mb{d}}^1, \,\, \mb{d}^2=\hat{\mb{d}}^2, \\
\!\!\!\!\!\!&\mb{t}^1=\hat{\mb{t}}^1, \,\,\,\,\, \mb{t}^2=\hat{\mb{t}}^2, \\
\!\!\!\!\!\!& \hat{\mbd{x}}_{TF}\!=\!\hat{\mbd{d}}_{TF}\!:=\!(\hat{\mb{d}}^{1T}\!,\hat{\mb{t}}^{1T}\!,\hat{\mb{d}}^{2T}\!,\hat{\mb{t}}^{2T}\!)^T.\!\!\!\!\!\!
\end{split}
\end{align}
For the TF beam element, the required number of degrees of freedom is visualized in Figure~\ref{fig:fig_nodedof_TF}.

The resulting element residual vector and stiffness matrix for the dynamic case are summarized in~\cite{meier2015b}. Due to the complete avoidance of any rotational degrees of freedom, standard procedures such as spatial discretization (in a manner that preserves objectivity of deformation measures), linearization, configuration updates, or time integration via finite differences in dynamics are considerably simplified as compared to standard geometrically exact beam elements, while still inheriting the high degree of computational accuracy and efficiency of these formulations. In contrast to standard geometrically exact formulations, the TF element will result in a symmetric tangent stiffness matrix (as long as no external moment contributions are considered, which are known to be non-conservative) and a symmetric and constant mass matrix (see also Section~\ref{sec:beams_temporal_TF}). The torsion-free model also seems to provide an ideal tool for the mechanical investigation of cable-like structures or 1D quasi-continua such as chains. In pure cable formulations, artificial bending terms are often necessary in order to stabilize compressional modes. On the contrary, the torsion-free formulation
naturally provides such a stabilization in a mechanically consistent manner.

\subsection{Temporal discretization}
\label{sec:beams_temporal}

While spatial discretization has exclusively been based on the finite element method (FEM), finite difference (FD) schemes will be employed for temporal discretization. Thereto, the considered time interval of interest $t \in [0,T]$ is subdivided into equidistant subintervals $[t_n,t_{n+1}]$ with constant time step size $\Delta t$, where $n \in \mathbb{N}_0$ is the time step index. Consequently, the solution for the primary variable fields describing the current configuration $C(s,t) \!:=\! (\mb{r}(s,t), \mb{\Lambda}(s,t))$ is computed at a series of discrete points in time with configurations $C(s,t_n) \!:=\! (\mb{r}(s,t_n), \mb{\Lambda}(s,t_n))\!=:\! (\mb{r}_n(s), \mb{\Lambda}_n(s))$. In the following, the temporal discretization of the TF beam element presented in the last section, as well as of the Simo-Reissner and Kirchhoff-Love elements presented previously, will be considered.

\subsubsection{Torsion-free beam element}
\label{sec:beams_temporal_TF}

For the torsion-free beam element, only the second time derivative of the primary variable field $\mb{r}(s,t)$ appears (linearly) in the weak form~\eqref{torsionfree_weakform}. Consequently, the assembled global residual vector~\eqref{global_system} of the spatially discretized, time-continuous beam problem can be simplified for this formulation on the basis of a constant global mass matrix $\mb{M}$:
\begin{align}
\label{global_system_tf}
\begin{split}
\!\!\!\!\!\!\mb{R}_{tot}\!&=\! \mb{M} \ddot{\mb{X}}(t) \!+\!\mb{R}_{int}(\mb{X}(t))\!-\!\mb{R}_{ext}(\mb{X}(t))\\
&+ \mb{R}_{con}(\mb{X}(t))\!=\!\mb{0}.\!\!\!\!\!\!
\end{split}
\end{align}
The structure of this system of residual equations is identical to the structure resulting from standard 3D volume finite elements. Consequently, also a standard time discretization scheme typically combined with this category of 3D finite elements can be chosen. In the present work, the well-known (standard) generalized$-\alpha$ method~\cite{chung1993} based on the parameters $\beta,\gamma,\alpha_m$ and $\alpha_f$ is employed for temporal discretization of the global vector of nodal primary variables $\mb{X}(t)$. The temporal discretization process is indentical to standard 3D elements and the required constant mass matrix is given in~\cite{meier2015b}.

\subsubsection{Simo-Reissner and Kirchhoff-Love elements}
\label{sec:beams_temporal_SRKL}

For temporal discretization of the Simo-Reissner and Kirchhoff-Love beam elements considered in this work, an extension of the generalized$-\alpha$ scheme recently proposed by~\cite{arnold2007,bruels2010,bruels2012} for the treatment of large rotations will be applied. Compared to the standard generalized$-\alpha$ scheme, the extended variant allows for temporal discretization based on multiplicative rotation increments. Such a procedure is independent from the specific rotation parametrization, which not only leads to considerably simplified discrete expressions, but also to a very general scheme that can be directly applied to arbitrary Simo-Reissner or Kirchhoff-Love beam element formulations without the need for additional adaptions. Also this extended generalized$-\alpha$ scheme can be identified as an implicit, one-step finite difference scheme inheriting the desirable properties of the standard generalized$-\alpha$ scheme such as second-order accuracy, unconditional stability (within the linear regime), controllable damping of the high-frequency modes and minimized damping of the low-frequency modes. Remarkably, the parameter choice leading to this optimal behavior is identical to that of the standard generalized$-\alpha$ scheme. Due to the inertia moment term $\mb{S}(\mb{w}) \mb{c}_{\rho} \mb{w} \!+\! \mb{c}_{\rho} \mb{a}$ in~\eqref{stress_resultants_and_inertia}, the inertia residual contributions will in general depend on the primary variables and their time derivatives in a nonlinear manner and does consequently not allow for a simplification as in~\eqref{global_system_tf}. More details on the combination of this extended generalized$-\alpha$ scheme with the geometrically exact beam element formulations considered herein as well as the inertia residual and stiffness contributions can be found in~\cite{meier2016}. The corresponding inertia residual and stiffness contributions of the Simo-Reissner element proposed in Section~\ref{sec:beams_SR} are presented in~\ref{anhang:sr_element}.

\section{Beam-to-beam contact formulation}
\label{sec:contact}

In this section, the essential constituents of the all-angle beam contact (ABC) formulation proposed in~\cite{meier2015c} (see Section~\ref{sec:contact_abc}), as well as the underlying point-to-point (Section~\ref{sec:contact_point}) and line-to-line (Section~\ref{sec:contact_line}) contact models will be repeated. While for most of the finite element formulations considered in the present work, the contact residual and stiffness contributions derived in~\cite{meier2015c} can directly be applied, the Kirchhoff-Love element formulation with a rotation vector-based parametrization of the nodal triads according to~\eqref{lambda_kirchhoff_tan} requires some additional transformations. These transformations will be presented in Section~\ref{sec:contact_adaption}.

In the following, two beams $1$ and $2$ with cross-section radii $R_1$ and $R_2$ are considered. The beam centerlines are represented by two parametrized curves~$\mb{r}_{1}(\xi)$ and $\mb{r}_{2}(\eta)$ with curve parameters~$\xi$ and $\eta$. Furthermore, $\mb{r}_{1,\xi}(\xi)= \mb{r}_1^{\shortmid}(\xi)$ and $\mb{r}_{2,\eta}(\eta)=\mb{r}_2^{\shortmid}(\eta)$ denote the tangents to these curves at positions $\xi$ and $\eta$, respectively. It is assumed that for the spatially discretized space curves a unique tangent vector exists at every position $\xi$ and $\eta$. This requirement is satisfied by means of the employed $C^1$-continuous Hermite interpolation~\eqref{interpolation}.

\subsection{Point contact model}
\label{sec:contact_point}

\begin{figure}[t]
 \centering
  \includegraphics[width=0.4\textwidth]{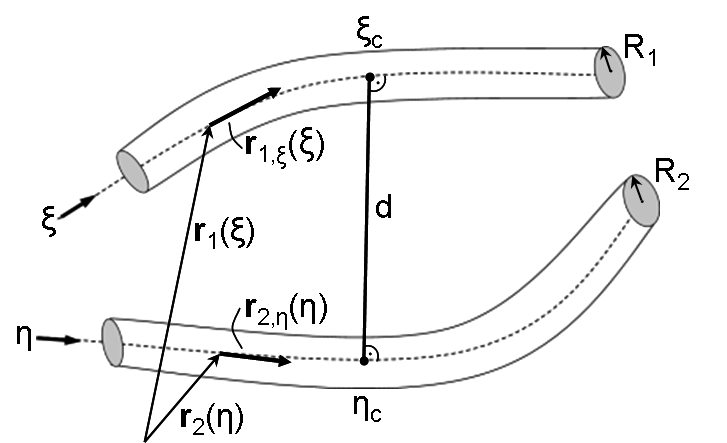}
  \caption{Point-to-point contact problem of two beams.}
  \label{fig:point_problemdescription1}
\end{figure}

In the following, the basics of the well-known point contact formulation~\cite{wriggers1997} are presented. The point-to-point beam contact model enforces the contact constraint by prohibiting penetration of the two beams at the closest point positions $\xi_c$ and $\eta_c$. The coordinates of these points are defined as solution of the bilateral closest point projection:
\begin{align}
\label{point_mindist}
\begin{split}
  d_{bl}&:=\min_{\xi, \eta} d(\xi, \eta) = d(\xi_c, \eta_c), \\
   d(\xi, \eta)&:=||\mb{r}_{1}(\xi)-\mb{r}_{2}(\eta)||.
\end{split}
\end{align}
This leads to two orthogonality conditions to be solved for the unknown coordinates $\xi_c$ and $\eta_c$:
\begin{align}
\label{point_orthocond}
\begin{split}
  \mb{r}^T_{1,\xi}(\xi)\left( \mb{r}_{1}(\xi)-\mb{r}_{2}(\eta) \right) &\dot{=} 0, \\
  \mb{r}^T_{2,\eta}(\eta)\left( \mb{r}_{1}(\xi)-\mb{r}_{2}(\eta) \right) &\dot{=} 0.
\end{split}
\end{align}
The non-penetration condition at the closest point is formulated by means of the inequality constraint
\begin{align}
  g \geq 0 \quad \text{with} \quad g:=d_{bl}-R_1-R_2,
\end{align}
where $g$ is the gap function. This constraint is enforced by means of the penalty potential
\begin{align}
\label{point_penaltypotential}
     \!\!\!\!\!\!
     \quad \Pi_{c\varepsilon}=\frac{1}{2} \varepsilon \langle g\rangle^2 \quad \text{with} \quad 
           \langle x\rangle=\left\{\begin{array}{ll}
                                 x, & x \leq 0 \\
                                 0, & x > 0
                    \end{array}\right. . \!\!\!\!\!\!
\end{align}
Variation of the penalty potential~\eqref{point_penaltypotential} yields the point contact contribution to the weak form:
\begin{align}
\label{point_weakform}
\begin{split}
  \delta \Pi_{c\varepsilon} &=
      \varepsilon \langle g\rangle \delta g =
      \varepsilon \langle g\rangle \left(\delta \mb{r}_{1c} - \delta \mb{r}_{2c} \right)^T \! \mb{n}, \\
      \mb{f}_{c\varepsilon}&= \underbrace{-\varepsilon \langle g\rangle}_{=:f_{c\varepsilon}} \mb{n}, \,\,\,\,\,\,  \mb{n}:=\frac{\mb{r}_{1}(\xi_c)-\mb{r}_{2}(\eta_c)}{||\mb{r}_{1}(\xi_c)-\mb{r}_{2}(\eta_c)||} \,\,.
\end{split}
\end{align}
According to~\eqref{point_weakform}, the point-to-point beam contact formulation models the contact force $\mb{f}_{c\varepsilon}$ that is transferred between the two beams 
as a discrete point force acting at the closest points of the beam centerlines in normal direction $\mb{n}$. The kinematic quantities introduced above are illustrated in Figure~\ref{fig:point_problemdescription1}. For later use, also the so-called contact angle shall be defined as the angle between the tangent vectors at the contact points:
\begin{align}
\label{point_contactangle}
    \alpha = \arccos{ \left(z \right) },
  \quad z=\frac{ ||\mb{r}_1^{\shortmid T}(\xi_c) \mb{r}_2^{\shortmid}(\eta_{c})|| }{ ||\mb{r}_1^{\shortmid}(\xi_c)|| \cdot ||\mb{r}_2^{\shortmid}(\eta_{c})|| }.
\end{align}
The contact contribution to the weak form is completely determined by the two beam centerline curves. Thus, insertion of~\eqref{interpolation} into~\eqref{point_weakform} followed by a consistent linearization yields the contact contributions to the element residual vector and stiffness matrix, which are e.g. summarized in~\cite{wriggers1997,meier2015c}. 

\subsection{Line contact model}
\label{sec:contact_line}

\begin{figure}[ht]
 \centering
  \subfigure[Space continuous problem setting.]
   {
    \includegraphics[width=0.4\textwidth]{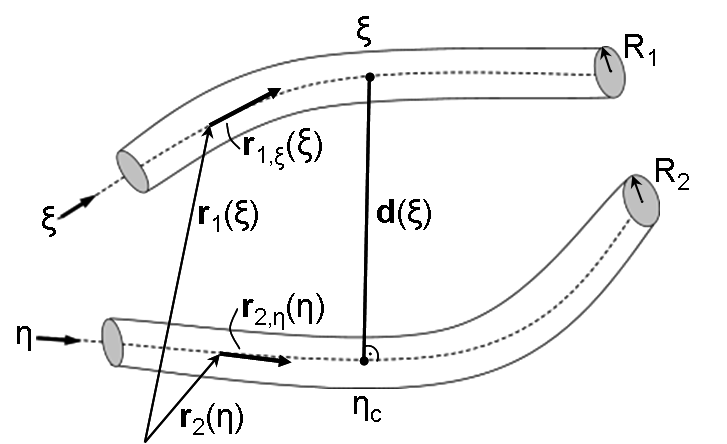}
    \label{fig:line_problemdescription_contiuous}
   }
   \hspace{0.05 \textwidth}
   \subfigure[Discretized problem setting.]
   {
    \includegraphics[width=0.4\textwidth]{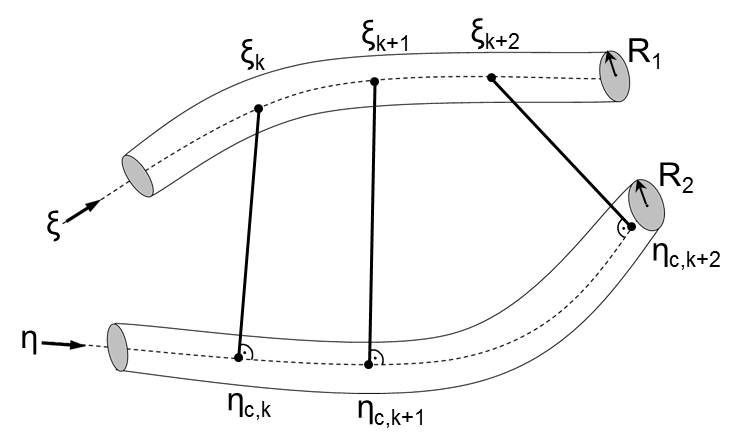}
    \label{fig:line_problemdescription_discrete}
   }
  \caption{Line-to-line contact problem of two beams.}
  \label{fig:line_problemdescription}
\end{figure}

In this section, the most important aspects of the line-to-line contact formulation proposed in \cite{meier2015b} shall be repeated. In contrary to the point-to-point contact model, this formulation is based on a line constraint enforced along the entire beam length. The relevant kinematic quantities of this approach are illustrated in Figure~\ref{fig:line_problemdescription_contiuous}.\\

In case of line contact, a distinction has to be made between a master beam (beam $1$) and a slave beam (beam $2$). The closest master point~$\eta_c$ to a given slave point~$\xi$ is determined as solution of the unilateral (``ul'') minimal distance problem
\begin{align}
\label{line_mindist}
\begin{split}
  d_{ul}(\xi)&:=\min_{\eta} d(\xi,\eta)= d(\xi,\eta_c), \\
   d(\xi,\eta) &:= ||\mb{r}_{1}(\xi)-\mb{r}_{2}(\eta)||.
\end{split}
\end{align}
Condition~\eqref{line_mindist} leads to one orthogonality condition to be solved for the unknown coordinate $\eta_c$:
\begin{align}
\label{line_orthocond}
  \mb{r}^T_{2,\eta}(\eta)\left( \mb{r}_{1}(\xi)-\mb{r}_{2}(\eta) \right) \dot{=} 0.
\end{align}
Thus, in contrary to the point contact model, the normal vector is still perpendicular to the master beam $2$, but not to the slave beam $1$ anymore. The non-penetration condition becomes
\begin{align}
\label{line_constraint}
  \!\!\!\!\!\! g(\xi) \geq 0 \, \forall \, \xi \,\,\, \text{with} \,\,\, g(\xi):=d_{ul}(\xi)-R_1-R_2,\!\!\!\!\!\!
\end{align}
and is integrated into the variational problem formulation by means of the penalty potential
\begin{align}
\label{line_pen_totalpotential}
     \Pi_{c\varepsilon}=\frac{1}{2} \varepsilon \int \limits_0^{l_1} \langle g(\xi) \rangle^2 ds_1.
\end{align}
Variation of the penalty potential defined in~\eqref{line_pen_totalpotential} leads to the contact contribution to the weak form:
\begin{align}
\label{line_pen_weakform}
\begin{split}
  \delta \Pi_{c\varepsilon} &= \varepsilon \int \limits_0^{l_1} \langle g(\xi)\rangle \delta g (\xi)  ds_1 \\
   \text{with} \quad \delta g(\xi) &= ( \delta \mb{r}_1(\xi) - \delta \mb{r}_2(\xi) )^T \mb{n}(\xi).
\end{split}
\end{align}
In~\eqref{line_pen_weakform}, the contact force vector $\mb{f}_{c\varepsilon}(\xi)$ as well as the normal vector $\mb{n}(\xi)$ can be identified:
\begin{align}
\label{line_pen_contactforce}
\begin{split}
  \mb{f}_{c\varepsilon}(\xi) &= \underbrace{- \varepsilon \langle g(\xi)\rangle }_{=:f_{c\varepsilon}(\xi)} \mb{n}(\xi),\\
    \mb{n}(\xi)&:=\frac{\mb{r}_{1}(\xi)-\mb{r}_{2}(\eta_c)}{||\mb{r}_{1}(\xi)-\mb{r}_{2}(\eta_c)||}.
  \end{split}
\end{align}
According to \eqref{line_pen_contactforce}, a line-to-line beam contact formulation models the contact force $\mb{f}_{c\varepsilon}(\xi)$ that is transferred between the beams as a distributed 
line force. Also in the line contact case, the contact angle field can be defined according to:
\begin{align}
\label{line_contactangle}
  \!\!\!\!\!\!
  \alpha(\xi) \!=\! \arccos{ \left(z(\xi) \right) },
  \,\,\,  z(\xi) \!= \!\frac{ ||\mb{r}_1^{\shortmid T}(\xi) \mb{r}_2^{\shortmid}(\eta_{c})|| }{ ||\mb{r}_1^{\shortmid}(\xi)|| \, ||\mb{r}_2^{\shortmid}(\eta_{c})|| }.\!\!\!\!\!\!
\end{align}
Eventually, spatial discretization has to be carried out by inserting the discretization \eqref{interpolation} into equation \eqref{line_pen_weakform} and replacing the analytical integral by a Gaussian quadrature (see Figure~\ref{fig:line_problemdescription_discrete}). The element residual and stiffness contributions resulting from this line-to-line contact scheme are summarized in~\cite{meier2015b}. Furthermore, in this reference, a special integration interval segmentation procedure has been proposed in order to avoid the numerical integration across strong discontinuities at master beam endpoints, which, in turn, leads to a considerable reduction of the overall discretization error.

\subsection{All-angle beam contact formulation}
\label{sec:contact_abc}

In~\cite{meier2015b,meier2015c}, it has been shown that line contact formulations applied to slender beams provide very accurate and robust mechanical models in the range of small contact angles, whereas the computational efficiency considerably decreases with increasing contact angles. On the other hand, point contact formulations serve as sufficiently accurate and very efficient models in the regime of large contact angles, while they are not applicable for small contact angles as a consequence of non-unique closest point projections. In~\cite{meier2015c}, the ABC formulation has been proposed in order to combine the advantages of these two basic types of formulations. This formulation applies a point contact formulation in the range of large contact angles and a line contact formulation in the range of small contact angles. The smooth model transition between these two regimes within a prescribed angle interval
\begin{align}
\label{ABC_shiftangles}
[\alpha_1;\alpha_2], \quad \alpha_1,\alpha_2 \in [0^{\circ};90^{\circ}], \quad \alpha_1 < \alpha_2,
\end{align}
is realized by defining the following angle-dependent transition factor $k(z)$, with $z=\cos(\alpha)$:
\begin{align}
\label{ABC_forcebased_transition}
     \!\!\!\!\!\!
     k(z) \!=\! \left\{\begin{array}{lll}
                                    \!\!\!1, & \alpha < \alpha_1 \\
                                    \!\!\!0.5 \! \left(1\!-\!\cos\left(\pi\frac{z\!-\!z_2}{z_1\!-\!z_2}\right)\right), & \alpha_2 \!\geq\! \alpha \!\geq\! \alpha_1 \!\!\!\!\!\! \\
                                    \!\!\!0, & \alpha > \alpha_2
        \end{array}\right.\!\!\!\!\!\!
\end{align}
In~\cite{meier2015c}, two different variants of model transition have been investigated, one on penalty force level,
\begin{align}
\label{ABC_forcebased_weakform}
\begin{split}
\delta \Pi_{c\varepsilon} &= \underbrace{\left[1-k(z_c)\right] \varepsilon_{\perp} \langle g \rangle}_{=:-f_{c\varepsilon  \perp}} \delta g \\ 
&+ \int \limits_0^{l_1} \underbrace{k(z(\xi)) \varepsilon_{\parallel} \langle g(\xi)\rangle}_{=:-f_{c \varepsilon  \parallel}(\xi)} \delta g (\xi) ds_1,
\end{split}
\end{align}
as well as one variant on penalty potential level, viz.
\begin{align}
\label{ABC_potentialbased_potential}
\begin{split}
\Pi_{c\varepsilon} &= \frac{1}{2}\varepsilon_{\perp} (1-k(z_c)^2) \langle g \rangle ^2 \\
&+ \frac{1}{2}\varepsilon_{\parallel}\int \limits_0^{l_1} k^2(z(\xi)) \langle g(\xi)\rangle^2 ds_1.
\end{split}
\end{align}
Variation of the potential-based variant~\eqref{ABC_potentialbased_potential} leads to additional contact moment contributions as compared to~\eqref{ABC_forcebased_weakform}. 
These contributions are required for a variationally consistent formulation that allows for exact conservation of energy, which is not necessarily guaranteed by the force-based model transition~\eqref{ABC_forcebased_weakform}. However, in~\cite{meier2015c}, an optimal ratio of the point penalty parameter $\varepsilon_{\perp}$ and the line penalty parameter $\varepsilon_{\parallel}$ has been derived such that the non-conservative work contributions of the force-based variant are minimized and this considerably simpler formulation can be applied to most practically 
relevant problem classes. A spatial discretization procedure similar to the ones for the underlying basic contact models again allows to derive the contact residual and stiffness contributions, see~\cite{meier2015c}. In that reference, further important algorithmic aspects can be found concerning contact search algorithm, smoothed penalty laws, consistent treatment of beam endpoint contacts or the application of a step size control that avoids an undetected crossing of beams in the range of large time step sizes.

\subsection{Rotation vector-based Kirchhoff-Love element}
\label{sec:contact_adaption}

In general the element residual contributions stemming from the contact interaction of two finite elements $1$ and $2$ can be written as:
\begin{align}
\label{residual_contact}
\mbd{r}_{con,1}(\hat{\mbd{x}}_{1},\hat{\mbd{x}}_{2})\quad \text{and} \quad \mbd{r}_{con,2}(\hat{\mbd{x}}_{1},\hat{\mbd{x}}_{2})
\end{align}
In a similar fashion, also the contact contributions to the element stiffness matrices can be formulated:
\begin{align}
\label{stiffness_contact}
\mbd{k}_{con,11}:=\frac{d \mbd{r}_{con,1}}{d \hat{\mbd{x}}_{1}}, \quad  \mbd{k}_{con,12}:=\frac{d \mbd{r}_{con,1}}{d \hat{\mbd{x}}_{2}},\\
\mbd{k}_{con,21}:=\frac{d \mbd{r}_{con,2}}{d \hat{\mbd{x}}_{1}}, \quad  \mbd{k}_{con,22}:=\frac{d \mbd{r}_{con,2}}{d \hat{\mbd{x}}_{2}},
\end{align}
As shown in~\cite{meier2015c}, the beam-to-beam contact schemes considered in this work yield residual and stiffness contributions that solely depend on the 
sets of degrees of freedom $\hat{\mbd{d}}_1$ and $\hat{\mbd{d}}_2$ defining the beam centerline. Since this set $\hat{\mbd{d}}_i$ defining 
the beam centerline is identical for the Simo-Reissner element~\eqref{lambda_reissner} (SR), the Kirchhoff-Love element with tangent vector-based nodal rotation parametrization~\eqref{lambda_kirchhoff_tan} (KL-TAN) and the torsion-free element~\eqref{lambda_torsionfree} (TF), the residual and stiffness contributions derived in~\cite{meier2015c} can directly be applied to these elements without the need for any further adaption. Only a correct assembly of the corresponding rows and columns based on the arrangement of the centerline degrees of freedom $\hat{\mbd{d}}$ within the vector $\hat{\mbd{x}}$ has to be considered.

On the other hand, for the rotation vector-based Kirchhoff-Love element~\eqref{lambda_kirchhoff_rot} (KL-ROT), the nodal position and tangent vectors $\mb{d}^1,\mb{d}^2,\mb{t}^1$ and $\mb{t}^2$ of the centerline interpolation~\eqref{interpolation} represent no primary variables, and, thus, the set $\hat{\mbd{d}}$ is different for this element. However, in~\cite{meier2016} it has been shown how standard element residual vectors and stiffness matrices can be transformed between the element variants KL-TAN and KL-ROT. Since the contact contributions of the KL-TAN element can directly be taken from~\cite{meier2016}, the contributions of the KL-ROT element can be determined via a similar transformation, which shall in the following be given for the beam-to-beam contact case. In~\cite{meier2016}, the variations $\delta \hat{\mbd{x}}_{TAN}$ and $\delta \hat{\mbd{x}}_{ROT}$ as well as the iterative increments $\Delta \hat{\mbd{x}}_{TAN}$ and $\Delta \hat{\mbd{x}}_{ROT}$ of the primary variable sets~\eqref{lambda_kirchhoff_tan} and~\eqref{lambda_kirchhoff_rot} are given, and the required (linear) transformation rule has been stated as:
\begin{align}
\label{anhang_lintanrot3}
\begin{split}
\delta \hat{\mbd{x}}_{TAN}\!=\!\tilde{\mbd{T}}_{\hat{\mbd{x}}} \delta \hat{\mbd{x}}_{ROT}, \quad
\Delta \hat{\mbd{x}}_{TAN}\!=\!\mbd{T}_{M\hat{\mbd{x}}} \Delta \hat{\mbd{x}}_{ROT}.
\end{split}
\end{align}
The deformation-dependent transformation matrices $\tilde{\mbd{T}}_{\hat{\mbd{x}}}$ and $\mbd{T}_{M\hat{\mbd{x}}}$ are specified in~\cite{meier2016}. Now, the transformation between element residual vectors~\eqref{residual_contact} and stiffness matrices~\eqref{stiffness_contact} associated with the KL-TAN element formulation (index $(.)_{TAN}$) and the KL-ROT element formulation (index $(.)_{ROT}$) will be conducted. For simplicity, the index $(.)_{con}$ in~\eqref{residual_contact} and~\eqref{stiffness_contact} will be dropped. According to~\cite{meier2016}, the residual transformations yield:
\begin{align}
\label{anhang_lintanrot6}
   \mbd{r}_{ROT,1} \!=\! \tilde{\mbd{T}}_{\hat{\mbd{x}},1}^T \mbd{r}_{TAN,1}, \quad
   \mbd{r}_{ROT,2} \!=\! \tilde{\mbd{T}}_{\hat{\mbd{x}},2}^T \mbd{r}_{TAN,2}.
\end{align}
Now, the linearization of the residual $\mbd{r}_{TAN,1}$ of the first contact element shall exemplarily be derived:
\begin{align}
\label{anhang_lintanrot7}
\begin{split}
   \!\!\!\!\!\!
   \Delta \mbd{r}_{ROT,1} \!&=\!\! \Delta \tilde{\mbd{T}}_{\hat{\mbd{x}},1}^T \mbd{r}_{TAN,1} \!+\! \tilde{\mbd{T}}_{\hat{\mbd{x}},1}^T \! \Delta \mbd{r}_{TAN,1} \!\!\!\!\!\! \\
   &\dot{=}\mbd{k}_{ROT,11} \Delta \hat{\mbd{x}}_{ROT,1} \!+\! \mbd{k}_{ROT,12} \Delta \hat{\mbd{x}}_{ROT,2}, \!\!\!\!\!\!
\end{split}
\end{align}
where the linearization $\Delta \mbd{r}_{TAN,1}$ can be written as
\begin{align}
\label{anhang_lintanrot7b}
\begin{split}
   \Delta \mbd{r}_{TAN,1}\!&=\! \mbd{k}_{TAN,11} \Delta \hat{\mbd{x}}_{TAN,1} \\ 
   &+\!\mbd{k}_{TAN,12}\Delta \hat{\mbd{x}}_{TAN,2} \\
   \!&=\! \mbd{k}_{TAN,11} \mbd{T}_{M\hat{\mbd{x}},1} \Delta \hat{\mbd{x}}_{ROT,1} \\ 
   &+\!\mbd{k}_{TAN,12} \mbd{T}_{M\hat{\mbd{x}},2} \Delta \hat{\mbd{x}}_{ROT,2},
\end{split}
\end{align}
and the linearization $\Delta \tilde{\mbd{T}}_{\hat{\mb{x}},1}^T\mbd{r}_{TAN,1}$ is given by
\begin{align}
\label{anhang_lintanrot8}
   \!\!\!\!\!\! \Delta \tilde{\mbd{T}}_{\hat{\mbd{x}},1}^T \mbd{r}_{TAN,1}=: \tilde{\mbd{H}}_{\hat{\mbd{x}},1} (\mbd{r}_{TAN,1})  \Delta \hat{\mbd{x}}_{ROT,1},\!\!\!\!\!\!
\end{align}
according to the definitions introduced in~\cite{meier2016}. Combining equations~\eqref{anhang_lintanrot7}-\eqref{anhang_lintanrot8} and applying the same procedure 
to $\mbd{r}_{ROT,2}$ eventually yields:
\begin{align}
\label{anhang_lintanrot9b}
\begin{split}
   \!\!\!\!\!\!
   \mbd{k}_{ROT,11}\!&=\!\tilde{\mbd{H}}_{\hat{\mb{x}},1} (\mbd{r}_{TAN,1})\!+\!\tilde{\mbd{T}}_{\hat{\mb{x}},1}^T \mbd{k}_{TAN,11} \mbd{T}_{M\hat{\mb{x}},1},\\
   \mbd{k}_{ROT,12}\!&=\!\tilde{\mbd{T}}_{\hat{\mb{x}},1}^T \mbd{k}_{TAN,12} \mbd{T}_{M\hat{\mb{x}},2}, \\
      \mbd{k}_{ROT,22}\!&=\!\tilde{\mbd{H}}_{\hat{\mb{x}},2} (\mbd{r}_{TAN,2})\!+\!\tilde{\mbd{T}}_{\hat{\mb{x}},2}^T \mbd{k}_{TAN,22} \mbd{T}_{M\hat{\mb{x}},2},\\
   \mbd{k}_{ROT,21}\!&=\!\tilde{\mbd{T}}_{\hat{\mb{x}},2}^T \mbd{k}_{TAN,21} \mbd{T}_{M\hat{\mb{x}},1}.\!\!\!\!\!\!
\end{split}
\end{align}
Equations~\eqref{anhang_lintanrot6} and~\eqref{anhang_lintanrot9b} allow to determine the corresponding element residual vectors and stiffness matrices of the KL-ROT element for given contributions of the KL-TAN element.


\section{Numerical examples}
\label{sec:examples}

In this section, the previously described beam element formulations from Section~\ref{sec:beams} as well as the smooth beam contact formulation from Section~\ref{sec:contact} will be applied to several exemplary fiber-based structures of varying system size and geometrical complexity. The applicability and accuracy of the methods will be verified numerically in quasi-static as well as dynamic test scenarios. Moreover, these examples aim at comparing the results of TF, KL and SR elements based on the characteristics of the structures and the applied boundary conditions. 

In all following simulations, a Newton-Raphson scheme is used in order to solve the set of nonlinear equations resulting from the temporally and spatially discretized weak form of the balance equations. Unless otherwise stated, the Euclidean norms of the displacement increment vector and of the residual vector are used as convergence criteria. Typically, the corresponding tolerances were chosen as~$10^{-10}$ and $10^{-7}$ respectively. Additionally, a load step control (in quasi-static simulations) or time step control (in dynamic simulations) algorithm is applied. The scheme starts with an initial value of $N_0$ steps. If the Newton-Raphson scheme has not converged within a prescribed number of iterations, the step size is halved and the step is repeated. This procedure is repeated until convergence is achieved. Then, after four subsequent steps with low step size level, the step size is doubled again. Also this procedure of successively doubling the step size after four converging steps at the current step size level is repeated until the original step size is reached again. This procedure will not only drastically increase the overall computational efficiency, it also allows for comparatively objective and fair comparisons of the performance of the Newton-Raphson scheme for different element formulations. In the context of beam contact simulations, additionally a step size control is applied which limits the maximal value of the displacement value per Newton step to a previously defined upper bound (see~\cite{meier2015c}). This upper bound is typically chosen as half of the minimal cross-section radius present in the respective example. On the one hand, this procedure prevents an undetected crossing of two beams within one iteration and on the other hand, it yields a more robust path to convergence.

\subsection{Complex fiber-based microstructures}
\label{sec:examples_micro}

\begin{figure*}[ht!!!]
 \centering
    \includegraphics[width=0.7\textwidth]{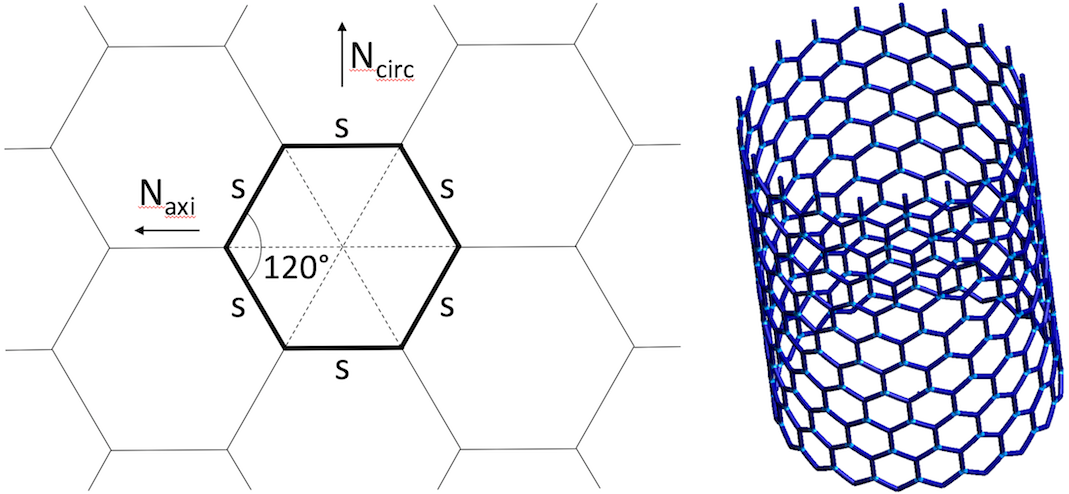}
   \caption{Schematic of the hexagonal microstructure geometry (left) and exemplary cylinder structure (right).}
  \label{fig:examples_micro_1}
\end{figure*}

Within this section, the applicability of the geometrically exact beam elements introduced in Section~\ref{sec:beams} and the smooth contact formulation presented
in Section~\ref{sec:contact} to the modeling of rather complex fiber-based structures and microstructures shall be illustrated. For this purpose, we exemplarily
investigate a generic cylindrical tube that is formed by curved hexagonal microstructures along its circumference and axis, see Figure~\ref{fig:examples_micro_1}.
The resulting cylindrical structures somewhat resemble the chemical bond layout of carbon nanotubes~\cite{Wong1997,Li2003} and could possibly be applied for
their mechanical analysis. However, within our present work these microstructures shall merely serve as typical representative of complex fiber-based assemblies,
and the focus of our analysis clearly lies on the accuracy and efficiency of the proposed SR and KL beam element formulations, respectively.

\begin{figure}[ht]
 \centering
    \includegraphics[width=0.3\textwidth]{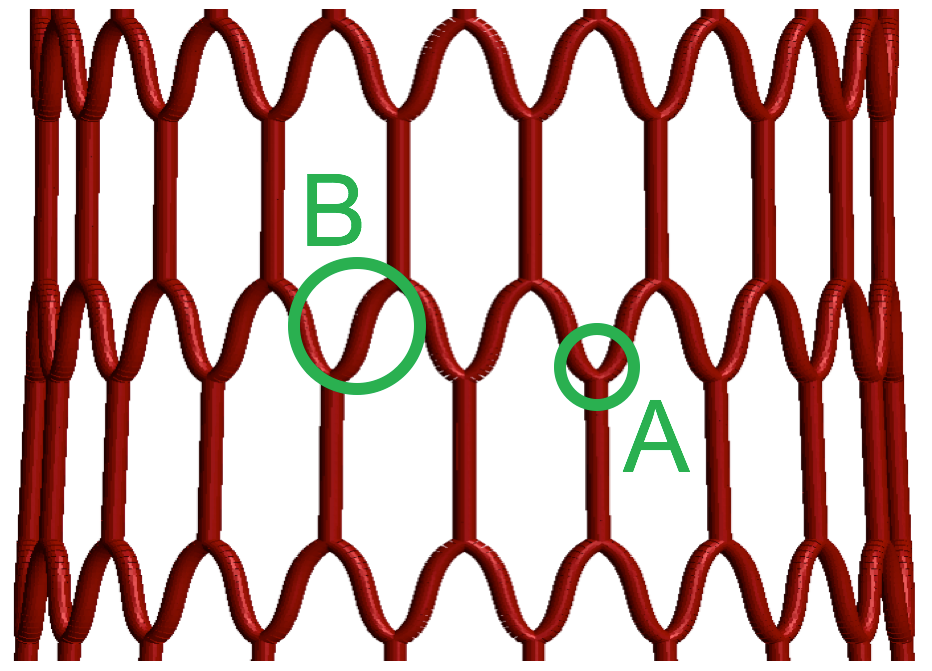}
   \caption{Detailed view of the microstructural deformation behavior with $\tilde{R}=25$, $R=0.5$, $N_{circ}=20$ and $N_{axi}=10$: Proper representation of the rigid
            connections (A) and physically meaningful deformation of the hexagon sides (B).}
  \label{fig:examples_micro_3}
\end{figure}

As can be seen in Figure~\ref{fig:examples_micro_1}, the fundamental building block for the chosen structures are regular hexagonal unit cells with all internal
angles being~$120^\circ$ and the side length~$s$ being the only free parameter. Models at different microstructural refinement levels can easily be generated by
setting the cylinder radius~$\tilde{R}$ as well as the number of hexagons along the circumference~$N_{circ}$ and along the cylinder axis~$N_{axi}$. This set of parameters
is complemented by the cross-section radius~$R$ of the beam segments representing the hexagon sides, while the finite element discretization is based on one beam element per hexagon side. However, owing to the~$C^1$-continuous beam centerline interpolation presented in Section~\ref{sec:beams_centerline}, this modeling approach is already sufficient for a smooth external
geometry representation of the cylinder structures. Concretely, the employed KL-ROT elements enable a straight-forward modeling of the rigid joints between the hexagon segments (see also~\cite{meier2016}) such that the nodal centerline tangents of finite elements connected at these joints (see e.g. point A in Figure~\ref{fig:examples_micro_3}) all lie within one tangent plane. Such a smooth representation of the enveloping cylinder hull turned out to be very beneficial for the robustness of beam-to-beam contact schemes. An exemplary structure resulting from this pre-processing framework is also
illustrated in Figure~\ref{fig:examples_micro_1}. Within the following numerical investigations we apply two beam element types: the already mentioned KL-ROT element from Section~\ref{sec:beams_KL} and the SR element from Section~\ref{sec:beams_SR}. The cylinder radius is set to~$\tilde{R}=25$ and, at least implicitly, the cylinder length is defined by requiring that~$N_{circ}/N_{axi}=2$. With regard to microstructural refinement, the four different levels~$N_{circ}=10,20,40,80$ are considered, and in terms of beam segment slenderness ratio we compare the three cases~$R=0.5$ (rather thick beams), $R=0.1$ (average slenderness) and~$R=0.02$ (rather thin beams). The boundary conditions for the numerical analysis are chosen as follows: all nodes at the cylinder bottom are simply
supported,~i.e.~$u_x=u_y=u_z=0$, while a prescribed displacement~$u_z=100$ (still with~$u_x=u_y=0$) is applied at the cylinder top in the direction of the cylinder axis.
As mentioned above, all interior joints are modeled as rigid connections. The simulations are carried out using a quasi-static load-stepping scheme and standard
Newton--Raphson iterations for solving the nonlinear problem. 

\begin{figure*}[ht!!!]
 \centering
    \includegraphics[width=0.72\textwidth]{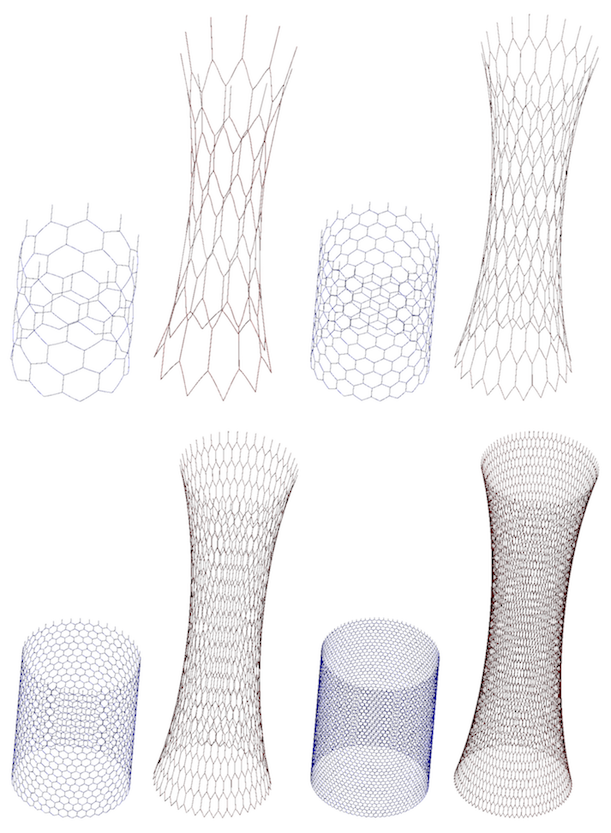}
   \caption{Axial tension test on cylindrical tube with hexagonal microstructure: Initial and deformed geometries for $\tilde{R}=25$, $R=0.1$ and $N_{circ}=10$ (top left), $N_{circ}=20$ (top right),
            $N_{circ}=40$ (bottom left) as well as $N_{circ}=80$ (bottom right).}
  \label{fig:examples_micro_2}
\end{figure*}

Some exemplary initial and final deformed configurations for the chosen setup are visualized in Figure~\ref{fig:examples_micro_2}. It can be seen that the prescribed
displacements lead to very large deformations both at the macroscopic structural level as well as at the microstructural level. In particular, we would like to
point out the deformation behavior in the vicinity of the hexagon joints, where the higher-order $C^1$-continuous centerline interpolation allow for
an accurate representation of the rigid connections and a physically meaningful deformation of the hexagon sides themselves at the same time, see Figure~\ref{fig:examples_micro_3}.

\begin{figure}[h!!!]
 \centering
  \subfigure[beam radius $R=0.5$.]
   {
    \includegraphics[width=0.45\textwidth]{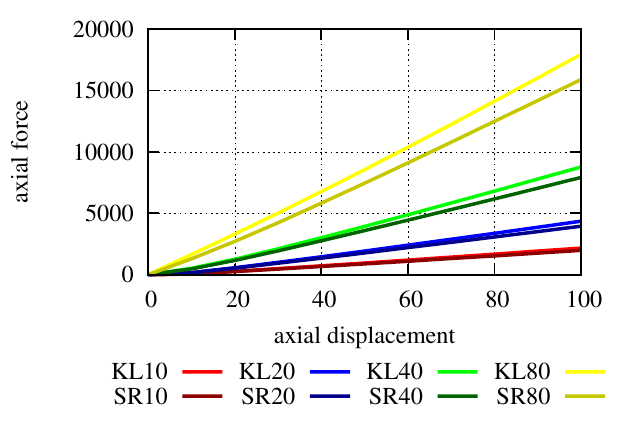}
    \label{fig:examples_micro_4a}
   }
  \subfigure[beam radius $R=0.1$.]
   {
    \includegraphics[width=0.45\textwidth]{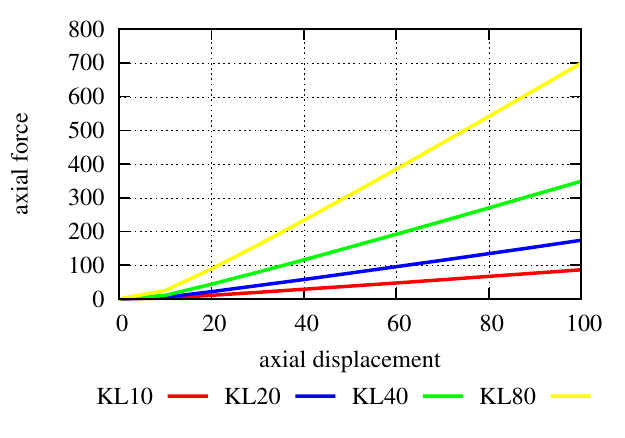}
   }
  \subfigure[beam radius $R=0.02$.]
   {
    \includegraphics[width=0.45\textwidth]{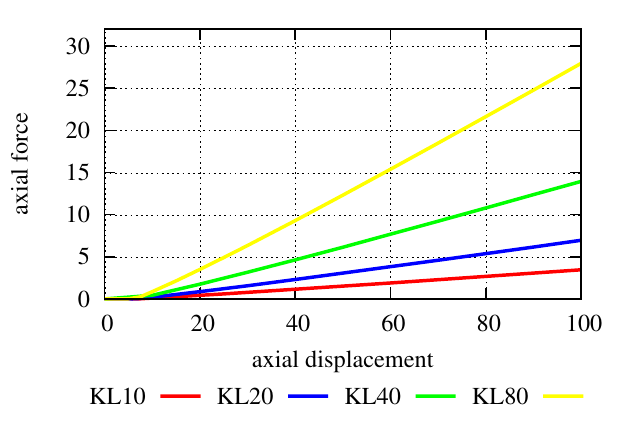}
   }
\caption{Axial force-displacement curves resulting from axial tension test on cylinder with hexagonal microstructure.}
\label{fig:examples_micro_4}
\end{figure}

In terms of quantitative analysis, axial force-displacement curves were measured for all simulation setups described above. The results are shown in
Figure~\ref{fig:examples_micro_4}. Each subfigure summarizes the results for one particular value of the beam radius, i.e.~$R=0.5,0.1,0.02$, while 
the four different microstructural refinement level~$N_{circ}=10,20,40,80$ are compared within each subfigure. All parameter combinations have 
been simulated by means of the KL(-ROT) element, whereas only for the case $R=0.5$, which is most critical in terms of shear deformation, a verification by 
means of the shear-deformable SR element has been performed.

Several observations can be made that are in excellent agreement with mechanical theory. Firstly, the force-displacement
curves scale linearly with the number of hexagons along the circumference and quadratically with the beam radius, i.e.~$F \sim N_{circ} * R^2$.
This makes perfect sense when thinking of the hexagonal microstructure as a mechanical continuum in the limit of high refinement levels, with
an effective cross-section area~$A \sim N_{circ} * R^2$. Secondly, as consequence of shear deformation, a certain deviation can be identified between the numerical solutions
for the SR beam element and the KL beam element in Figure~\ref{fig:examples_micro_4a}. The shear-free KL elements result in a slightly stiffer response visible by higher values of the axial reaction forces. As expected, this deviation decreases with a decreasing value of $N_{circ}$ leading to an increasing beam segment length and slenderness ratio. However, also for the most critical of the investigated cases represented by the combination $R=0.5$ (largest cross-section radius) and $N_{circ}=80$ (shortest segment length) the deviation remains below~$10\%$. Eventually, also some considerations concerning the behavior of the performance of the Newton-Raphson scheme shall be made. Thereto, the total number of accumulated Newton iterations required to solve the problem based on the load step control described in the beginning of this section and an initial value of $N_0=1$ load steps has been recorded for the KL and SR elements. Exemplarily, the results for the variants~$N_{circ}=40$ and~$R=0.5,0.1,0.02$ shall be investigated. Based on the KL element, the problem could be solved in one load step and $14-15$ iterations for all three slenderness ratios. On the contrary, the required number of load steps and accumulated iterations resulting from the SR element increases considerably with increasing slenderness ratio leading to values of $107$, $180$ and $360$ iterations for the cross-section radii $R=0.5,0.1,0.02$. This observation is in complete accordance with the results already derived in~\cite{meier2016}. Even though the variants with~$N_{circ}\!=\!40$ represent the case where this effect was most pronounced, also for all the other parameter combinations considered in this work, the KL element exhibited a considerably lower number of Newton iterations as the SR element. Furthermore, according to the investigations made in~\cite{meier2016}, the Newton-Raphson performance of the KL-TAN elements can be expected to be even better than for the applied KL-ROT elements. However, the latter element formulation has been preferred here since it simplifies the formulation of joint conditions as required for the present example.

\begin{figure}[h!!!]
 \centering
    \includegraphics[width=0.4\textwidth]{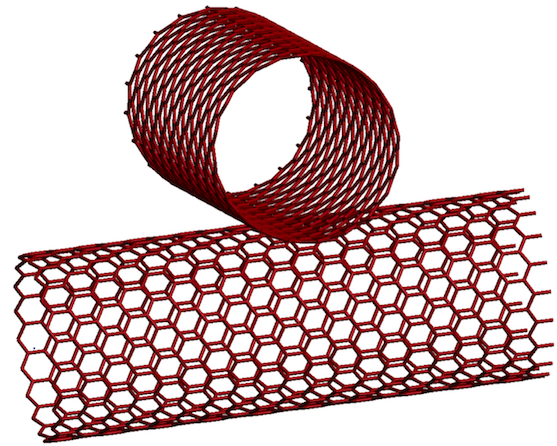}
   \caption{Contact of two cylindrical tubes with hexagonal microstructures: 3D view of initial geometry.}
  \label{fig:examples_micro_5}
\end{figure}

\begin{figure*}[ht!!!]
 \centering
    \includegraphics[width=0.85\textwidth]{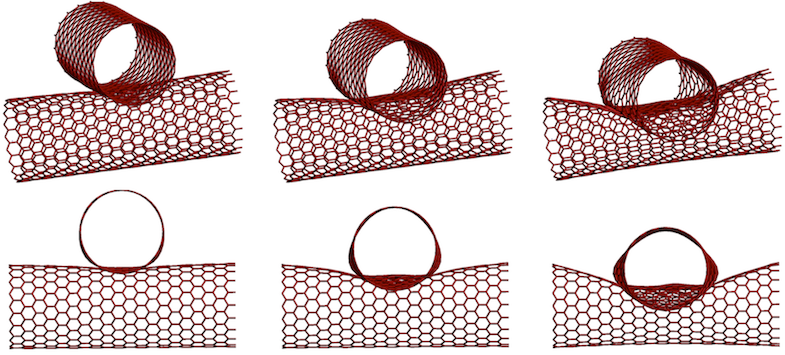}
   \caption{Contact of two cylindrical tubes with hexagonal microstructures: 3D and 2D view of deformed geometry at different load steps: $u_x=5$ (left), $u_x=10$ (middle) and $u_x=25$ (right).}
  \label{fig:examples_micro_6}
\end{figure*}

Finally, the applicability of the beam-to-beam contact algorithms presented in Section~\ref{sec:contact} to fiber-based microstructures such
as the ones investigated above shall be demonstrated in a qualitative manner. For this purpose, two cylinders with hexagonal microstructures
($\tilde{R}=25$, $N_{circ}=40$, $N_{axi}=20$, $R=0.5$, KL beam elements) are placed next to each other, their axes being orthogonal at a distance of~$2\tilde{R}+2R$,
see Figure~\ref{fig:examples_micro_5}. The two ends of the lower cylinder are simply supported, i.e.~$u_x=u_y=u_z=0$, while the two ends of the upper cylinder
are moved towards the lower cylinder with a prescribed displacement~$u_x=25$ (still with~$u_y=u_z=0$).
The course of deformation is illustrated in Figure~\ref{fig:examples_micro_6}. It can be seen that the two microstructures come into contact
with the active contact zone progressively becoming larger and substantial deformations occurring in the microstructures themselves. While being qualitative
in nature, these results underline the superior robustness of the devised geometrically exact beam elements and beam-to-beam contact algorithms.
As elaborated in Section~\ref{sec:contact} the robustness of the contact formulation can particularly be attributed to the smooth~beam
centerline interpolation (even for microstructures with crosspoints) and the unified treatment of point- and line-based
beam contact within a variationally consistent model transition approach, see~\cite{meier2015c} for details.

\subsection{Static twisting process of a rope}\label{sec:examples_cabletwisting}

\begin{figure*}[ht!!!]
 \centering
   \subfigure[Undeformed initial configuration.]
   {
    \includegraphics[width=0.31\textwidth]{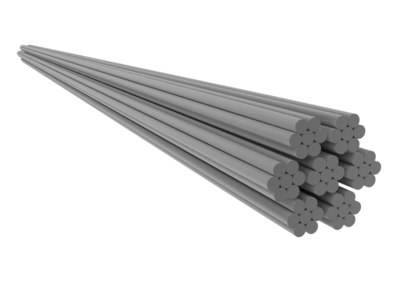}
    \label{fig:manybeams_twisting_step000}
   }
      \subfigure[Configuration at load step $20$.]
   {
    \includegraphics[width=0.31\textwidth]{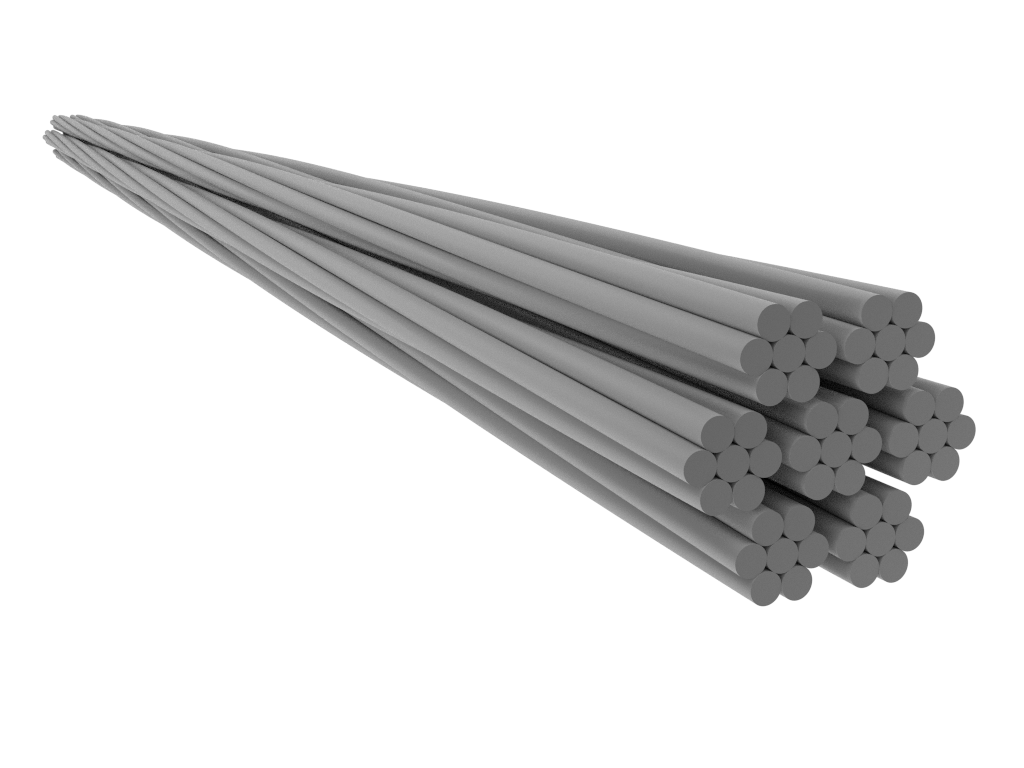}
    \label{fig:manybeams_twisting_step020}
   }
      \subfigure[Configuration at load step $40$.]
   {
    \includegraphics[width=0.31\textwidth]{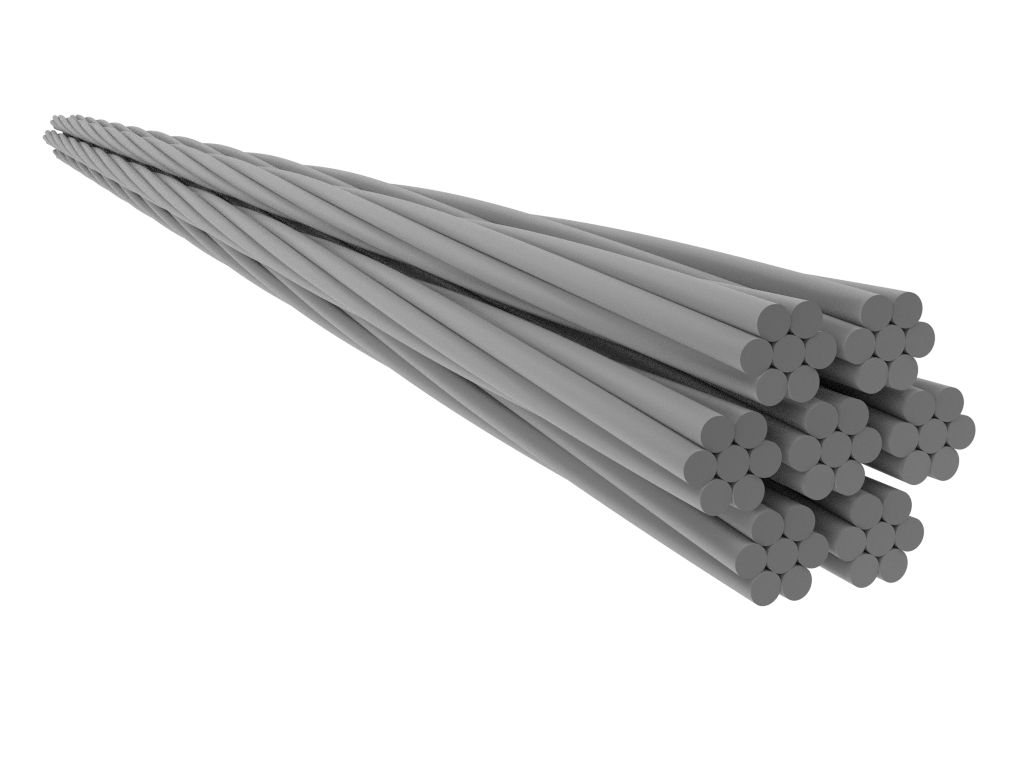}
    \label{fig:manybeams_twisting_step030}
   }
      \subfigure[Configuration at load step $60$.]
   {
    \includegraphics[width=0.31\textwidth]{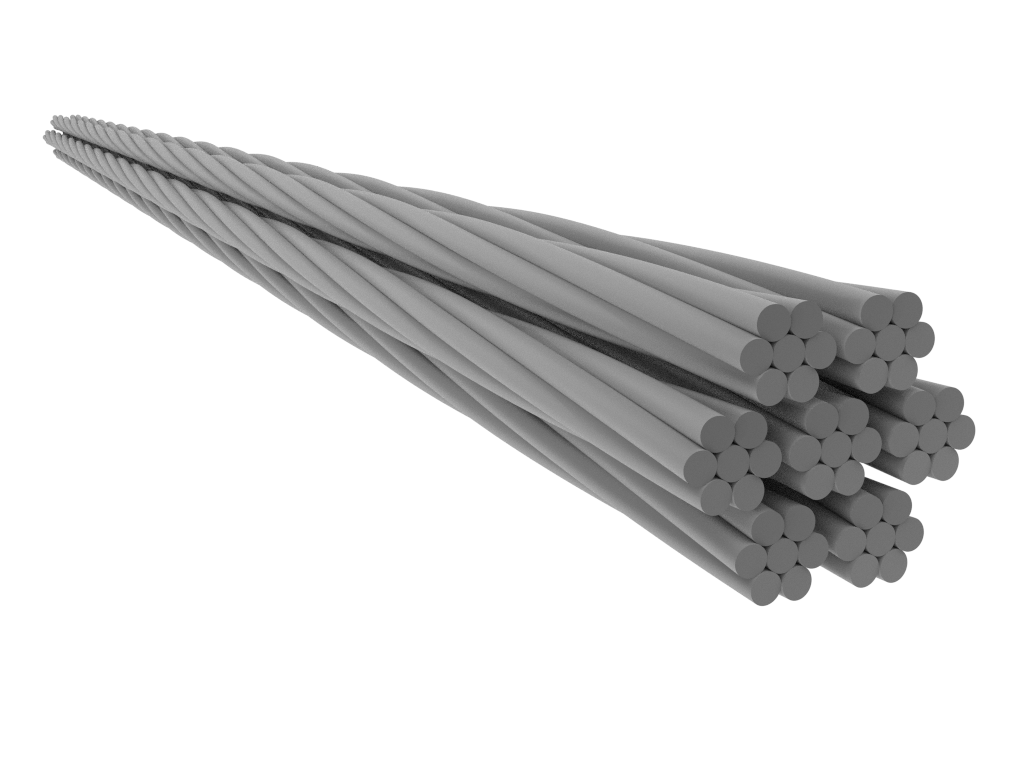}
    \label{fig:manybeams_twisting_step060}
   }
      \subfigure[Configuration at load step $80$.]
   {
    \includegraphics[width=0.31\textwidth]{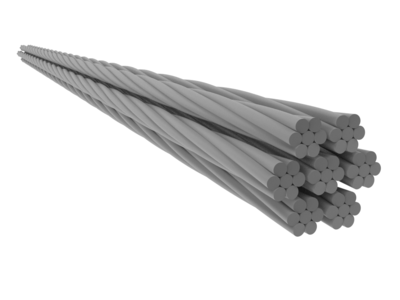}
    \label{fig:manybeams_twisting_step080}
   }
      \subfigure[Configuration at load step $100$.]
   {
    \includegraphics[width=0.31\textwidth]{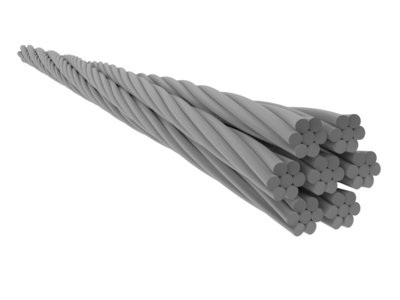}
    \label{fig:manybeams_twisting_step100}
   }
  \caption{Static simulation of the twisting process of a rope consisting of $7\! \times \! 7$ fibers.}
  \label{fig:manybeams_twisting}
\end{figure*}

In this example, the static twisting process of a rope will be investigated. The considered rope is built from $7\! \times \!7$ individual fibers with length $l\!=\!5$, circular cross-section of radius $R\!=\!0.01$ and Young's modulus $E\!=\!10^9$. The arrangement of the initially straight fibers in seven sub-bundles with seven fibers per sub-bundle is illustrated in Figure~\ref{fig:manybeams_twisting_step000}.

While this example has originally been considered in~\cite{meier2015b} on the basis of the TF beam element formulation, here the numerical results derived by spatial discretizations with $10$ TF, KL and SR beam elements, respectively, shall be compared. Since the enclosed angle is small for all contacting fibers throughout the entire simulation, only the line contact formulation from Section~\ref{sec:contact_line} is applied. The contact parameters are chosen identical to~\cite{meier2015b}. In the first stage of the twisting process, each of the seven sub-bundles is twisted by four full rotations within $80$ static load steps. The twisting process is performed in a Dirichlet-controlled manner, such that the cross-section center points at one end of the sub-bundles (front side in Figure~\ref{fig:manybeams_twisting}) are moving on a circular path with respect to the individual sub-bundle center points, while the corresponding points at the other end of the sub-bundles (back side in Figure~\ref{fig:manybeams_twisting}) remain fixed. The deformed configurations at characteristic load steps after one, two, three and four full rotations are illustrated in Figures~\ref{fig:manybeams_twisting_step020}-\ref{fig:manybeams_twisting_step080}. In the second stage of the twisting process, all seven sub-bundles together are twisted by one further rotation within $20$ additional static load steps. This time, the cross-section center points are moving on a circular path with respect to the center point of the entire $7 \times 7$-rope.

For the simulations, a Newton-Raphson scheme with load step adaption scheme and step size control as explained in the beginning of this section is applied. The given standard values are used for the tolerances of the convergence criteria. The upper bound of the admissible step size (displacement increment) is set to a value of $0.5R\!=\!0.005$ within the step size control and the maximum number of iterations per step is chosen as $50$.

The deformed configuration at the end of this twisting process is illustrated in Figure~\ref{fig:manybeams_twisting_step100}. While the cross-section center points of all fiber endpoints at one  end of the rope (front side in Figure~\ref{fig:manybeams_twisting}) are fixed in axial direction, the cross-section center points of all fiber endpoints at the other end of the rope (back side in Figure~\ref{fig:manybeams_twisting}) are free to move in axial direction. Additionally, a constant axial tensile force $\bar{f}_{ax}=1000$ acting on each of these axially freely movable fiber endpoints provides axial pre-stressing during the entire twisting process. All fiber endpoints are simply supported but not clamped. Consequently, Dirichlet conditions are only applied to the positional degrees of freedom at the endpoints but not to the tangential degrees of freedom. In the case of KL and SR elements, the rigid body mode associated with a rotation around the fiber axis is blocked by additional Dirichlet conditions at one end of the rope (back side in Figure~\ref{fig:manybeams_twisting}). This is not necessary for the torsion-free TF element formulation where such a rigid body mode does not exist by definition.

\begin{figure}[ht]
 \centering
    \includegraphics[width=0.45\textwidth]{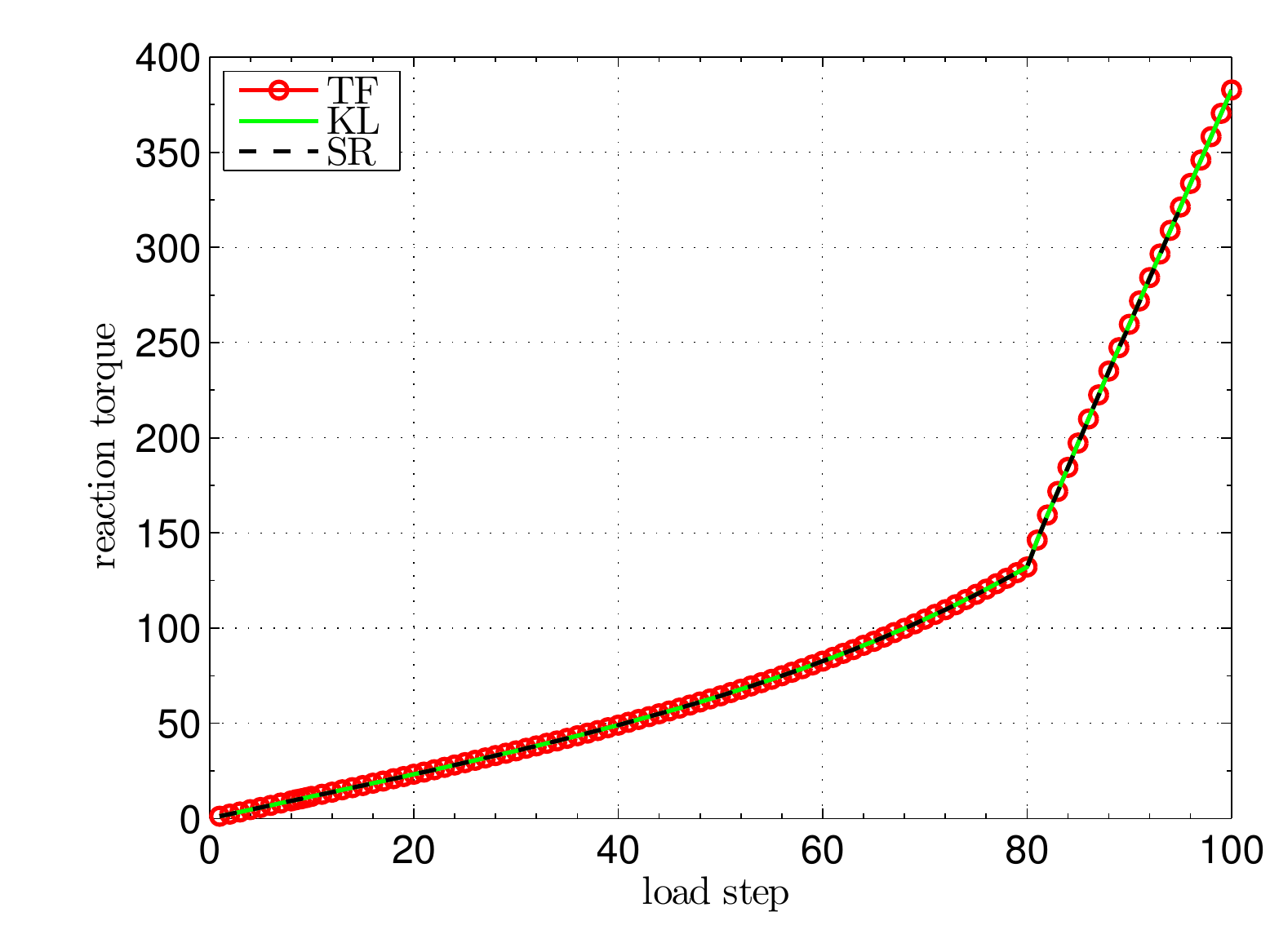}
   \caption{Axial reaction torque during twisting process.}
  \label{fig:manybeams_twisting_static_torque}
\end{figure}

The straight initial geometry of the individual fibers as well as the chosen loading and Dirichlet boundary conditions are compatible with the requirements discussed in Section~\ref{sec:beams_TF}. As a 
result, each individual fiber remains torsion-free and the TF beam elements yield exact results for the considered static example. Given the global twisting state of the rope as illustrated in Figure~\ref{fig:manybeams_twisting_step100}, this result might contradict first intuition. Nevertheless, of course, an overall external axial torque resulting from the moment contributions of 
the reaction forces at the beam endpoints with respect to the centerline of the rope is necessary in order to guarantee for static equilibrium of the twisted rope at different load steps.
The corresponding evolution of this external axial torque during the deformation process is plotted in Figure~\ref{fig:manybeams_twisting_static_torque}.

As expected due to (exactly) vanishing torsional deformation and small shear deformation expected for the considered fiber slenderness ratio of~$\zeta\!=\!500$, the results for the three different applied beam formulations are in excellent agreement. Interestingly, the evolution of the twisting torque over the twisting angle is almost linear within the two stages of deformation, i.e. the behavior of the rope is similar to the twisting response of a slender continuum. The higher slope in the second twisting stage, where all sub-bundles are twisted uniformly with respect to the centerline of the rope, results from the increased overall elastic stiffness. The external work required in order to perform the considered twisting process in a quasi-static manner is proportional to the area enclosed by the graph of the twisting torque and the horizontal axis of Figure~\ref{fig:manybeams_twisting_static_torque}.

By taking into account the different number of degrees of freedom per element, the computational cost in order to derive these results is smallest for TF, followed by KL and highest for SR elements. These findings underline the superiority of torsion and/or shear-free beam formulations in cases where the underlying assumptions are met.

\subsection{Static and dynamic loading of a helical spring}\label{sec:examples_helicalspring}
\begin{figure*}[ht!!!]
  \centering
  \subfigure[$\!u_z\!=\!-168\!$: max. pressure]
   {
    \includegraphics[width=0.1\textwidth]{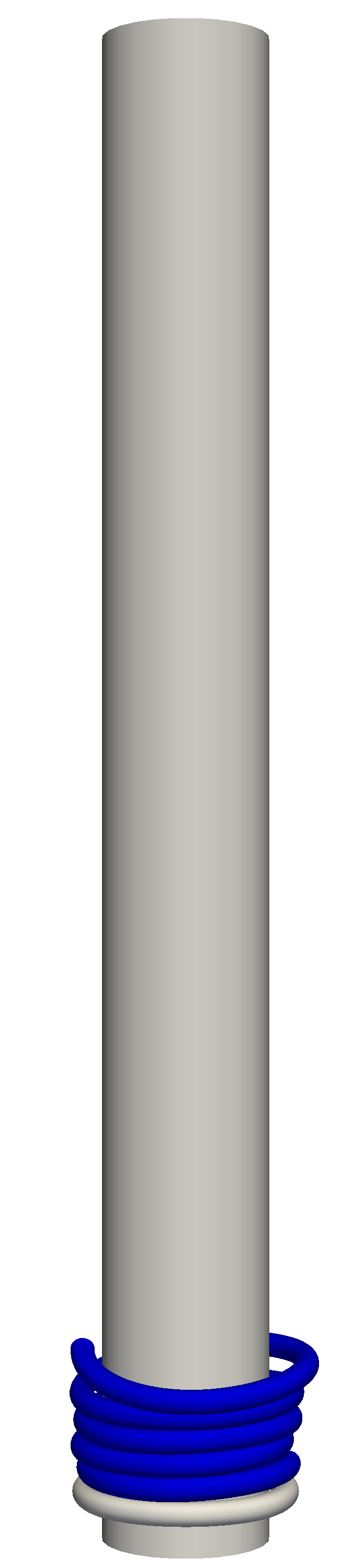}
   }\hfill
  \subfigure[$u_z\!=\!-150$]
   {
    \includegraphics[width=0.1\textwidth]{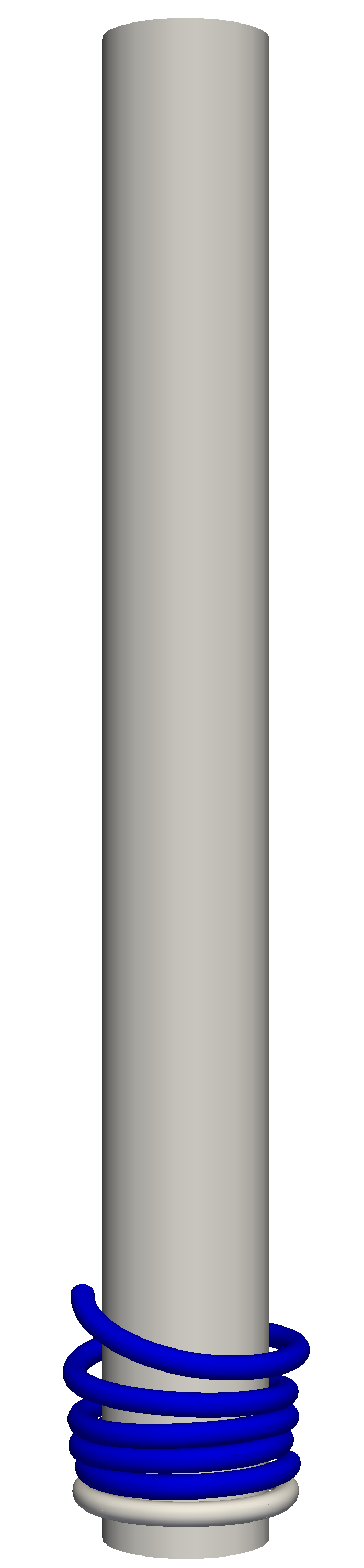}
   }\hfill
  \subfigure[$u_z\!=\!-100$]
   {
    \includegraphics[width=0.1\textwidth]{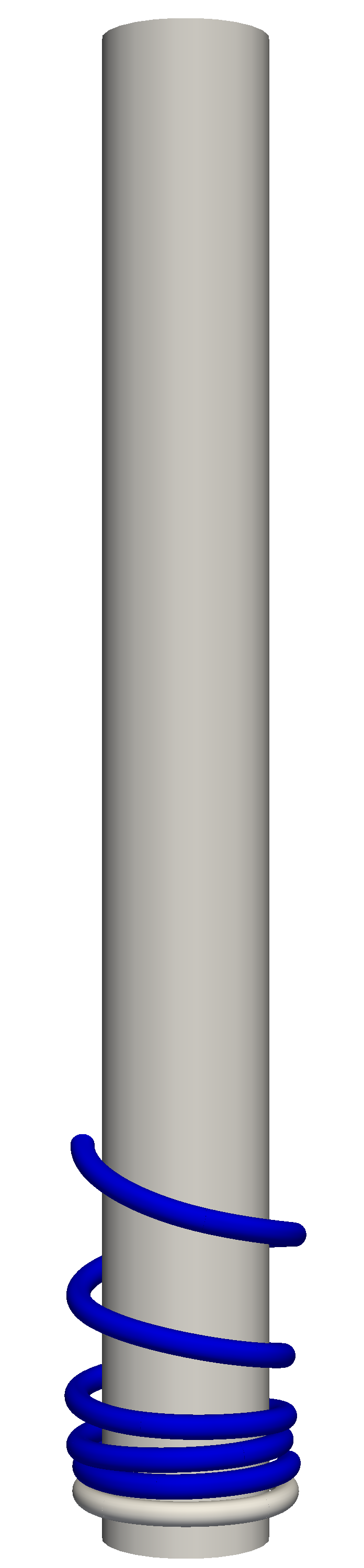}
   }\hfill
  \subfigure[$u_z\!=\!0$: undeformed]
   {
   \label{fig:helical_spring_static_time0}
    \includegraphics[width=0.1\textwidth]{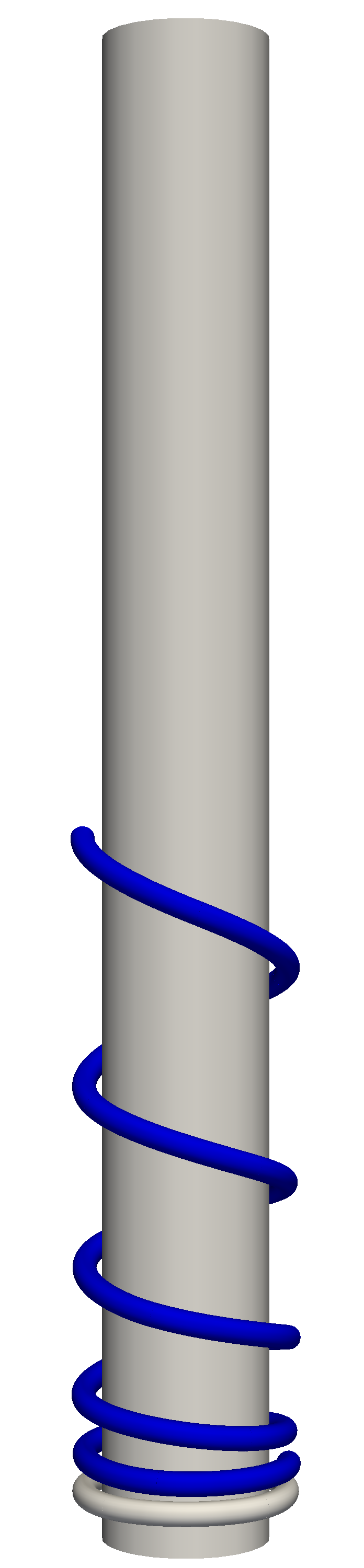}
   }\hfill
  \subfigure[$u_z\!=\!125$]
   {
    \includegraphics[width=0.1\textwidth]{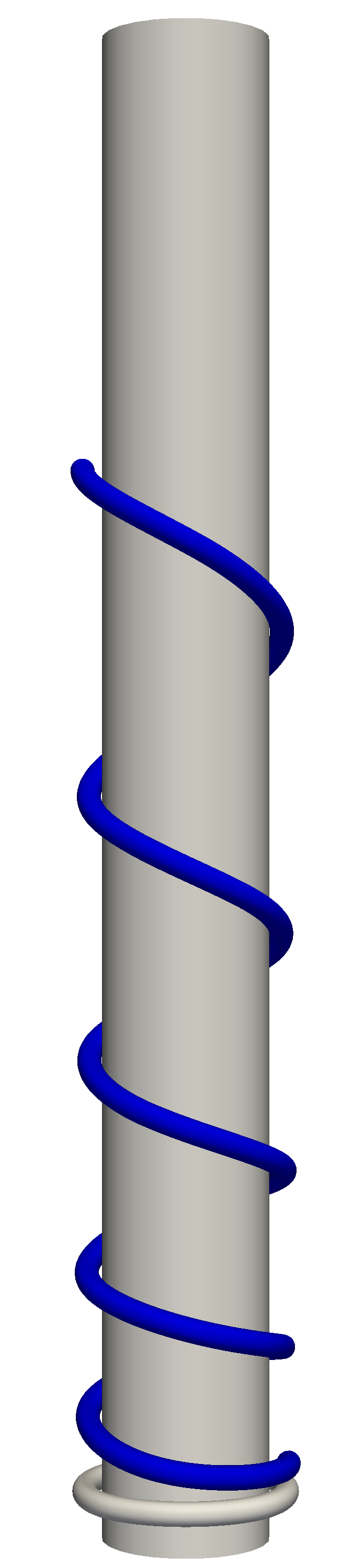}
   }\hfill
  \subfigure[$u_z\!=\!250$: max. tension]
   {
    \label{fig:helical_spring_static_fullcompression}
    \includegraphics[width=0.1\textwidth]{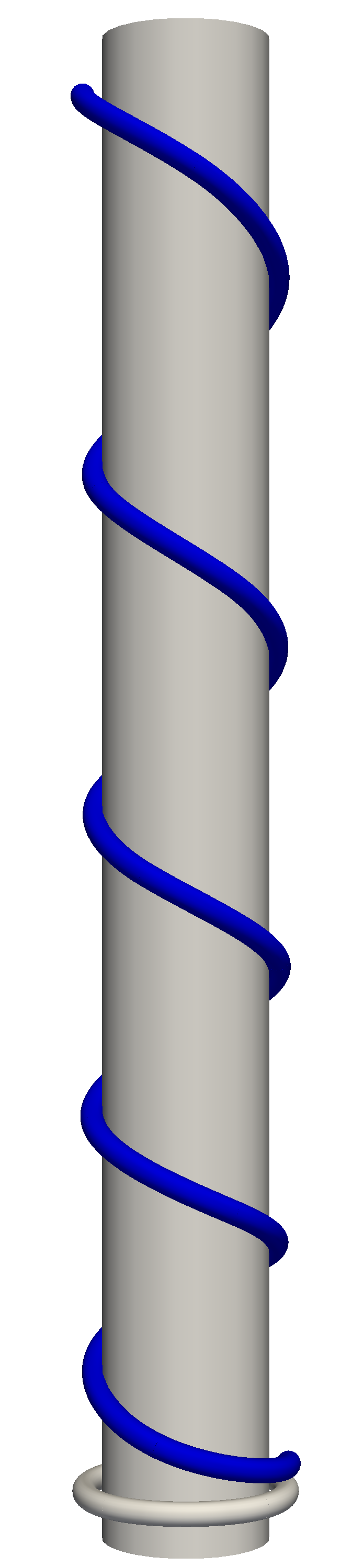}
   }\\
  \subfigure[detail views at $u_z\!\approx\!-168$ (central rigid cylinder hidden)]
   {
    \label{fig:helical_spring_static_details}
    \includegraphics[width=0.2\textwidth]{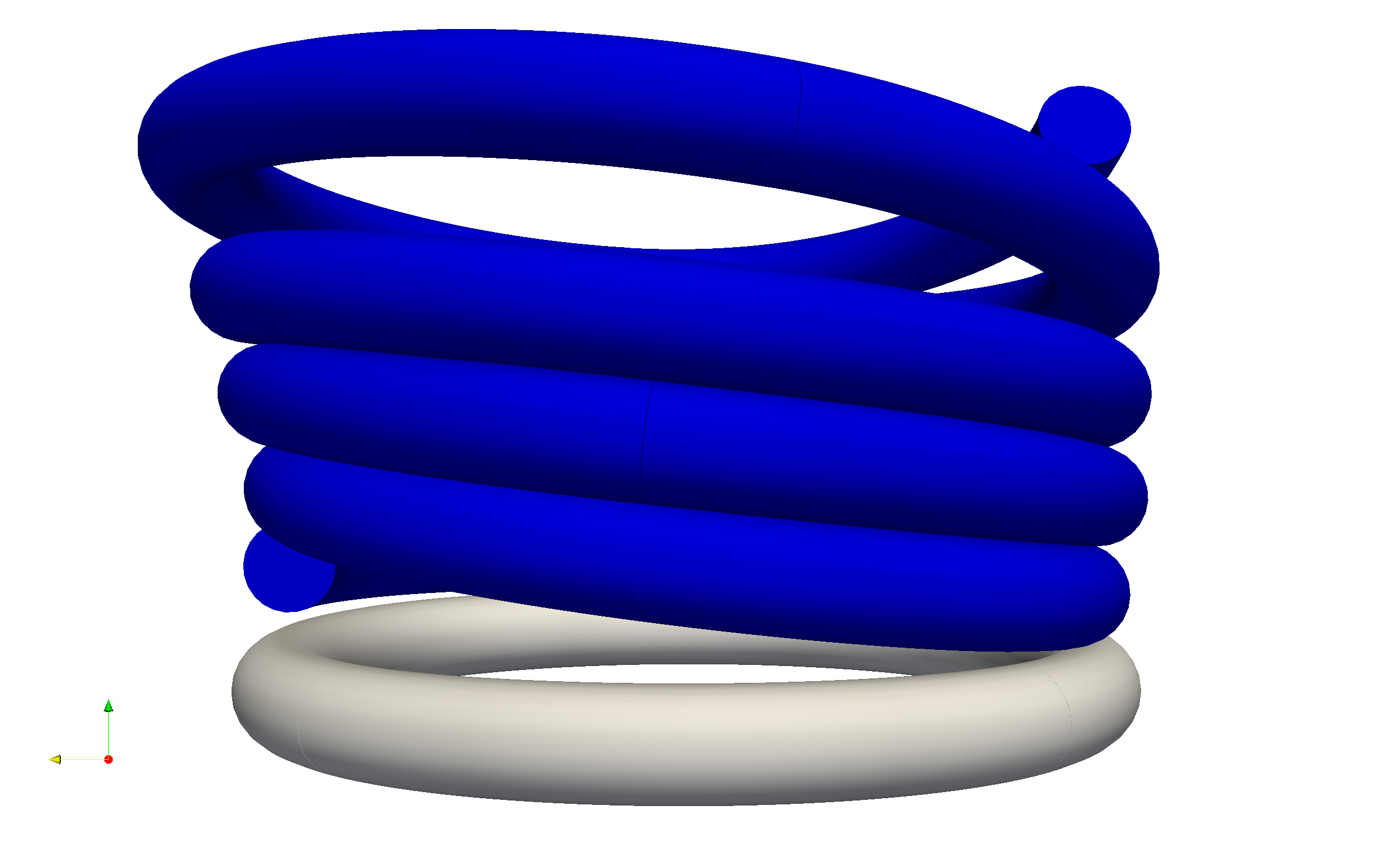}
    \includegraphics[width=0.2\textwidth]{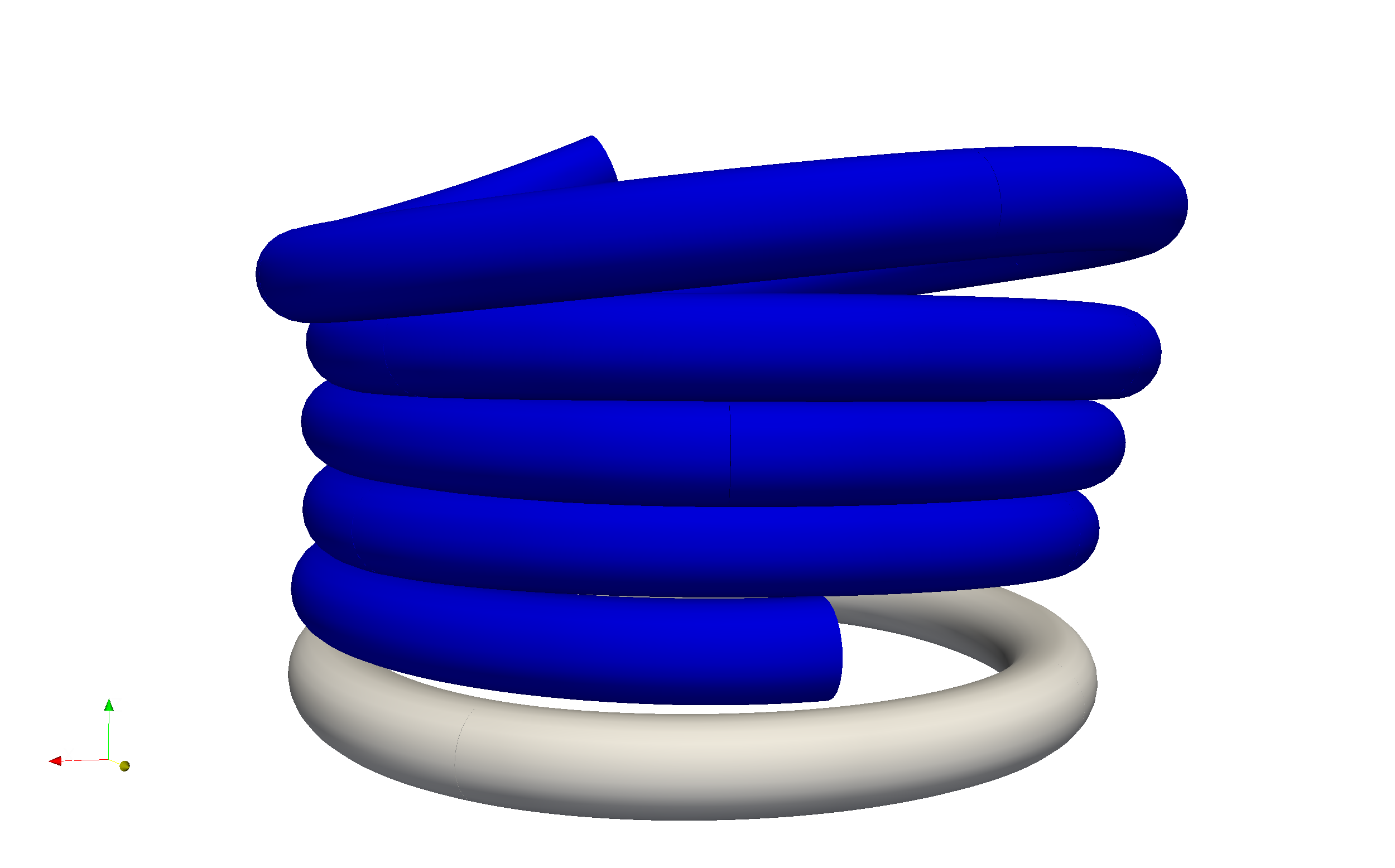}
    \includegraphics[width=0.2\textwidth]{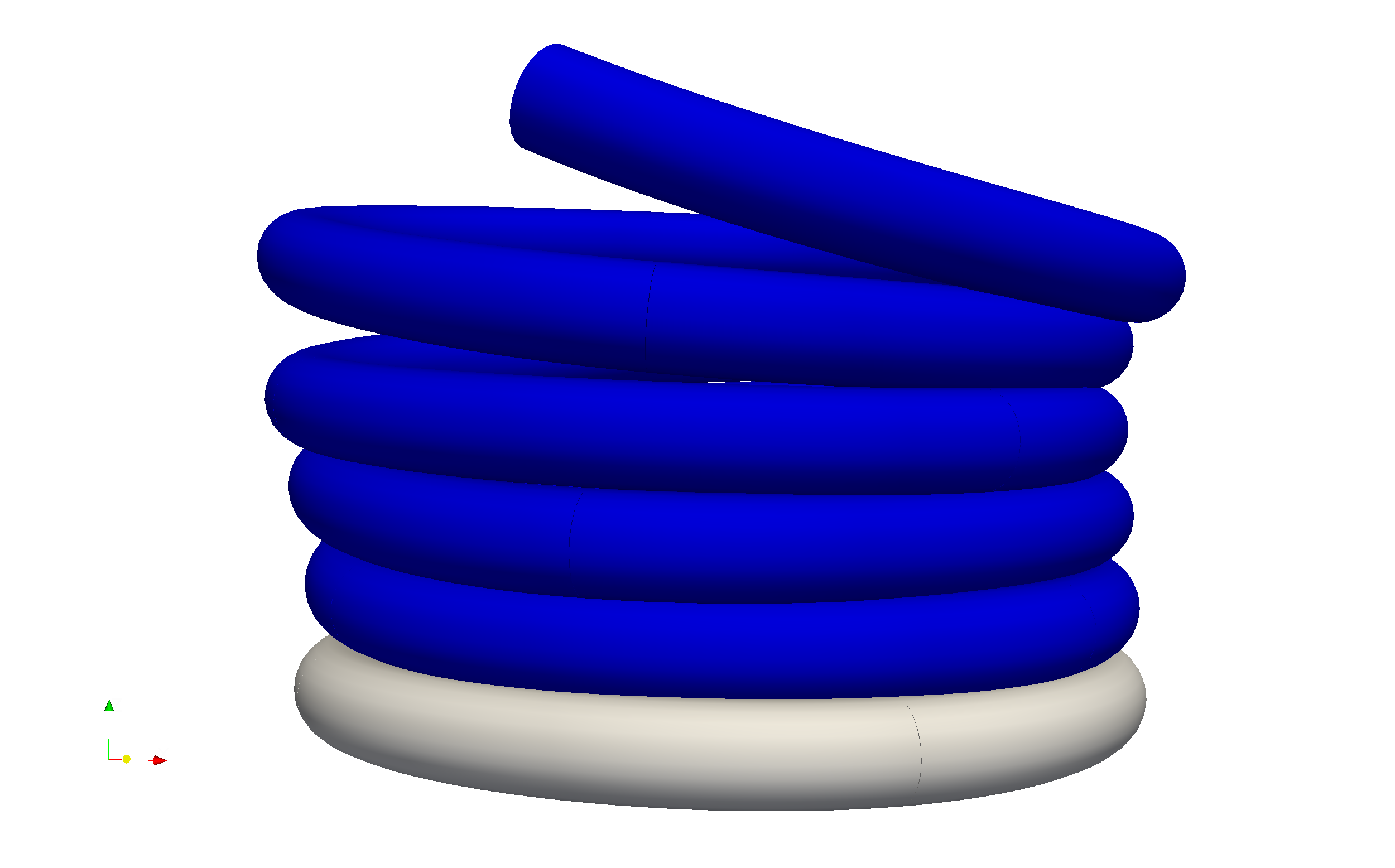}
    \includegraphics[width=0.2\textwidth]{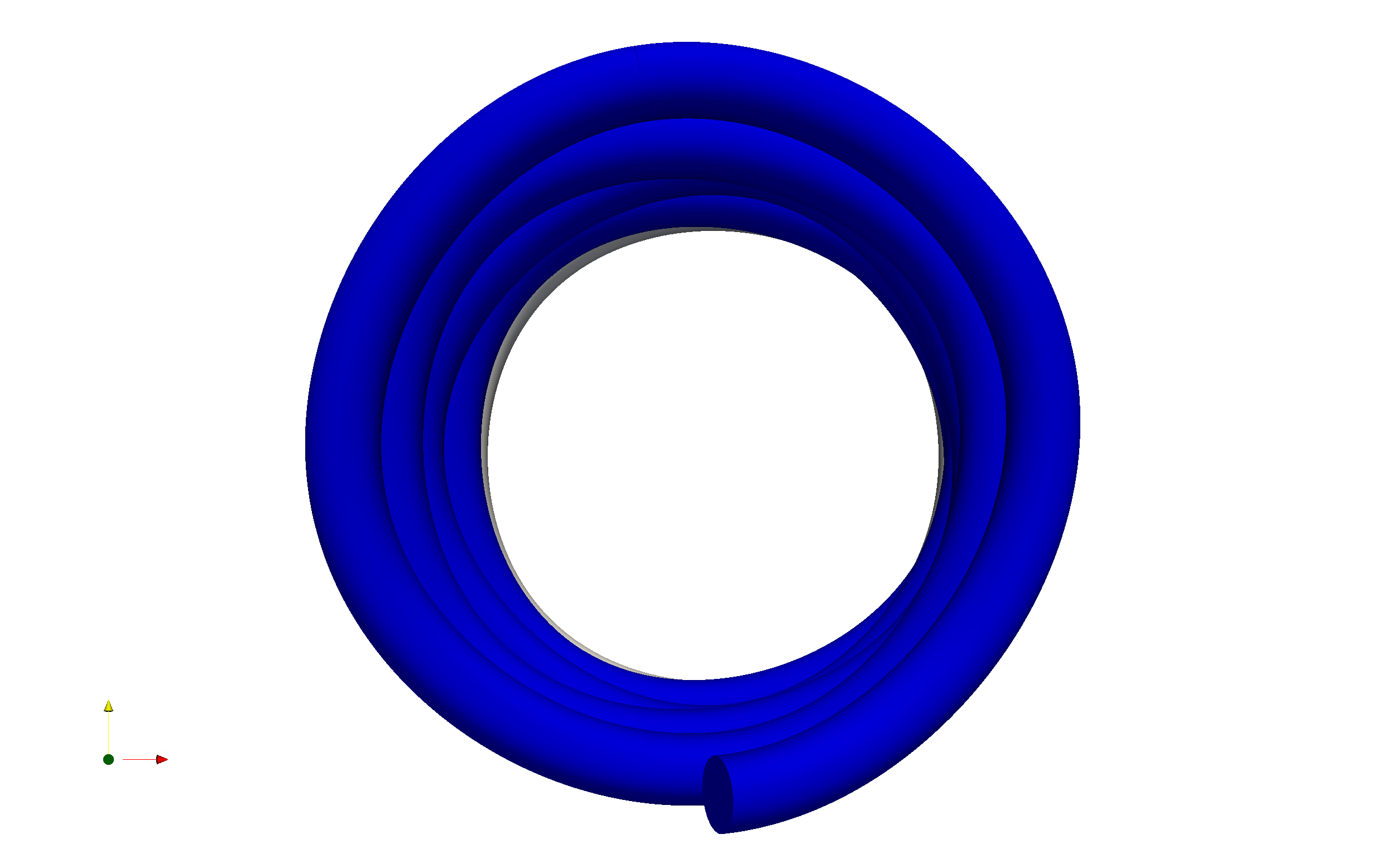}
   }
\caption{Quasi-static simulation of a helical spring with linearly increasing slope.}
\label{fig:helical_spring_staticconfigs}
\end{figure*}

The following numerical example investigates the response of a spring to static as well as dynamic loading. Its initial and stress-free geometry is defined as a helix with linearly increasing slope (see Figure~\ref{fig:helical_spring_static_time0}) via the analytic representation
\begin{align}
\label{r0_analyt_3Dhelix}
  \mb{r}_0(\beta) \!&=\!
  \left(
   \begin{array}{c}
   r_{0x} \\
   r_{0y} \\ 
   r_{0z}
   \end{array}
   \right)\!=\! R_0 \!
  \left(
   \begin{array}{c}
   \sin{\beta} \\
   \cos{\beta} \!-\!1 \\ 
    \frac{6}{81 \pi^2} \beta^2
   \end{array}
   \right)\!,
   \,\,\,\\
   R_0 \!&=\! \frac{l}{6\sqrt{\left(\frac{3 \pi}{4}\right)^2 \!+\! 1} \!+\! \frac{27 \pi^2}{8} \ln{\left(\frac{4}{3\pi}\!+\!\sqrt{1\!+\!\left(\frac{4}{3\pi}\right)^2}\right)}}\\
   &\approx 34.36.
\end{align}
The radius $R_0$ of the enveloping cylinder of the helix is chosen such that the helix exactly consists of $4.5$ coils, i.e. $\beta \in[0;9 \pi]$, along the length of $l\!=\!1000$. A value of $R=4$ is chosen for the cross-section radius of the helix which results in a slenderness ratio of $\zeta\!=\!250$. The spring is clamped at one end (bottom in Figure~\ref{fig:helical_spring_staticconfigs}) and load will be applied to the other end point (top in Figure~\ref{fig:helical_spring_staticconfigs}). In order to mimic a realistic application of this spring, two additional rigid structures were added in the surrounding of the helix to guide its large deformation under the applied load (see grey colored structures in Figure~\ref{fig:helical_spring_staticconfigs}). A rigid cylinder of radius $R_b\!=\!28$ and length $l_b\!=\!500$ is placed along the centerline of the helix, providing a close guidance with an initial closest distance of approximately~$2.36$ from the surface of the spring. Moreover, a rigid torus with the same cross-section area as the helix is placed such that its centerline contour lies in the $xy-$plane at the vertical position $z\!=\!-9.1$.

The material of the spring is modeled by a St. Venant Kirchhoff law with Young's modulus $E\!=\!1.0$, Poisson's ratio $\nu\!=\!0$ and density $\rho\!=\!1.0\cdot 10^{-8}$. In case of the shear-deformable SR formulation, the shear correction factor is chosen as $\kappa \!=\! 1.0$.

Since small contact angles are prevailing in this example, only the line-to-line contact formulation of Section~\ref{sec:contact_line} with a linear penalty law and quadratic regularization (see~\cite{meier2015b}) is applied here. Note that a point contact formulation would anyway not be able to model the contact between spring and the central rigid cylinder (aligned with the helix axis)  because no unique bilateral closest point exists, which is a prerequisite for this type of formulation. Eventhough the ABC formulation would automatically resolve this geometrical property in a correct and robust manner, here use has been made of the a priori knowledge that only small contact angles will ocur which makes the point-to-point contact contribution of the ABC formulation obsolete. The corresponding line penalty parameter is chosen as $\varepsilon_{||}\!=\!10^{-3}$ in the quasi-static simulations and $\varepsilon_{||}\!=\!10^{-2}$ in the dynamic simulations, and the regularization parameter of the quadratically regularized penalty law are $\bar{g}\!=\!0.05R\!=\!0.2$ and $\bar{g}\!=\!0.25R\!=\!1.0$ respectively (see~\cite{meier2015b}). Each beam element is subdivided in twenty $5$-point integration segments (see also~\cite{meier2015b}).

\begin{figure}[h!!!]
 \centering
    \includegraphics[width=0.45\textwidth]{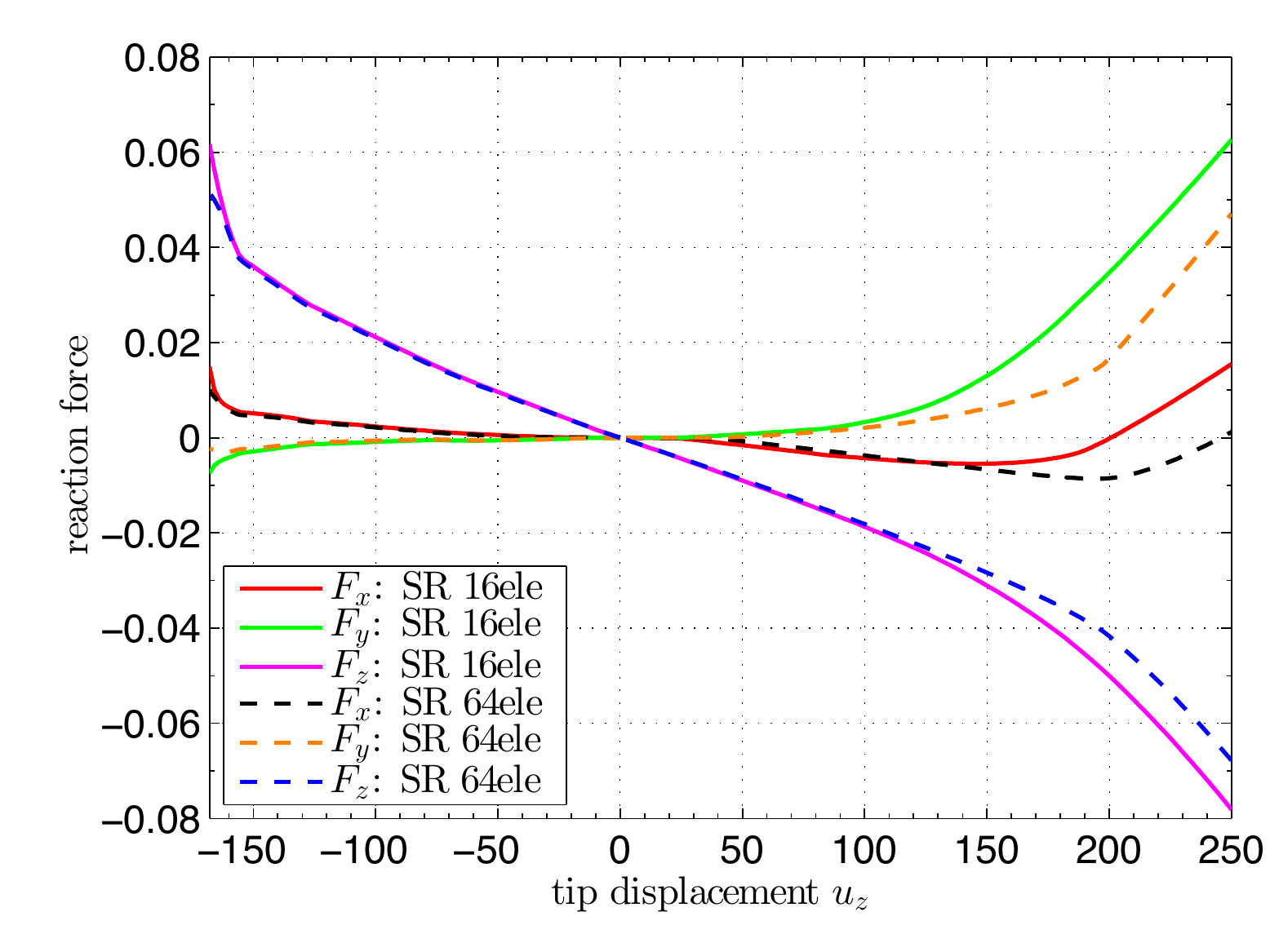}
 \caption{Displacement-reactions curves: $16$ SR vs $64$ SR}
 \label{fig:helical_spring_displreactions1}
\end{figure}

\begin{figure}[h!!!]
 \centering
  \includegraphics[width=0.45\textwidth]{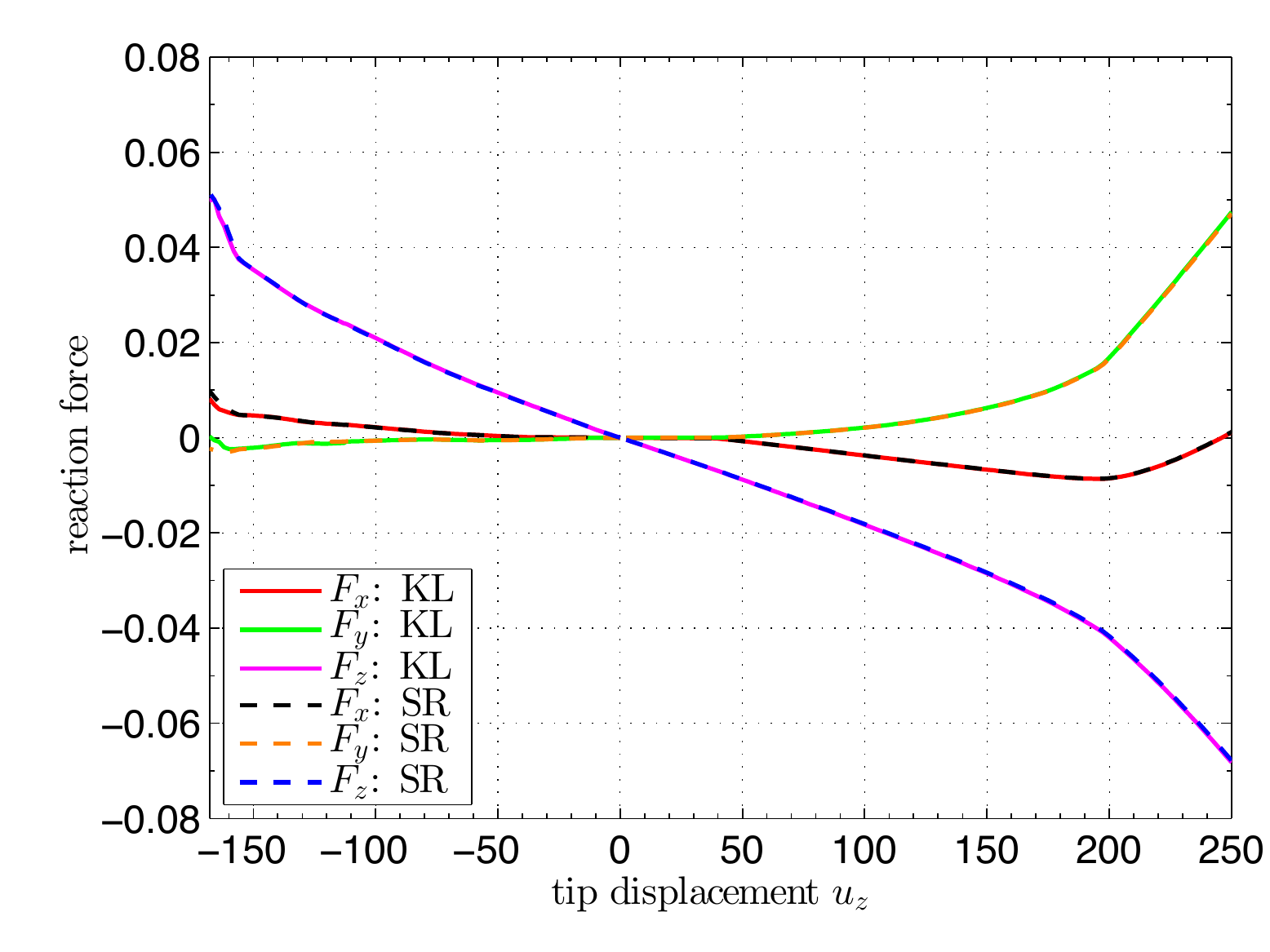}
 \caption{Displacement-reactions curves: $64$ KL vs $64$ SR}
 \label{fig:helical_spring_displreactions2}
\end{figure}

First, a quasi-static loading process will be analyzed. For this purpose, the helical structure is discretized by $16$ or $64$ beam elements of tangent vector-based KL(-TAN) type or SR type respectively. The tip of the spring is displaced in a Dirichlet controlled way prescribing zero displacement in $x$- and $y$-direction $u_x\!=\!u_y\!=\!0$ and a piecewise linear displacement in $z-$direction: $u_z\!=\!0\ldots250$ for $0\!<\!t\!<\!1$, $u_z\!=\!250\ldots0$ for $1\!<\!t\!<\!2$ and $u_z\!=\!0\ldots-\!200$ for $2\!<\!t\!<\!3$. Like for a clamped end, all rotations of the beam cross-section at $\beta\!=\!0$ are suppressed by appropriate Dirichlet conditions for the tangential and rotational degrees of freedom in case of KL and SR elements respectively.

Via the defined time curve for the tip displacement~$u_z$, first the tensile and then the compression range of the characteristic displacement-reactions curve of the spring is measured by tracking the reaction forces at the tip. The simulations are conducted with a time step size of $\Delta t\!=\! 0.01$ using a Newton-Raphson scheme with step size control that limits the displacement increment per iteration to a maximum value of $0.25R\!=\!0.1$ (see beginning of Section~\ref{sec:examples} for details). Again, a step adaptivity is employed that repeats a step with half of the original (pseudo-) time step size if convergence fails within the prescribed maximum number of $30$ iterations. Tolerances for convergence criteria are again chosen as introduced in the beginning of Section~\ref{sec:examples}.

Some characteristic configurations are illustrated in Figure~\ref{fig:helical_spring_staticconfigs}. The simulation ends at the fully compressed state with~$u_z\approx-168$ (see Figure~\ref{fig:helical_spring_static_details}). The resulting displacement-reactions curves are plotted in Figures~\ref{fig:helical_spring_displreactions1} and~\ref{fig:helical_spring_displreactions2}. As expected from Hooke's law, the relationship between displacement in axial direction of the spring~$u_z$ and axial reaction force~$F_z$ is linear in the regime of moderate displacements (approx. $-100\!<\!u_z\!<\!100$). For large tensile deformations, the stiffness of the spring increases smoothly but steadily due to the strong geometrical nonlinearity and the growing influence of stiff axial tension deformation modes. In the compression regime, a strong nonlinear effect, in form of a kink in the load-displacement curve illustrated in Figure~\ref{fig:helical_spring_displreactions2}, due to the contacting coils can be observed. This compression stiffness is mainly determined by the number of coils in contact and the cross-section stiffness represented by the penalty regularization of the applied beam contact model.

Also for this example of an initially curved elastic fibrous structure, the results from KL and SR elements are in excellent agreement (see Figure~\ref{fig:helical_spring_displreactions2}). This meets the expectation of negligible shear-deformation for highly slender bodies as it is the case in this example. Furthermore, Figure~\ref{fig:helical_spring_displreactions2} reveals that the comparatively rough discretization with 16 SR elements leads to a still visible discretization error as compared to the discretization with 64 finite elements, however, the qualitative behavior is already captured very well. 

\begin{figure}[h!!!]
 \centering
  \includegraphics[width=0.45\textwidth]{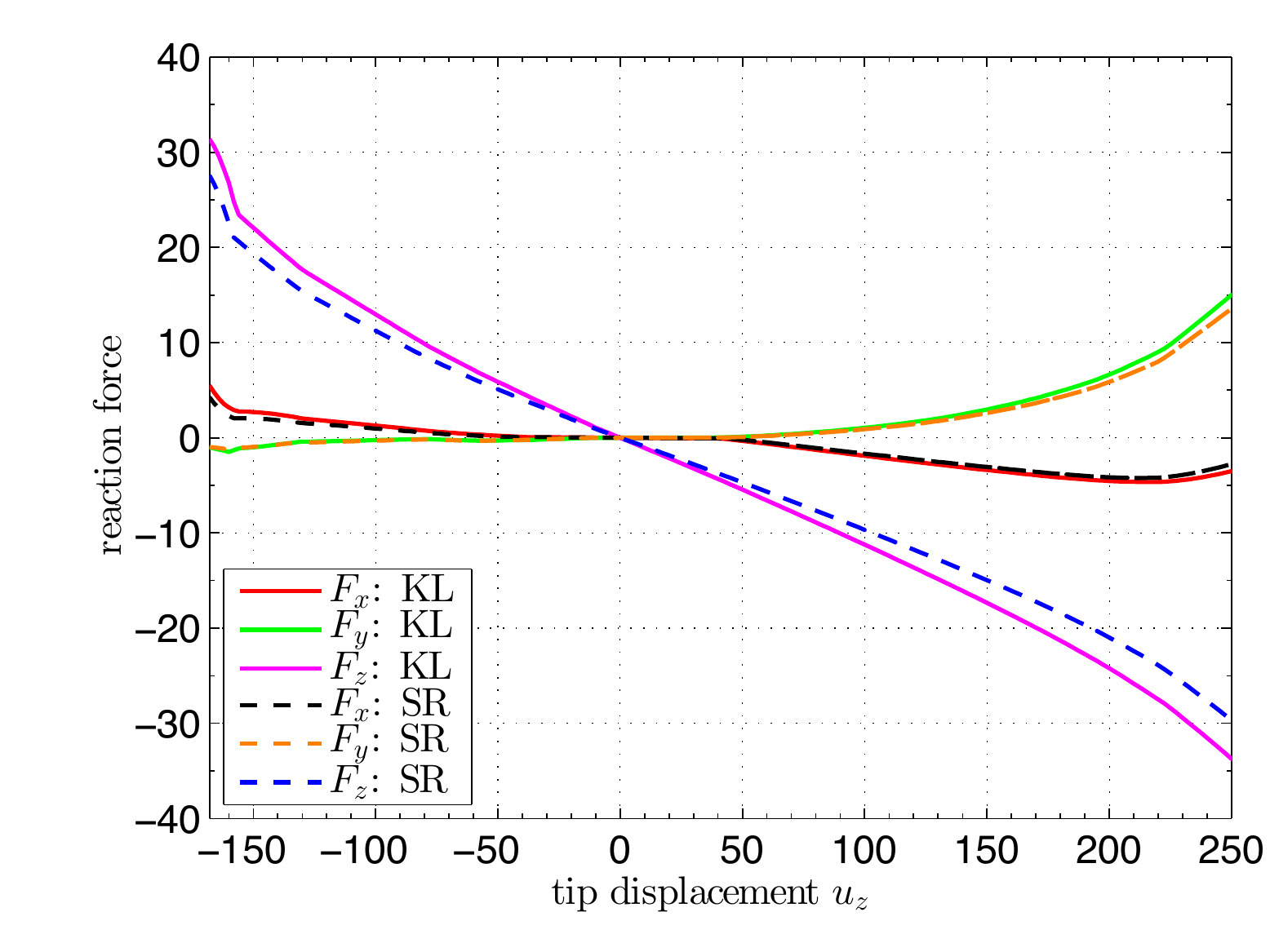}
 \caption{Displacement-reactions curves in case of an increased cross-section radius $\bar{R}\!=\!5R$: $64$ KL vs $64$ SR}
 \label{fig:helical_spring_displreactions3}
\end{figure}

In order to investigate the influence of shear deformation, a further variant of this example with increased beam cross-section radius shall be investigated. For better comparability, only the value of the cross-section radius occurring in the section constitutive law~\eqref{storedenergyfunctionkirchhoff2} is scaled by a factor of $5$, i.e. $\bar{R}\!=\!5R$, leading to correspondingly adapted values of $A,A_2,A_3,I_2,I_3,I_T$, while the external beam geometry, i.e. the cross-section radius occurring in the beam-to-beam contact model, remains unchanged. Again, the simulation is performed in a Dirichlet-controlled manner similar to the first variant. The resulting load-displacement curve is illustrated in Figure~\ref{fig:helical_spring_displreactions3}. Accordingly, due to this artificially increased cross-section radius a clear difference between the results of the shear-free KL elements and the shear-deformable SR elements becomes visible. Consequently, in such a case the Simo-Reissner theory of thick rods might be preferable as compared to the Kirchhoff-Love theory of thin rods.

\begin{figure*}[t!!!]
 \centering
  \subfigure[time $t=0$]
   {
    \includegraphics[width=0.18\textwidth]{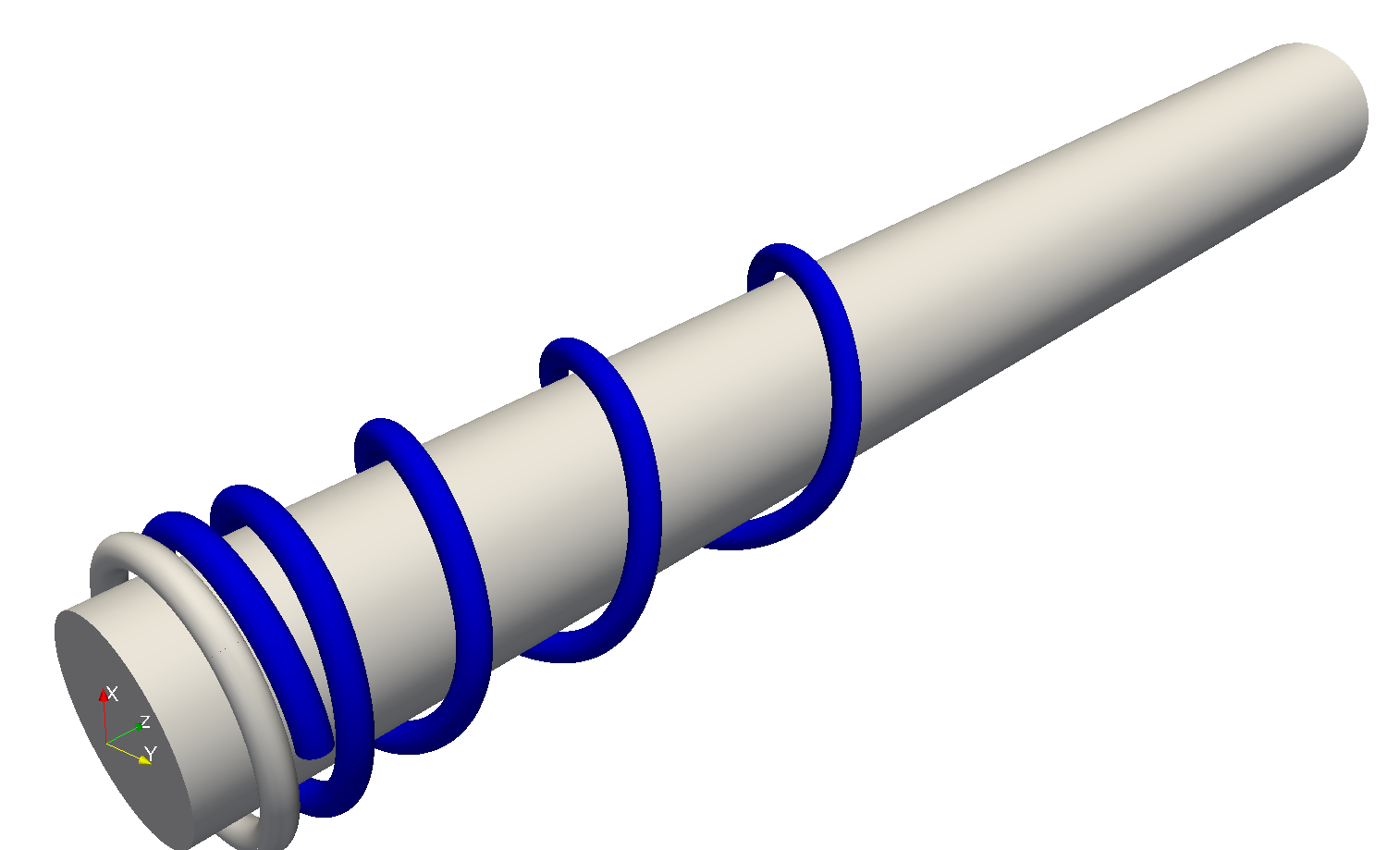}
   }
  \subfigure[time $t=2$]
   {
    \includegraphics[width=0.18\textwidth]{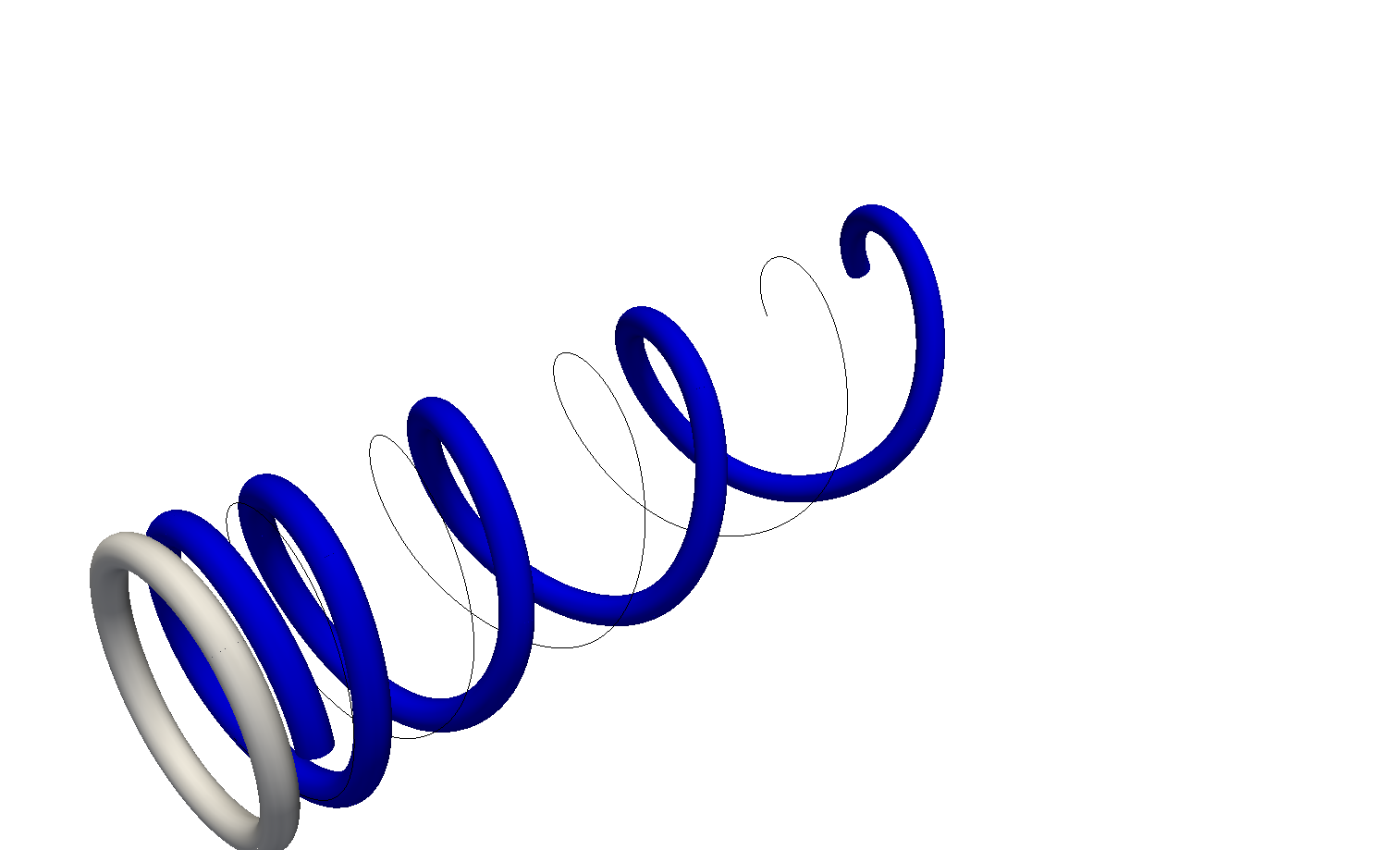}
   }
  \subfigure[time $t=4.5$]
   {
    \includegraphics[width=0.18\textwidth]{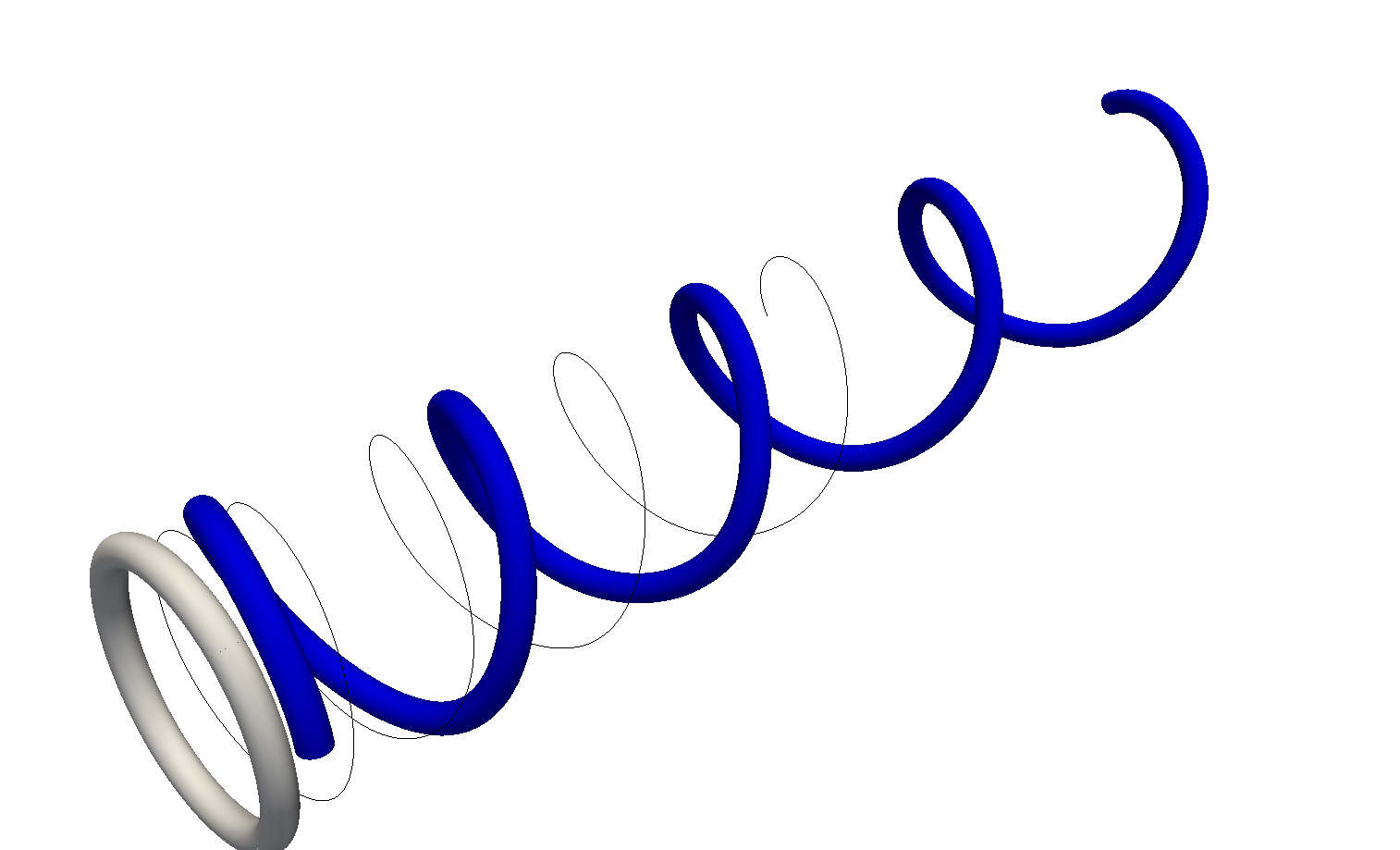}
   }
   \subfigure[time $t=6$]
   {
    \includegraphics[width=0.18\textwidth]{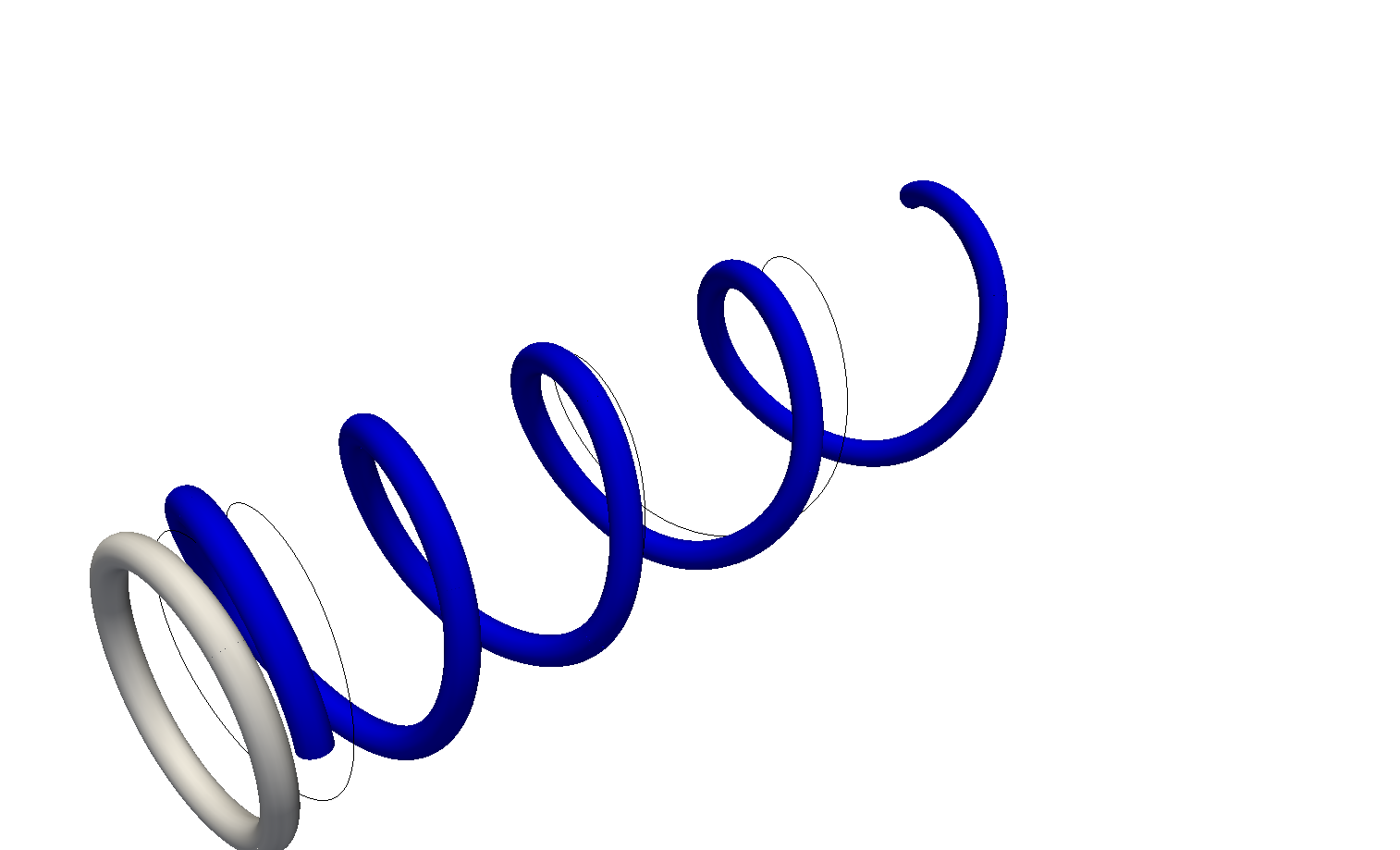}
   }
     \subfigure[time $t=6.5$]
   {
    \includegraphics[width=0.18\textwidth]{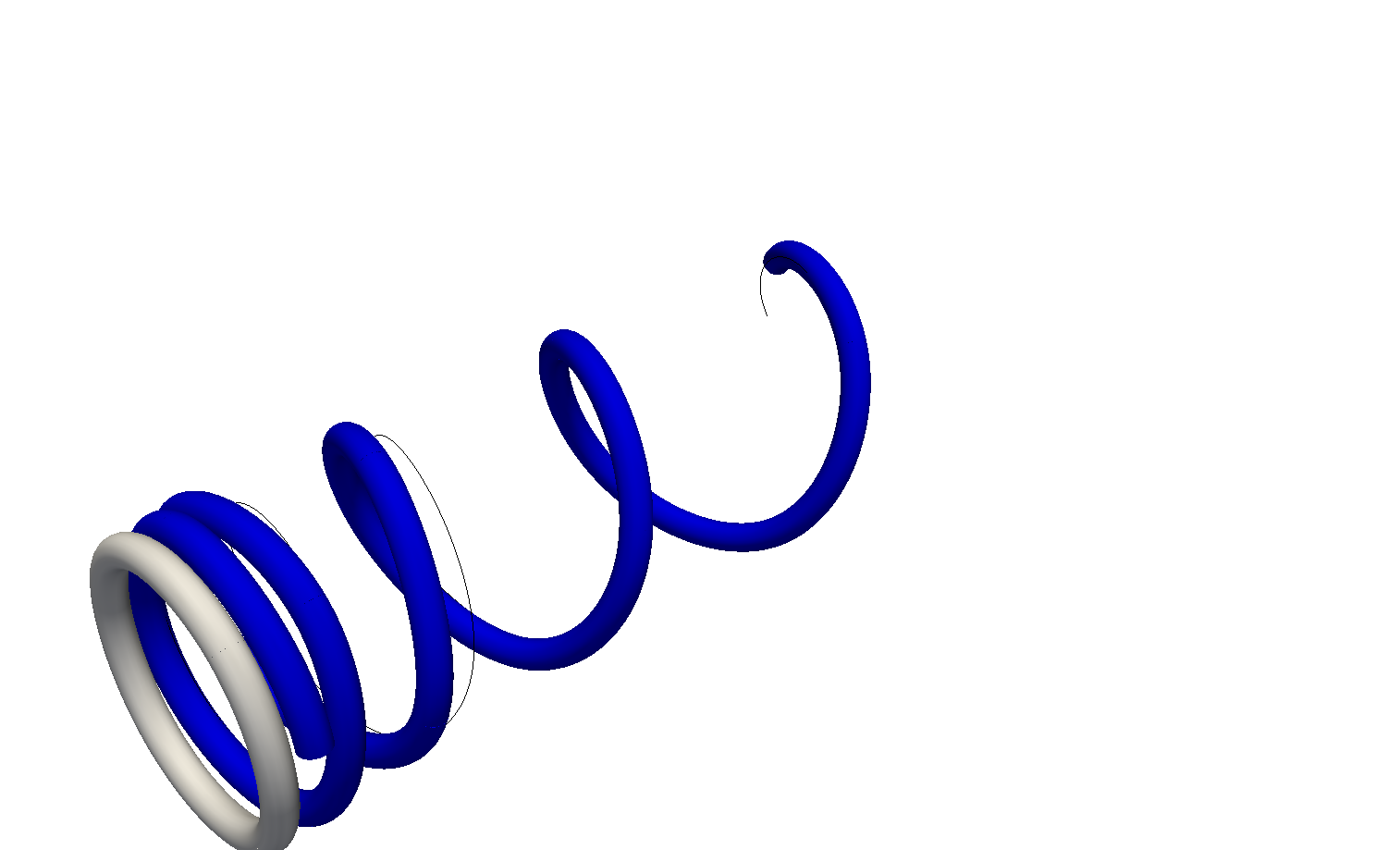}
   }
     \subfigure[time $t=7$]
   {
    \includegraphics[width=0.18\textwidth]{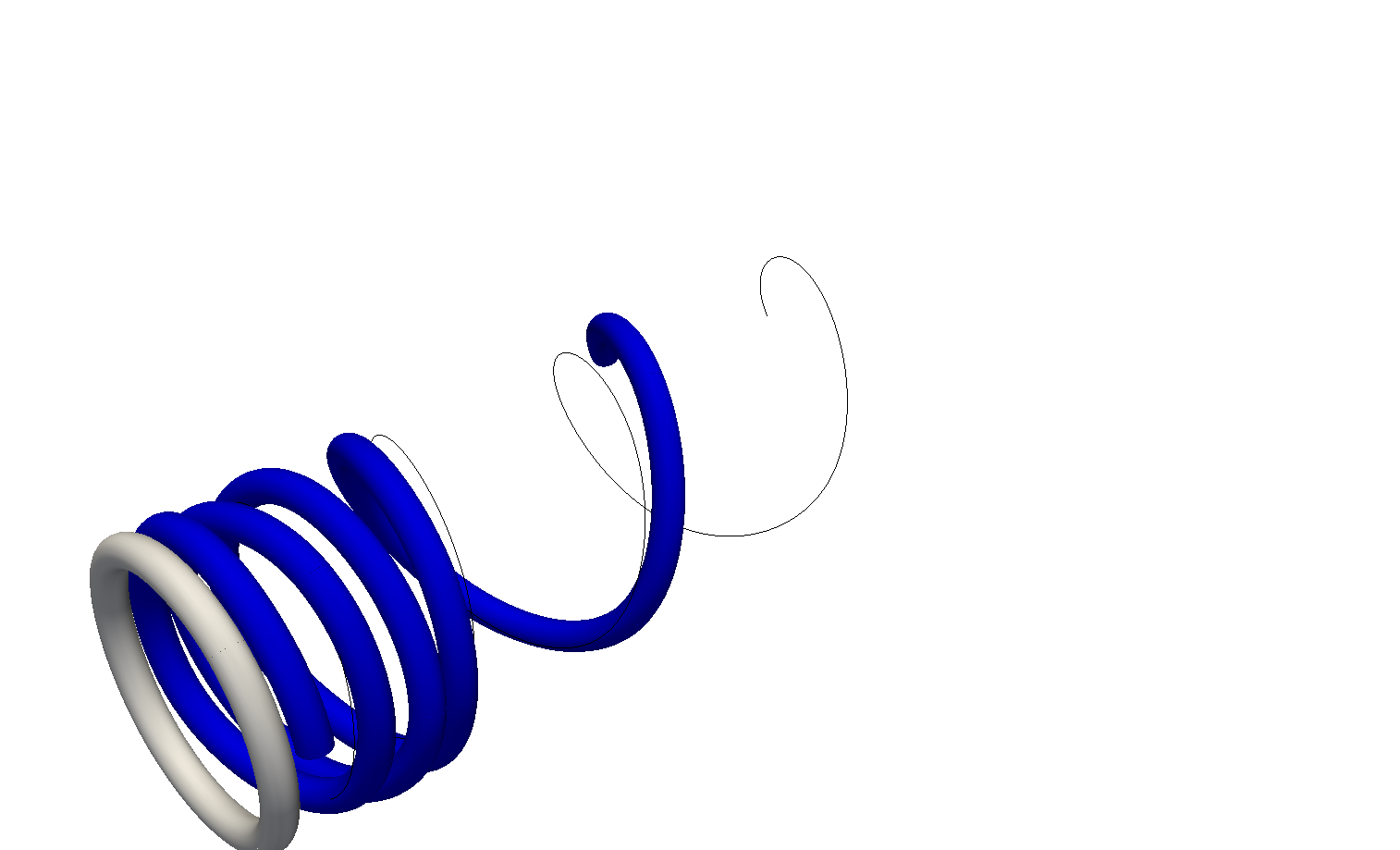}
   }
     \subfigure[time $t=7.5$]
   {
    \includegraphics[width=0.18\textwidth]{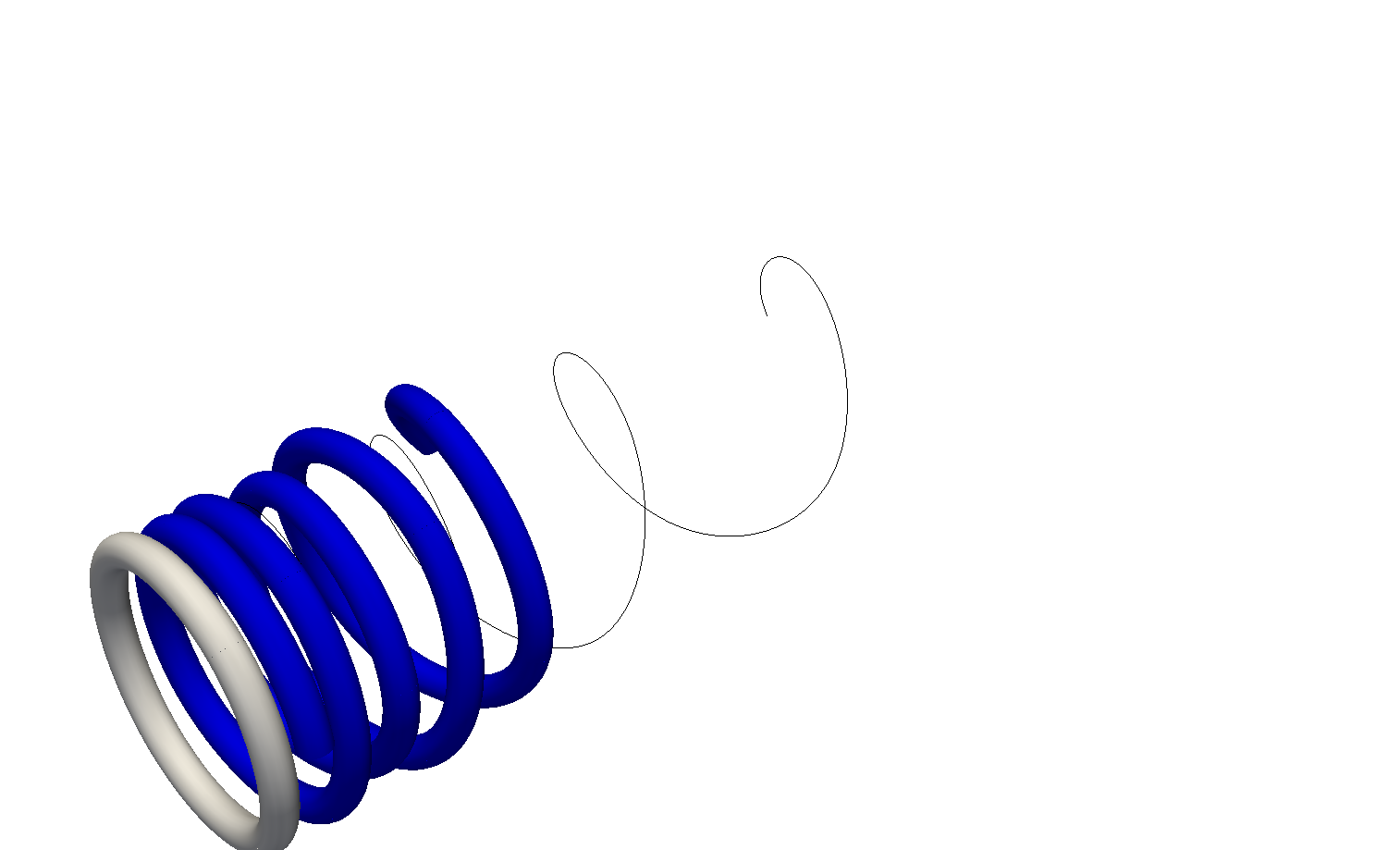}
   }
   \subfigure[time $t=8$]
   {
    \includegraphics[width=0.18\textwidth]{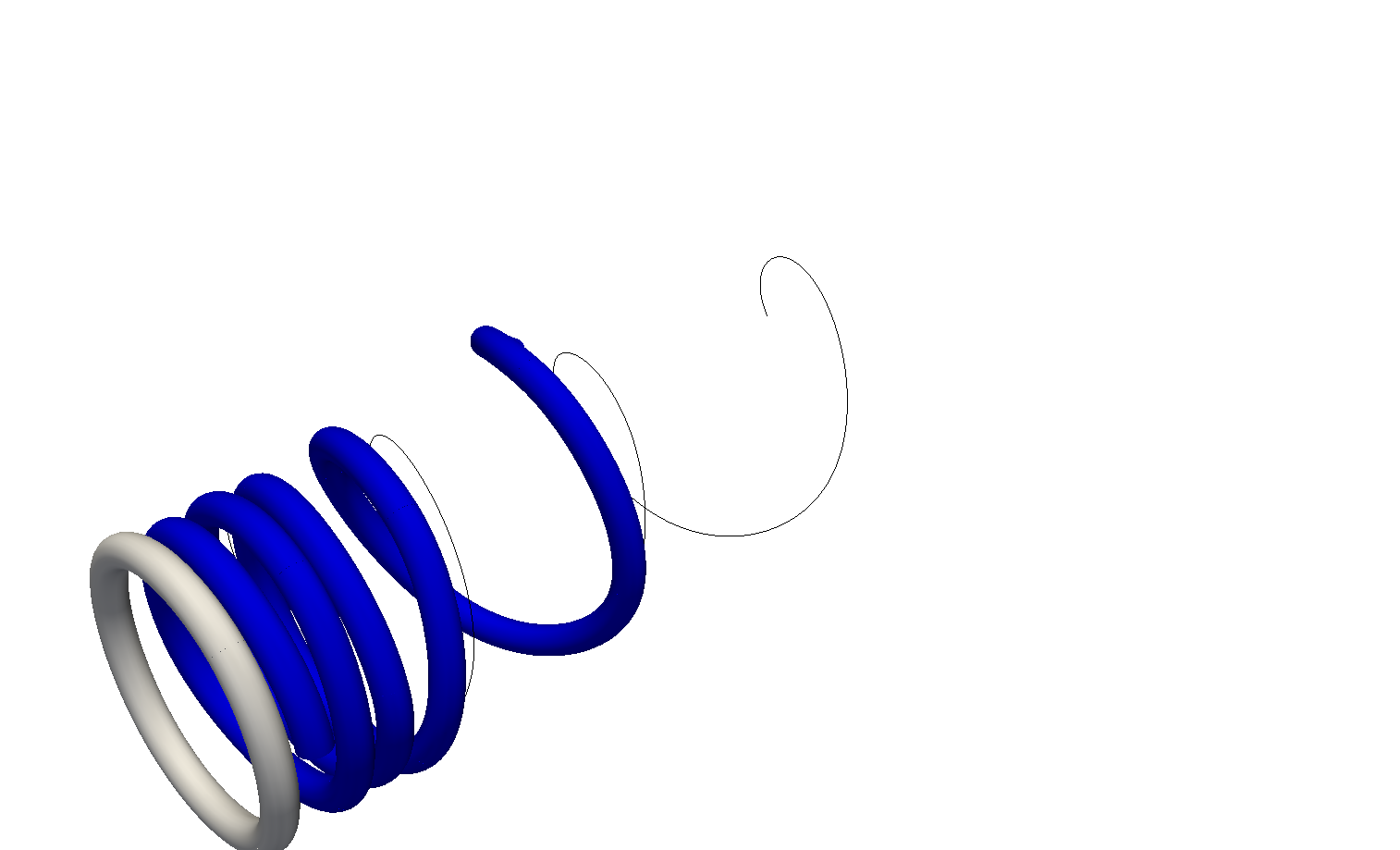}
   }
   \subfigure[time $t=9$]
   {
    \includegraphics[width=0.18\textwidth]{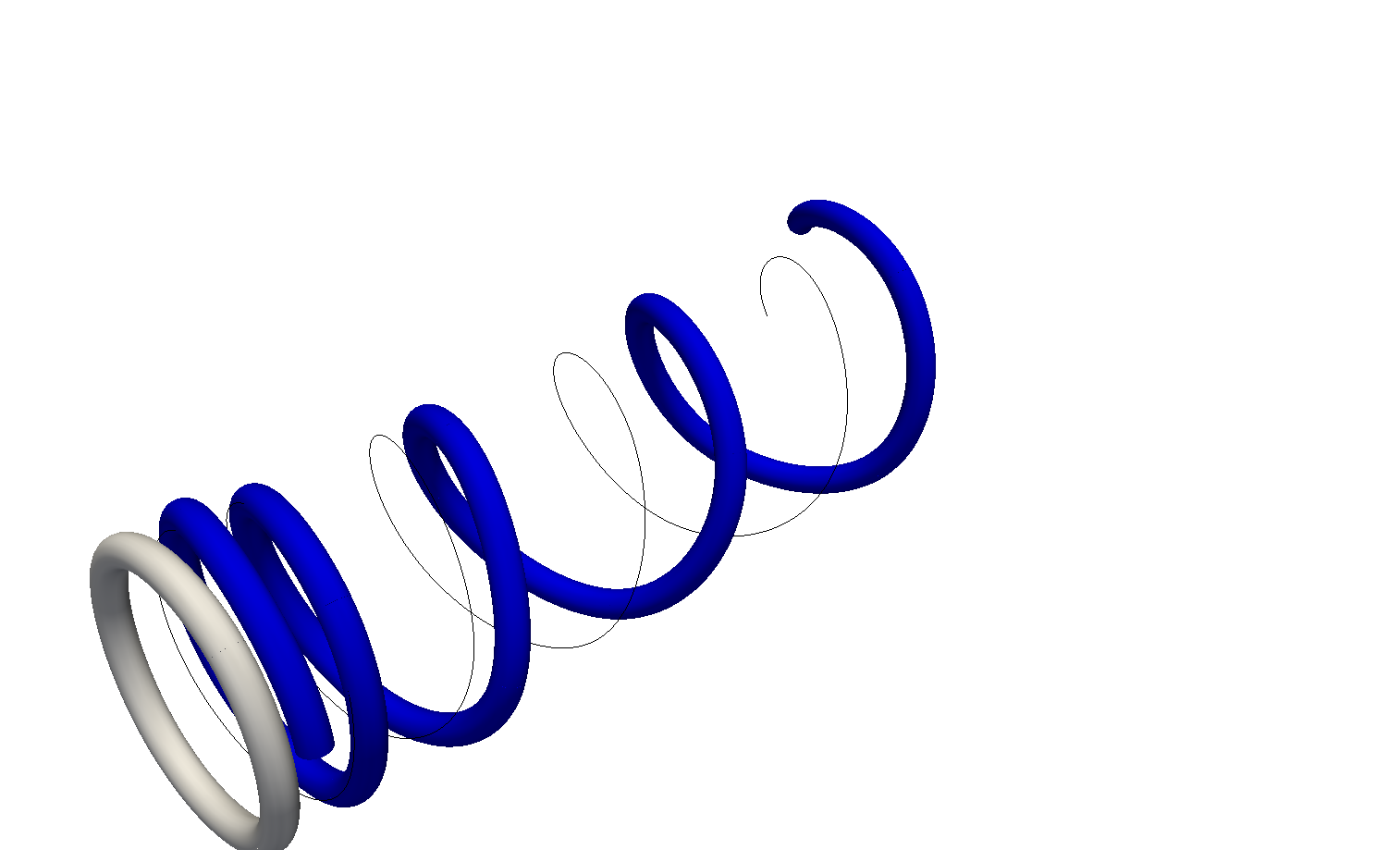}
   }
  \subfigure[time $t=10$]
   {
    \includegraphics[width=0.18\textwidth]{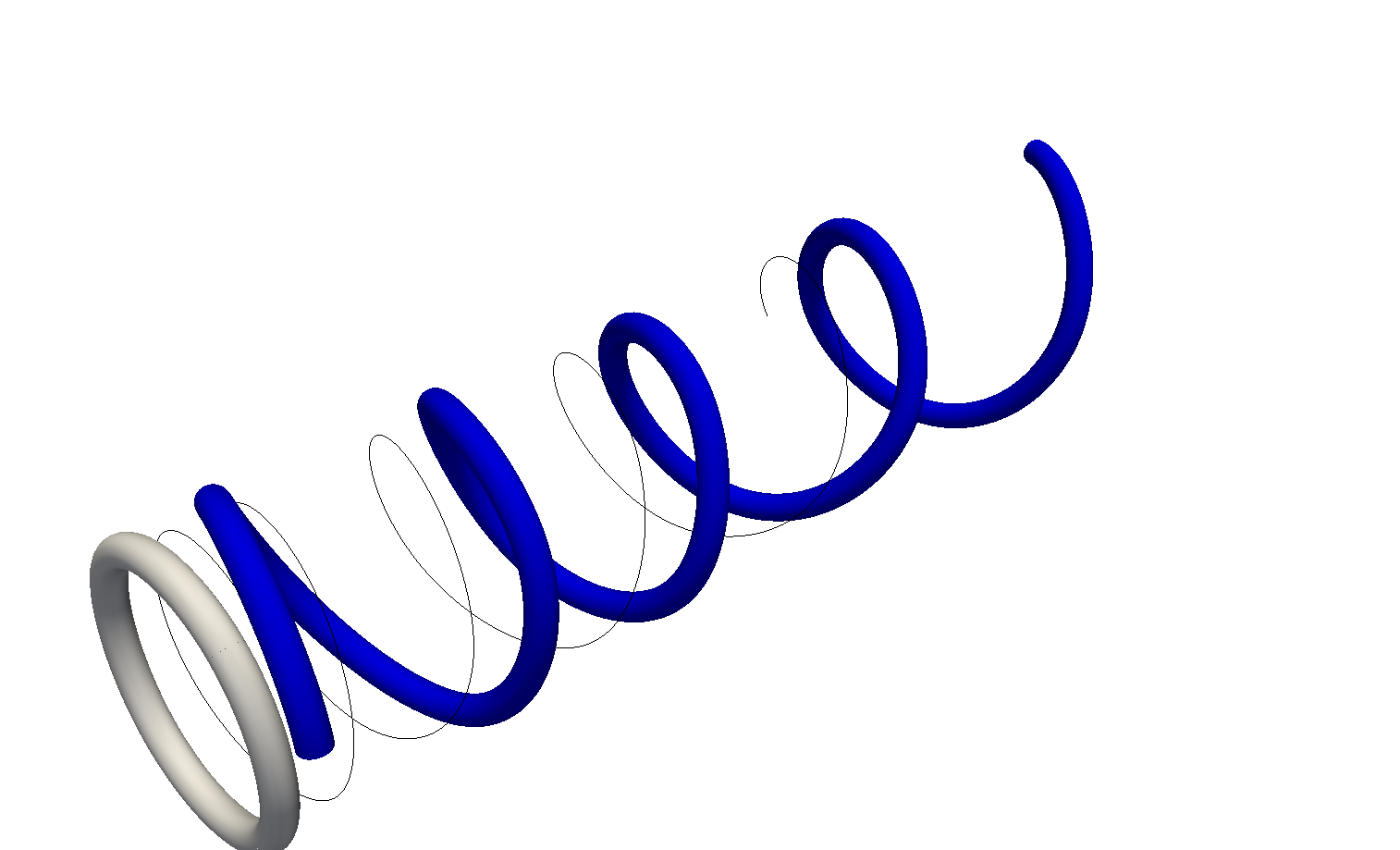}
   }
\caption{Dynamic simulation of a clamped helical spring loaded by a tip force.}
\label{fig:helical_spring_dynamicconfigs}
\end{figure*}

\begin{figure*}[ht!!!]
 \centering
  \subfigure[$16$ SR vs $64$ SR elements]
   {
   \label{fig:helical_spring_energy_16vs64ele}
    \includegraphics[width=0.45\textwidth]{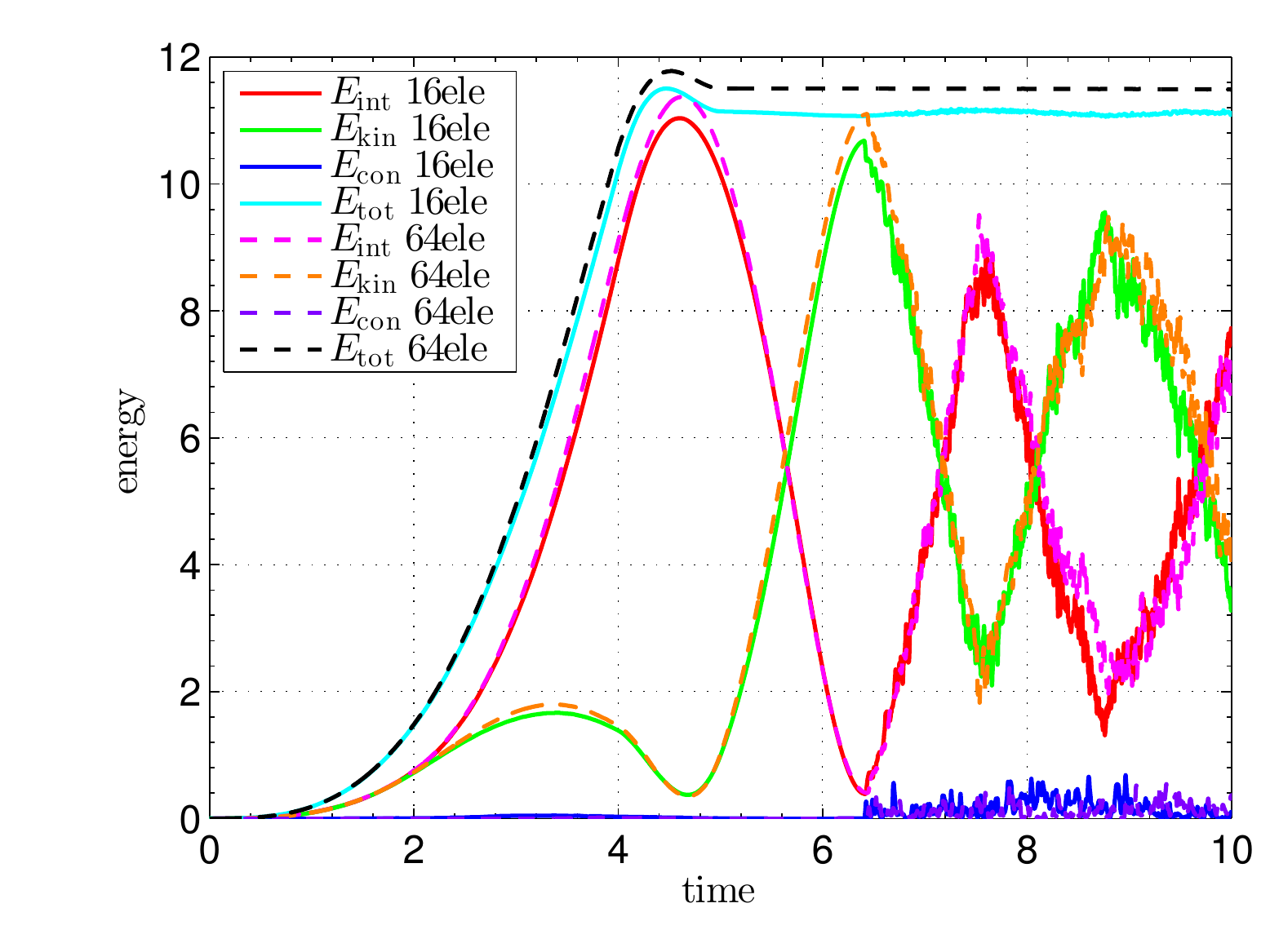}
   }
  \subfigure[$64$ KL vs $64$ SR elements]
   {
   \label{fig:helical_spring_energy_KLvsSR}
    \includegraphics[width=0.4\textwidth]{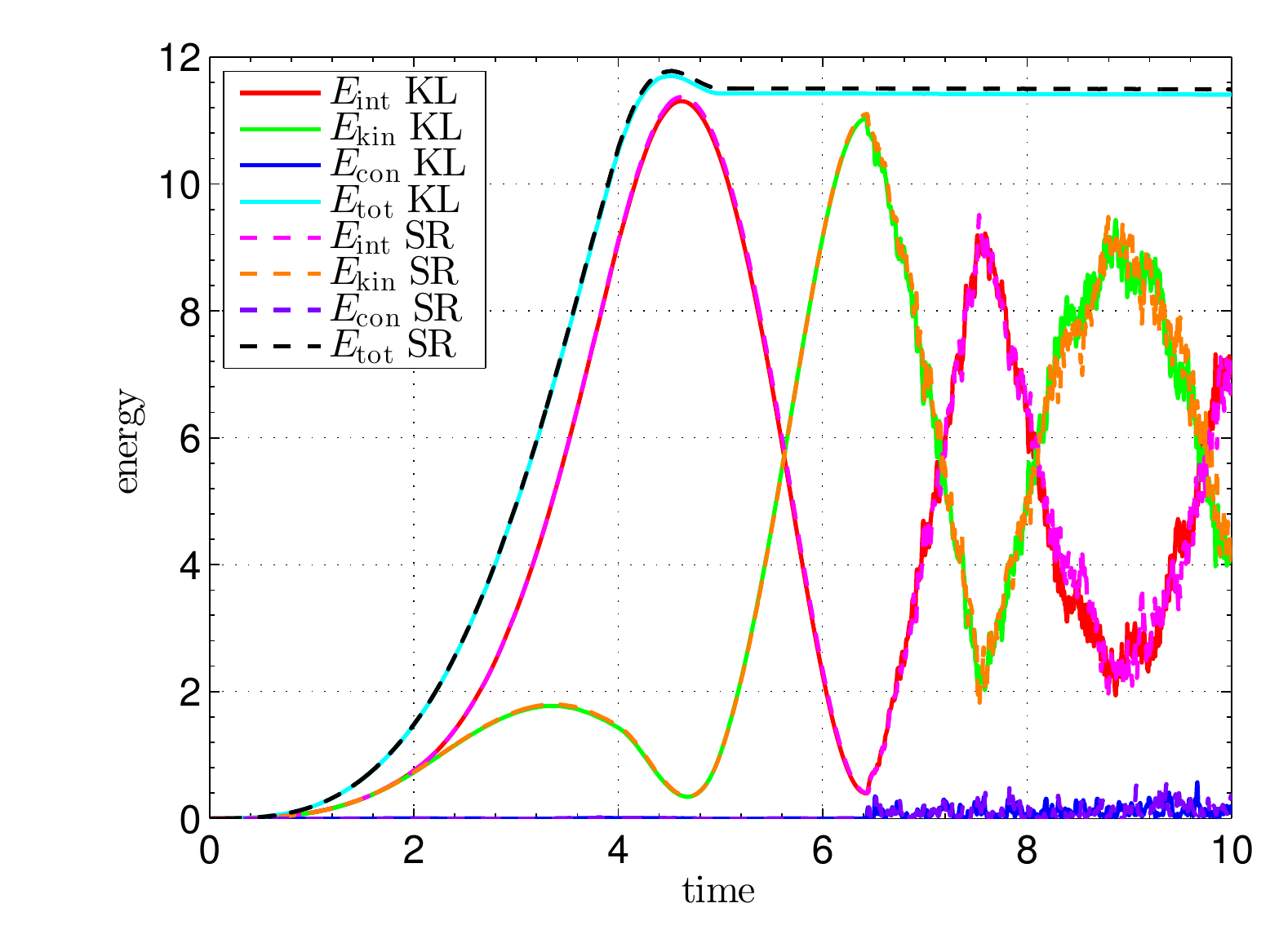}
   }
\caption{Conservation of total system energy.}
\label{fig:helical_spring_energy}
\end{figure*}

After this quantitative static analysis, also the dynamic response of the elastic spring shall be investigated in a qualitative as well as quantitative manner. For comparison purposes, the helical structure is discretized by $16$ or $64$ beam elements of tangent-vector based KL or SR type respectively. Again, one end of the spring is clamped but now the other end is loaded by a discrete external force~$\bar{f}_z$ in $z$-direction. As the location of load application, i.e. the tip, does not lie on the centerline of the helix, it causes a moment around the $x$-axis that is (partly, since the tip location changes in time) balanced by an additionally applied discrete external moment $\bar{m}_x\!=\!R_0\bar{f}_z$. Both loads are increased linearly within~$0\!<\!t\!<\!4$ and subsequently decreased linearly within~$4\!<\!t\!<\!5$. After the release of the applied load, the spring performs a free oscillation and the simulation covers a full cycle up to $t_{\text{end}}=10$.

The  extended generalized-$\alpha$ scheme presented in Section~\ref{sec:beams_temporal_SRKL} with a prescribed spectral radius $\rho_\infty\!=\!0.95$, i.e. a small amount of numerical dissipation, and a constant time step size of $\Delta t\!=\! 10^{-3}$ is applied for temporal discretization of the problem.

Figure~\ref{fig:helical_spring_dynamicconfigs} visualizes the deformed geometry at several points in time. In the course of the simulation, the spring undergoes large deformations in the tensile as well as the compression regime and shows a highly dynamic motion. Contact interactions with the rigid guides as well as self contact of the coils occur over the entire simulated time span. Note that in the state of high compression, also the beam endpoints come into contact with other coils such that the contact model for beam endpoints as mentioned in Section~\ref{sec:contact} and proposed in~\cite{meier2015b} is mandatory for this challenging example.

In terms of quantitative analysis, the elastic, kinetic and contact penalty energy is plotted over time in Figure~\ref{fig:helical_spring_energy}. The sum of these three contributions is plotted as total system energy and remains constant with only very little deviations after the external load is released (see Figure~\ref{fig:helical_spring_energy_16vs64ele}). Already for a rather coarse spatial discretization with $16$ beam elements, energy conservation is thus very well fulfilled by the applied beam and contact formulation and the time integration scheme. For a fine discretization of $64$ elements, energy is even better conserved and the fluctuations associated with the highly dynamic shocks from self-contacting coils in the compressed state are hardly visible anymore. Even though the total system energy is slightly higher for the SR elements due to the prevalence of additional shear deformation modes, the results for KL and SR elements match excellently well for a sufficiently fine discretization with small spatial discretization error (see Figure~\ref{fig:helical_spring_energy_KLvsSR}).

While this example represents the only dynamic test case within this work, further dynamic applications and investigations done on the important topic of total energy conservation in the context of the proposed ABC formulation can be found in~\cite{meier2015c}. There, for example the dynamic failure of two contacting ropes, similar to the one presented in Section~\ref{sec:examples_cabletwisting}, as well as the associated system energy have been investigated. Moreover, also the influence of mechanical contact on the Brownian dynamics of biopolymer networks, tight networks of highly slender filaments prevalent e.g. in biological cells, have been studied in this reference.

\subsection{Static load test on a webbing}
\label{sec:examples_webbing}

As final application, a static load test performed on a fibrous webbing shall be analyzed. The webbing consists of $10 + 20$ ribbons, each of them made out of $10$ individual fibers with circular cross-sections. The geometrical and constitutive parameters of an individual fiber are given by $l\!=\!500$, $R\!=\!1$, $E\!=\!2G\!=\! 1.0 \!\cdot\! 10^7$. Each fiber endpoint is simply supported and the positions of these supports are chosen such that the fibers are initially stress-free in case no beam-to-beam contact interaction is considered. Within a ribbon, two neighboring fibers pointing in global $x$-direction exhibit a vanishing initial gap $g_{0}\!=\!0$ while the fibers pointing in global $y$-direction are placed with an initial distance of $g_{0}\!=\!R$. Each fiber is discretized by $20$ TF elements yielding a global system that consists of $300$ fibers, $6000$ finite elements and approximately $38000$ degrees of freedom. In this example, the regimes of point and line contact are clearly separated. Thus, the shifting angles of the ABC formulation are chosen to $\alpha_1\!=\!40^{\circ}$ and $\alpha_2\!=\!45^{\circ}$. Since no active contacts lying within the transition interval are expected, the line and point penalty parameters do not necessarily have to be harmonized. Concretely, a quadratically regularized penalty law with $\varepsilon_{\perp}\!=\!2.4 \cdot 10^5$, $\varepsilon_{\parallel}\!=\!2.0 \cdot 10^4$ and $\bar{g}\!=\!0.1R$ in combination with one three-point integration interval per slave element is chosen. Again, the global Newton-Raphson scheme is supplemented by the step size control and load step adaption scheme as introduced in the beginning of Section~\ref{sec:examples}.

\begin{figure*}[ht!!!]
 \centering  
   {
    \includegraphics[width=0.46\textwidth]{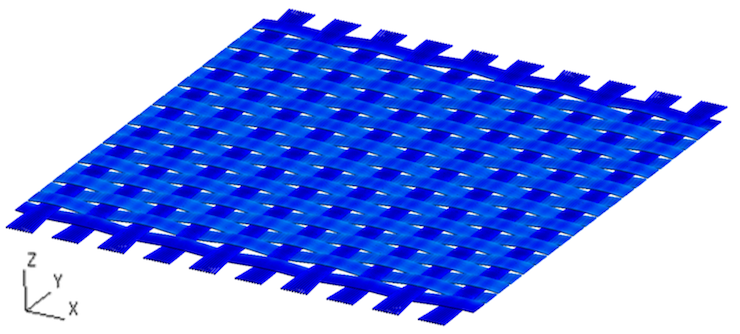}
    \label{fig:webbing_initial1c}
   }         
   {
    \includegraphics[width=0.46\textwidth]{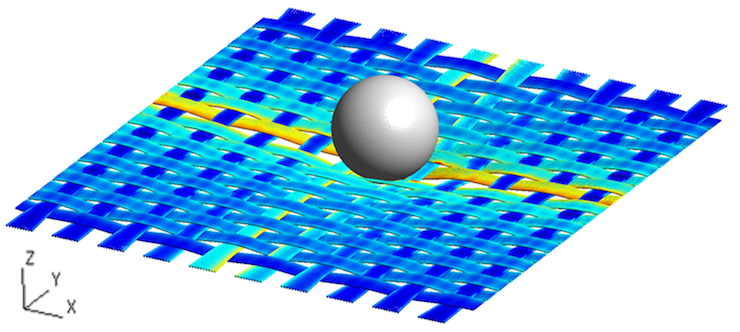}
    \label{fig:webbing_deformed1c}
   }
   \vspace{0.3cm}   
   {
    \includegraphics[width=0.46\textwidth]{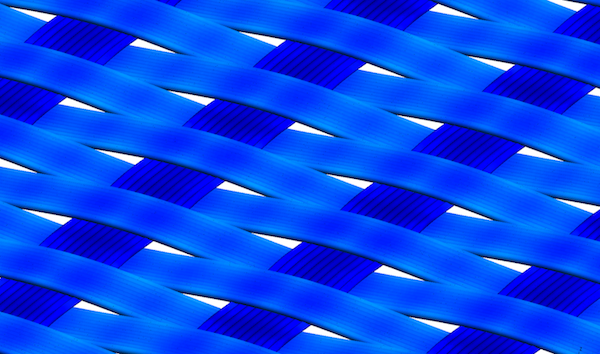}
    \label{fig:webbing_initial2c}
   }
   {
    \includegraphics[width=0.46\textwidth]{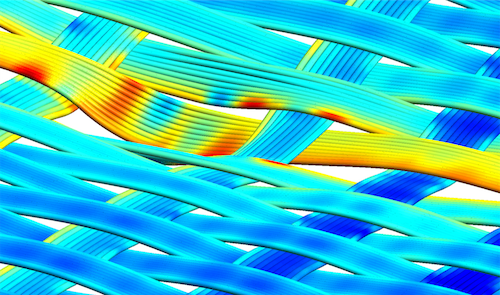}
    \label{fig:webbing_deformed3c}
   }
   \vspace{0.40cm}   
   {
    \includegraphics[width=0.46\textwidth]{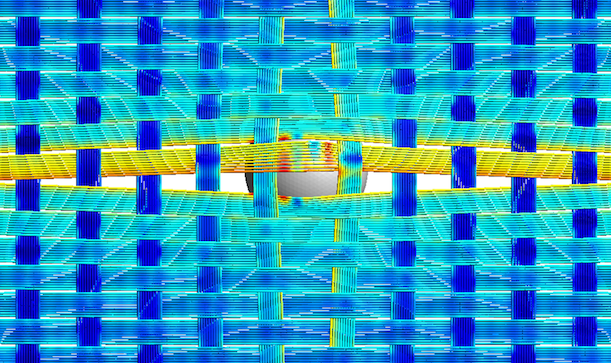}
    \label{fig:webbing_deformed4b}
   }
   {
    \includegraphics[width=0.46\textwidth]{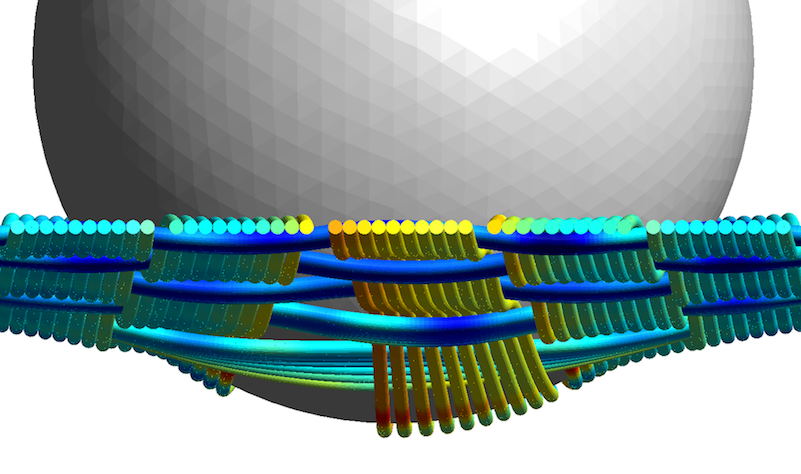}
    \label{fig:webbing_deformed7b}
   }   
   \vspace{0.40cm}      
   {
    \includegraphics[width=0.94\textwidth]{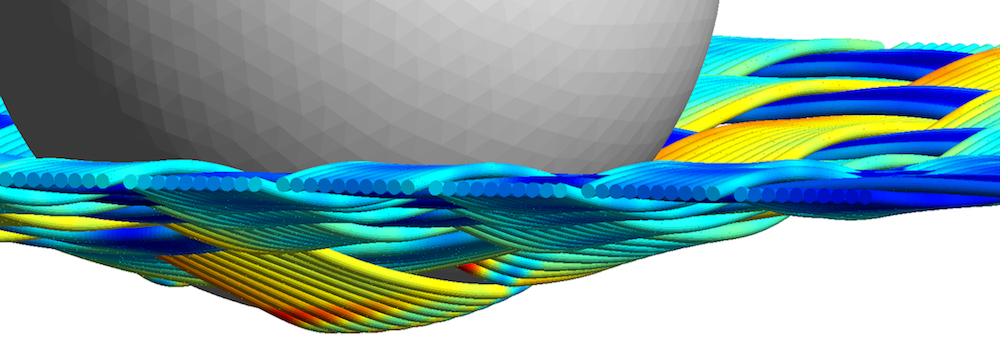}
    \label{fig:webbing_deformed6b}
   }
   \vspace{0.3cm}
   {
    \includegraphics[width=0.91\textwidth]{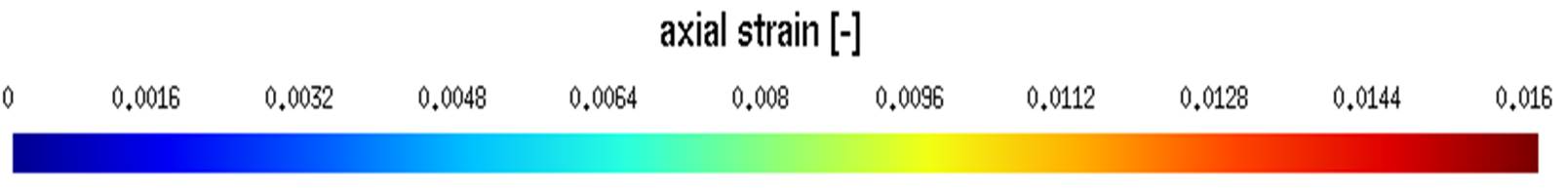}
    \label{fig:webbing_colorbarb}
   }
  \caption{Static load test on a webbing: Initial and deformed configurations.}
  \label{fig:webbing}
\end{figure*}

In order to determine the pre-stressed initial configuration, the fibers in $x$-direction are first loaded by a properly chosen sinusoidal line load. However, contact interaction is not considered in this first step. After activating the contact algorithm, the line load is reduced to zero in an incremental manner in order to finally yield the equilibrium configuration of the unloaded system as illustrated in Figure~\ref{fig:webbing} (first and second row, left). In a next step, the deformation of the resulting webbing when exposed to a point loading shall be investigated. Thereto, a test piece in form of a rigid sphere (radius $50$) is driven into the webbing. This process is performed in an incremental, Dirichlet-controlled manner. The modeling of the contact interaction between the rigid sphere and the individual fibers is similar to the procedure described in Section~\ref{sec:contact_point} in the context of beam point-to-point contact. Different perspectives of the final deformed configuration are again shown in Figure~\ref{fig:webbing}. This state is characterized by approximately $15000$ active line contact Gauss points, $16000$ active point contacts and maximal penetrations in the range of $10\%$ of the cross-section diameter for both regimes. The well-balanced number of active point and line contacts underlines the efficiency potential of the ABC formulation and demonstrates that the computationally expensive line contact contributions have been successfully reduced. In Figure~\ref{fig:webbing}, also the magnitudes of the resulting axial tension within the individual fibers are illustrated. Often, the mechanical fiber interaction in webbings of the type considered here is strongly determined by friction forces. Thereto, a future extension of the proposed beam-to-beam contact formulation by frictional effects seems to be very promising in order to improve the model quality and the significance of the generated simulation results. Nevertheless, for material pairings exhibiting low friction coefficients, the outcomes visualized in Figures~\ref{fig:webbing} already provide a first quantification of the expected fiber stresses and a first hint with respect to possible failure mechanisms. Furthermore, the perhaps more important purpose of this example is to demonstrate the robustness and scalability of the presented beam element and beam-to-beam contact formulations when applied to systems of practical relevance and size.

\section{Conclusion}
\label{sec:conclusion}

The focus of this contribution lay on the development of finite element formulations for the accurate modeling and efficient implicit dynamics simulation of slender fiber- or rod-like components and their contact interaction being embedded in complex systems of fiber-based materials and structures. 

While the vast majority of existing geometrically exact beam element formulations is based on the Simo-Reissner beam theory of thick (shear-deformable) rods, in the authors' recent contributions~\cite{meier2014,meier2015,meier2016} the first geometrically exact beam elements based on the Kirchhoff-Love theory of thin (shear-free) rods have been proposed that are capable of modeling general beam geometries with arbitrary initial curvatures and anisotropic cross-section shapes and that preserve important mechanical properties such as objectivity and path-independence. In~\cite{meier2016} it has been shown that the avoidance of the very high shear stiffness contributions achieved by these Kirchhoff-Love elements leads to considerable numerical advantages in the range of high beam slenderness ratios as compared to existing formulations of Simo-Reissner type. Exactly such high beam slenderness ratios are prevalent in many applications of practical interest, recommending the proposed Kirchhoff-Love beam element formulations as method of choice for the numerical simulation of such systems.

Concretely, four different Kirchhoff-Love element variants have been proposed and analyzed in~\cite{meier2016}. These basically differ in the applied rotation interpolation, either based on a strong or a weak enforcement of the Kirchhoff constraint, as well as in the parametrization of nodal rotations, either based on nodal rotation vectors or on nodal tangent vectors. Within the present work, only the two variants either being based on nodal rotation vectors (KL-ROT element) or on nodal tangent vectors (KL-TAN element) have been distinguished. Besides this general Kirchhoff-Love beam elements, also a reduced model leading to a special torsion-free beam element formulation has been proposed in~\cite{meier2015}. There, it has been shown that under certain restrictions concerning the initial beam geometry (straight beams with isotropic/circular cross-sections) and external loads (no torsional components of external moments), the Kirchhoff-Love theory yields solutions with vanishing torsion even for arbitrarily large displacements and rotations. This finding justified the development of a torsion-free beam element (TF element) that inherits the high accuracy well-known for geometrically exact beam element formulations, while simultaneously avoiding any rotational degrees of freedom typical for geometrically exact formulations. In turn, this leads to considerably simplified and consequently more efficient finite element formulations. Both, the general as well as the torsion-free Kirchhoff-Love elements mentioned so far are based on a $C^1$-continuous beam centerline representation, a distinctive property which enables smooth kinematics in the context of beam contact schemes.

While in~\cite{meier2016}, Kirchhoff-Love finite elements have been identified as the formulations of choice in the regime of high beam slenderness ratios, it is beyond all question that finite element formulations of Simo-Reissner type should be preferred for thick beam geometries where shear deformation may play an important role. For that reason, a novel geometrically exact Simo-Reissner beam element based on a third-order Hermite centerline interpolation has been proposed as alternative to the considered Kirchhoff-Love elements when modeling smooth contact interaction in the range of low beam slenderness ratios. For this formulation, optimal spatial convergence rates as well as the successful avoidance of membrane and shear locking, achieved via a Gauss-Lobatto reduced integration scheme, has been verified numerically.

In the mechanical modeling and numerical simulation of beam-to-beam contact interaction basically two different types of approaches can be distinguished:  point-to-point and line-to-line contact models. In the authors' recent works~\cite{meier2015b,meier2015c}, it has been shown that line contact formulations applied to slender beams provide very accurate and robust mechanical models in the range of small contact angles, whereas the computational efficiency considerably decreases with increasing contact angles. On the other hand, point contact formulations serve as sufficiently accurate and very elegant and efficient models in the regime of large contact angles, while they are not applicable for small contact angles as a consequence of non-unique closest point projections. In order to combine the advantages of these two basic types of formulations, while abstaining from their disadvantages, a novel all-angle beam contact (ABC) formulation has been proposed in~\cite{meier2015c}. This formulation applies a point contact formulation in the range of large contact angles and a line contact formulation in the range of small contact angles, the two being smoothly connected by means of a variationally consistent model transition approach. However, all examples presented in~\cite{meier2015c} have employed the TF element formulation~\cite{meier2015} mentioned above. In the present work, the all-angle beam contact formulation proposed in~\cite{meier2015c} has been extended to the different types of general Kirchhoff-Love beam elements proposed in~\cite{meier2016} and to the new SR element proposed in the present work. Whereas no adaption of the ABC formulation has been necessary for most of the considered beam elements, the combination of the ABC scheme with the KL-ROT element employing a rotation vector-based parametrization of nodal rotations (see also~\cite{meier2016}) required an additional transformation of the contact residual and stiffness contributions as compared to~\cite{meier2015c}, which has been derived in the present work.

Eventually, a series of practically relevant applications that pose important challenges to the applied beam element formulations have been investigated. Therein, the properties and performance of the different beam element formulations under consideration, i.e. of the torsion-free Kirchhoff-Love formulations, the general Kirchhoff-Love formulations and the Simo-Reissner formulations, have been compared. Especially the treatment of examples exhibiting initial curvatures as well as rigid joint connections, thus scenarios which can not be modeled by means of the torsion-free theory, have been in the focus. By means of a numerical test case considering the contact interaction between two cylindrical tubes with complex hexagonal microstructure, it has been shown that this KL-ROT element enables a simple and straight-forward formulation of mechanical joint conditions on the one hand, and that the $C^1$-continuous geometry representation enabled by the Hermite centerline interpolation can even be preserved at multi-element joints on the other hand. This property turned out to be very beneficial for modeling and robustly simulating the contact interaction between such microstructures. In a further example, the contact dynamics occurring during the free oscillation of an initially curved helical spring have been investigated on the basis of different spatial discretizations with KL and SR elements. Accordingly, the important property of energy conservation could be confirmed for the combination of these different spatial finite element discretizations, the proposed ABC formulation and the employed time integration scheme, a recently proposed extension of the well-known generalized$-\alpha$ scheme to problems involving large rotations. Furthermore, two examples fulfilling the requirements of the torsion-free theory, namely the twisting process of a rope and a static load test performed on an industrial webbing, have been considered allowing for a comparison with the general KL and SR elements.

In all these examples, the main tendencies observed in~\cite{meier2016} for problems without beam-to-beam contact could be confirmed under the presence of contact interaction. Correspondingly, the results obtained by Simo-Reissner and Kirchhoff-Love beam elements coincided very well in the range of moderate to high slenderness ratios, while for very low beam slendernesses a visible difference between these two models could be observed. However, whenever the considered beam slenderness ratio allowed for an application of the shear-free Kirchhoff-Love elements, these formulations turned out to be clearly favorable in terms of nonlinear solver performance and overall robustness leading to a considerably decreased number of accumulated Newton iterations as compared to the investigated shear-deformable Simo-Reissner formulation. Furthermore, it has been verified that, also for complex examples involving intensive mechanical contact interaction, the TF element formulation yields identical results as the general Kirchhoff-Love elements as long as the specified restrictions concerning external loads and initial geometry are fulfilled. Moreover, in~\cite{meier2015,meier2016} it has already been shown that the number of degrees of freedom required to achieve a certain discretization error level typically increases when switching from the TF element to the KL element and eventually to the SR element. By taking into account all these aspects, the overall computational cost required to guarantee for a specified solution quality is smallest for TF, followed by KL and highest for SR elements. These findings underline the superiority of torsion and/or shear-free beam element formulations in cases where the underlying assumptions are met.

\appendix

%
\section{Residual vector and stiffness of Simo-Reissner element}
\label{anhang:sr_element}
%

For the following derivations, the vector $\hat{\mbd{x}}$ containing the nodal degrees of freedom of the Simo-Reissner element (see Section~\ref{sec:beams_SR}) is split into the two parts $\hat{\mbd{d}}$ of degrees of freedom associated with the centerline interpolation and $\hat{\mbds{\psi}}$ of degrees of freedom associated with the rotation interpolation:
\begin{align}
\label{lambda_reissner_anhang}
\begin{split}
\hat{\mbd{x}}\!&:=\!(\hat{\mb{d}}^{1T}\!,\hat{\mb{t}}^{1T}\!,\hat{\boldsymbol{\psi}}^{1T}\!,\hat{\mb{d}}^{2T}\!,\hat{\mb{t}}^{2T}\!,\hat{\boldsymbol{\psi}}^{2T},\hat{\boldsymbol{\psi}}^{3T})^T, \\\hat{\mbd{d}}\!&:=\!(\hat{\mb{d}}^{1T}\!,\hat{\mb{t}}^{1T}\!,\hat{\mb{d}}^{2T}\!,\hat{\mb{t}}^{2T}\!)^T, \\
\hat{\mbds{\psi}}\!&:=\!(\hat{\boldsymbol{\psi}}^{1T}\!,\hat{\boldsymbol{\psi}}^{2T}\!,\hat{\boldsymbol{\psi}}^{3T}\!)^T.
\end{split}
\end{align}
In a similar manner, the variations $\delta (.)$ and increments $\Delta (.)$ occurring in a linearization are defined:
\begin{align}
\label{lambda_reissner_anhang2}
\begin{split}
\delta \hat{\mbd{d}}\!&:=\!(\delta \hat{\mb{d}}^{1T}\!,\delta \hat{\mb{t}}^{1T}\!,\delta \hat{\mb{d}}^{2T}\!,\delta \hat{\mb{t}}^{2T}\!)^T, \\
\Delta \hat{\mbd{d}}\!&:=\!(\Delta \hat{\mb{d}}^{1T}\!,\Delta \hat{\mb{t}}^{1T}\!,\Delta \hat{\mb{d}}^{2T}\!,\Delta \hat{\mb{t}}^{2T}\!)^T, \\
\delta \hat{\mbds{\theta}}\!&:=\!(\delta \hat{\boldsymbol{\theta}}^{1T}\!,\delta \hat{\boldsymbol{\theta}}^{2T}\!,\delta \hat{\boldsymbol{\theta}}^{3T}\!)^T, \\
\Delta \hat{\mbds{\theta}}\!&:=\!(\Delta \hat{\boldsymbol{\theta}}^{1T}\!,\Delta \hat{\boldsymbol{\theta}}^{2T}\!,\Delta \hat{\boldsymbol{\theta}}^{3T}\!)^T.
\end{split}
\end{align}
The different symbol $\boldsymbol{\theta}$ has been chosen for the variations / increments of $\boldsymbol{\psi}$, since these are of multiplicative nature (see e.g.~\cite{simo1985,jelenic1999,meier2016}), while $\delta \hat{\mbd{d}}$ and $\Delta \hat{\mbd{d}}$ represent additive variations and increments of the primary variables $\hat{\mbd{d}}$. In addition, the matrix $\mbd{H}$ is introduced. It contains the Hermite polynomials~\eqref{shapefunctions} such that the interpolation~\eqref{interpolation} and its variation / increment can equivalently be expressed by:
\begin{align}
\label{lambda_reissner_anhang3}
\begin{split}
\!\!\!\!\!\!
 \mb{r} \!&=\! \mbd{H} \hat{\mbd{d}}, \,\, \mb{r}^{\prime} \!=\! \mbd{H}^{\prime} \hat{\mbd{d}}, \\
 \Delta \mb{r} \!&=\! \mbd{H} \Delta \hat{\mbd{d}}, \,\, \Delta \mb{r}^{\prime} \!=\! \mbd{H}^{\prime} \Delta \hat{\mbd{d}}, \\
 \delta \mb{r} \!&=\! \mbd{H} \delta \hat{\mbd{d}}, \,\, \delta \mb{r}^{\prime} \!=\! \mbd{H}^{\prime} \delta \hat{\mbd{d}}.
 \!\!\!\!\!\!
\end{split}
\end{align}
Based on the nodal triads $\boldsymbol{\Lambda}^i\!=\!\boldsymbol{\Lambda}^i(\hat{\boldsymbol{\psi}}^i)$ for $i\!=\!1,2,3$, the rotation interpolation originally proposed in~\cite{crisfield1999,jelenic1999} and given in abstract manner by equation~\eqref{abstract_triad_interpolation}, shall now be detailed:
\begin{align}
\label{lambda_reissner_anhang4}
\begin{split}
\mb{\Lambda}(\xi) \!&=\! \mb{\Lambda}^2 \exp{\!(\mb{S}(\boldsymbol{\Phi}_{l}(\xi)))}, \\
\boldsymbol{\Phi}_{l}(\xi) \!&=\! \sum_{i=1}^{3} L^i(\xi) \boldsymbol{\Phi}_{l}^i, \\
\exp{\!(\mb{S}(\boldsymbol{\Phi}_{l}^i))}&=\mb{\Lambda}^{2T} \! \mb{\Lambda}^i.
\end{split}
\end{align}
The rotation vector variation field $\delta \boldsymbol{\theta}$ is interpolated in a Petrov-Galerkin manner as given in~\eqref{petrovspininterpolation}
\begin{align}
\label{lambda_reissner_anhang5}
\!\!\!\!\!\!
\delta \boldsymbol{\theta}(\xi) \!=\! \sum_{i=1}^{3} L^i(\xi) \delta \hat{\boldsymbol{\theta}}^i \!=:\! \mbd{L} \delta \hat{\mbds{\theta}}, \,\,
\delta \boldsymbol{\theta}^{\prime}(\xi) \!=\! \mbd{L}^{\prime} \delta \hat{\mbds{\theta}},
\!\!\!\!\!\!
\end{align}
where $\mbd{L}$ is a elementwise assembly of the third-order Lagrange polynomials $L^i(\xi)$. The rotation vector increment field $\Delta \boldsymbol{\theta}$ associated with a consistent linearization of~\eqref{lambda_reissner_anhang4} has been derived in~\cite{jelenic1999}:
\begin{align}
\label{lambda_reissner_anhang6}
\begin{split}
\!\!\!\!\!\!
 \Delta \boldsymbol{\theta}(\xi) \!&=\! \sum_{i=1}^{3} \tilde{\mb{I}}^i(\xi) \Delta {\boldsymbol{\theta}}^i\!=:\!\tilde{\mbd{I}} \Delta \mbds{\theta}, \,\,
 \Delta \boldsymbol{\theta}^{\prime}(\xi) \!=\! \tilde{\mbd{I}}^{\prime} \Delta \mbds{\theta}.
 \!\!\!\!\!\!
\end{split}
\end{align}
The deformation-dependent shape function matrices $\tilde{\mb{I}}^i$ are defined in the original work~\cite{jelenic1999} and can also be found in~\cite{meier2016} (in a notation identical to the one used here). The matrix $\tilde{\mbd{I}}$ represents again a proper elementwise assembly of the nodal shape function matrices $\tilde{\mb{I}}^i$. Based on this notation and the employed interpolations, the element residual vector can be derived from~\eqref{weakformspatial} according to:
\begin{align}
\label{lambda_reissner_anhang7}
\begin{split}
 \!\!\!\!\!\!
 \mbd{r}_{SR,\hat{\mb{d}}} &\!=\! \int \limits_{-1}^{1} \left( \mbd{H}^{\prime T} \mb{f} \!-\! \mbd{H}^{\!T} \tilde{\mb{f}}_{\rho} \right) J d \xi  \!-\! \Big[\mbd{H}^{\!T} \mb{f}_{{\sigma}} \Big]_{\varGamma_{\sigma}} \!\!\!\!\!\!\\ \!\!\!\!\!\!
\mbd{r}_{SR,\hat{\boldsymbol{\theta}}} &\!=\! \int \limits_{-1}^{1} \left( \mbd{L}^{\! \prime T} \mb{m} \!-\! \mbd{L}^{T} \! \mb{S}(\mb{r}^{\prime}) \mb{f}
\!-\!\mbd{L}^{\!T} \tilde{\mb{m}}_{\rho} \right) J d \xi \!\!\!\!\!\! \\ & \,\,\,\,\,\,\,\,\, - \Big[ \mbd{L}^{\!T} \mb{m}_{{\sigma}} \Big]_{\varGamma_{\sigma}}\!\!\!\!\!\!.
\end{split}
\end{align}
For simplicity, distributed external forces and moments as well as inertia forces and moments are collected according to $\mb{\tilde{f}}_{\rho}\!:=\!\mb{\tilde{f}}\!+\!\mb{f}_{\rho}$ and $\mb{\tilde{m}}_{\rho}\!:=\!\mb{\tilde{m}}\!+\!\mb{m}_{\rho}$. The linearization of the residual~\eqref{lambda_reissner_anhang7} yields:
\begin{align}
\label{lambda_reissner_anhang8}
\begin{split}
 \!\!\!\!\!\!
 \Delta \mbd{r}_{SR,\hat{\mb{d}}} &\!=\! \int \limits_{-1}^{1} \left( \mbd{H}^{\prime T} \Delta \mb{f} \!-\! \mbd{H}^{\!T} \Delta \mb{f}_{\rho} \right) J d \xi  \!\!\!\!\!\!\\ \!\!\!\!\!\!
\Delta \mbd{r}_{SR,\hat{\boldsymbol{\theta}}} &\!=\! \int \limits_{-1}^{1} \Big( \mbd{L}^{\! \prime T} \Delta \mb{m} 
\!+\! \mbd{L}^{T} \! \mb{S}(\mb{f}) \Delta \mb{r}^{\prime} \!\!\!\!\!\! \\
\!&\,\,\,\,\,\,\,\,\,\,\,-\! \mbd{L}^{T} \!  \mb{S}(\mb{r}^{\prime}) \Delta \mb{f}\!-\!\mbd{L}^{\!T} \Delta \mb{m}_{\rho} \Big) J d \xi\!\!\!\!\!\!.
\end{split}
\end{align}
The linearization $\Delta  \mb{m}$ of the internal moments is
\begin{align}
\label{lambda_reissner_anhang9}
\begin{split}
\!\!\!\!\!\!
  \Delta  \mb{m} \!&=\! -\mb{S}(\mb{m}) \Delta \boldsymbol{\theta} \!+\! \mb{c}_m \Delta \boldsymbol{\theta}^{\prime},
 \!\!\!\!\!\!
\end{split}
\end{align}
whereas the linearization $\Delta  \mb{f}$ of the force stress resultants is given by the expression:
\begin{align}
\label{lambda_reissner_anhang10}
\begin{split}
\!\!\!\!\!\!
 \Delta \mb{f}= \mb{c}_f  \Delta \mb{r}^{\prime}+\big(\mb{c}_f \mb{S}(\mb{r}^{\prime})-\mb{f}\big)\Delta \boldsymbol{\theta}.
 \!\!\!\!\!\!
\end{split}
\end{align}
The extended generalized$-\alpha$ scheme (see Section~\ref{sec:beams_temporal_SRKL}), which espresses the translational as well as rotational velocities and accelerations as occurring in the inertia forces $\mb{f}_{\rho}$ and moments $\mb{m}_{\rho}$ (see~\eqref{stress_resultants_and_inertia}), is presented in~\cite{meier2016}. It has to be emphasized that temporal discretization is performed \textit{after} spatial discretization (as typical for geometrically exact beam formulations). Thus, the corresponding finite difference relations of the time integration scheme are \textit{not} applied to the vector $\hat{\mbd{x}}$ of discrete nodal variables and its time derivatives, but rather to spatially interpolated quantities evaluated at the Gauss points required for numerical integration of~\eqref{lambda_reissner_anhang7}. Applying the procedure described in~\cite{meier2016}, the following linearization of the inertia forces and moments can be formulated:
\begin{align}
\label{lambda_reissner_anhang11}
\begin{split}
  \!\!\!\!\! 
  -&\Delta \mb{f}_{\rho}\!=\! k_1 \rho A \Delta \mb{r}, \!\!\!\!\!\!\\\!\!\!\!\!\! 
  -&\Delta \mb{m}_{\rho}\!=\! \mb{S}(\mb{m}_{\rho})\Delta \boldsymbol{\theta} \!\!\!\!\!\!\\\!\!\!\!\!\!
  +&\boldsymbol{\Lambda}^T[k_2\{ \mb{S}(\mb{w})\mb{c}_{\rho} \!-\! \mb{S}(\mb{c}_{\rho}\mb{w}) \}\!+\! k_1\mb{c}_{\rho}] \mb{T} \Delta \boldsymbol{\theta}.\!\!\!\!\!\!
\end{split}
\end{align}
Here, the constants $k_1$ and $k_2$ are defined by the parameters $\beta,\gamma,\alpha_f$ and $\alpha_m$ of the the (extended) generalized$-\alpha$ scheme according to:
\begin{align}
\label{lambda_reissner_anhang12}
\begin{split}
 k_1\!=\! \frac{1\!-\!\alpha_m}{(1\!-\!\alpha_m)\beta \Delta t^2}, \quad
k_2\!=\! \frac{\gamma}{\beta \Delta t}.
\end{split}
\end{align}
Furthermore, the transformation matrix $\mb{T}$ is also given in~\cite{meier2016}. Eventually, the linearization~\eqref{lambda_reissner_anhang8} can be stated in the following general form:
\begin{align}
\label{lambda_reissner_anhang13}
\begin{split}
\Delta \mbd{r}_{SR,\hat{\mb{d}}}&=\mbd{k}_{SR,\hat{\mb{d}}\hat{\mb{d}}} \Delta \hat{\mb{d}} + \mbd{k}_{SR,\hat{\mb{d}}\hat{\mbds{\theta}}} \Delta \hat{\mbds{\theta}}, \\
\Delta \mbd{r}_{SR,\hat{\boldsymbol{\theta}}}&=\mbd{k}_{SR,\hat{\boldsymbol{\theta}}\hat{\mb{d}}} \Delta \hat{\mb{d}} + \mbd{k}_{SR,\hat{\boldsymbol{\theta}}\hat{\mbds{\theta}}} \Delta \hat{\mbds{\theta}}.
\end{split}
\end{align}
Inserting equations~\eqref{lambda_reissner_anhang3} and~\eqref{lambda_reissner_anhang6} into equations~\eqref{lambda_reissner_anhang9},~\eqref{lambda_reissner_anhang10} and~\eqref{lambda_reissner_anhang11} and the latter into~\eqref{lambda_reissner_anhang8} finally allows to determine the element stiffness contributions $\mbd{k}_{SR,\hat{\mb{d}}\hat{\mb{d}}},\mbd{k}_{SR,\hat{\mb{d}}\hat{\mbds{\theta}}}, \mbd{k}_{SR,\hat{\boldsymbol{\theta}}\hat{\mb{d}}}$ as well as $\mbd{k}_{SR,\hat{\boldsymbol{\theta}}\hat{\mbds{\theta}}}$. In order to avoid membrane and shear locking, all residual and stiffness terms containing the force stress resultant vector $\mb{f}$ are numerically integrated by means of a 3-point Gauss-Lobatto scheme. All the remaining residual and stiffness contributions are integrated on the basis of a 4-point Gauss-Legendre (full) integration scheme. The reduced Gauss-Lobatto integration scheme leads to a number of $n_{eq,c}\!=\!3(2n_{ele}+1)$ constraint equations for a discretization with $n_{ele}$ finite elements providing a total of $n_{eq}\!=\!12n_{ele}+9$ unknowns / equilibrium equations. Similar to the procedure in~\cite{meier2016}, it can be shown that this procedure leads to a discrete constraint ratio $r_h$ that equals the constraint ratio of the space continuous problem $r$, i.e. $r_h\!=\!r\!=\!2$, and consequently no locking effects are expected for this element formulation. Furthermore, since the proposed $C^1$-continous Simo-Reissner element formulation combines two interpolation schemes already investigated in~\cite{meier2016}, namely the Hermite centerline interpolation~\eqref{interpolation} and the rotation interpolation~\eqref{lambda_reissner_anhang4}, it is straight-forward to show that the applied spatial finite element discretization fulfills essential properties such as objectivity, path-independence as well as conservation of linear and angular momentum.

\bibliography{special_issue_durville.bib}

\end{document}